\documentclass[manuscript]{aastex}
\usepackage{amsmath}
\usepackage{amssymb}
\usepackage{longtable}
\usepackage[flushleft]{threeparttable}
\usepackage{rotating}
\usepackage{rotating,booktabs}
\usepackage{longtable}

\begin{document}

\title{On the incidence of \textit{WISE} infrared excess among solar analog, twin and sibling stars}

\author{ A. D. Costa\altaffilmark{1}, B. L. Canto Martins\altaffilmark{1}, I. C. Le\~ao\altaffilmark{2}, J. E. Lima Jr\altaffilmark{1}, D. Freire da Silva \altaffilmark{1}, D. B. de Freitas\altaffilmark{1}, J. R. De Medeiros\altaffilmark{1}}

\affil{\altaffilmark{1}Departamento de Fí­sica Teórica e Experimental, Universidade Federal do Rio Grande do Norte,
    Campus Universitário, Natal, RN, Brazil, 59072-970.}
\affil{\altaffilmark{2}European Southern Observatory, Karl-Schwarzschild-Str. 2, 85748 Garching, Germany.}

\email{dgerson@fisica.ufrn.br}

\begin{abstract}

This study presents a search for IR excess in the
3.4, 4.6, 12 and 22 $\mu$m bands in a sample of 216 targets, composed of solar sibling, twin and analog stars observed by the
\textit{WISE} mission. In general, an infrared excess suggests the existence of warm dust around a star.
We detected 12 $\mu$m and/or 22 $\mu$m excesses at the 3$\sigma$ level of confidence in five solar analog stars, corresponding to a frequency of 
4.1 $\%$ of the entire sample of solar analogs analyzed, and in one out of 29 solar sibling candidates,
confirming previous studies.
The estimation of the dust properties shows that the sources with infrared excesses possess circumstellar
material with temperatures that, within the uncertainties, are similar to that of the material found in the asteroid belt in our solar system.
No photospheric flux excess was identified at the W1 (3.4 $\mu$m) and W2
(4.6 $\mu$m) \textit{WISE} bands, indicating that, in the majority of stars of the present sample, no detectable dust is generated.
Interestingly, among the sixty solar twin stars analyzed in this work, no \textit{WISE} photospheric flux excess was detected. However, a null-detection excess does not necessarily indicate the absence of dust around a star because different causes, including dynamic processes and instrument limitations, can mask its presence.
\end{abstract}


\section{INTRODUCTION}

The search for stellar infrared excesses may offer important constraints 
for our understanding of the nature and evolution of circumstellar dust disks, 
which are the most clear sign of other planetary systems. These structures 
may be indicative of perturbing forces, revealing the presence of
planets that would otherwise remain undetected, 
or of other influences, including the presence of remnant gas, that may
sculpt the starlight-scattering materials in these systems into ring-like morphologies (e.g.: Aumann et al. 1984; 
Zuckerman \& Song 2004; Lagrange et al. 2009; Chen et al. 2009; Melis et al. 2013; Chen et al. 2014; Vican \& Schneider 2014; 
Su \& Rieke 2014; Meshkat et al. 2015). For instance, in the solar system zodiacal cloud, Earth has cleared out a region in its vicinity as a result of resonant tidal interactions (Dermott et al. 1994). The current literature reports debris disks composed of belts of rocks and dust around hundreds of 
solar-type stars (e.g., Aumann et al. 1984; Oudmaijer et al. 1992; Mannings and Barlow 1998; Chen et al. 2006; Cruz-Saenz de Miera et al. 2014; Patel et al. 2014; Chen et al. 2014), with some studies indicating that planets may be frequent in debris disks (e.g.: Morales et al. 2011; Ballering et al. 2013).  
Among these stars, a few dozen are also known to harbor planets (Lawler and Gladman 2012; Morales et al. 2012; Bonsor et al. 2013).

Infrared (IR) excess in main-sequence Sun-like stars is believed to result from the production of collisional dust during the final stages of planet formation, at least for relatively young stars, or produced around older stars as long as dust is liberated in higher-speed collisions (Wyatt 2008; Krivov 2010). The incidence of debris disks around main-sequence stars of spectral types A, F, G and K, based on the detection of IR excess, is reported by different authors (Habing et al. 2001; Rieke et al. 2005; Bryden et al. 2006; Chen et al. 2006; Su et al. 2006; Wyatt et al. 2007; Trilling et al. 2008; Meyer et al. 2008; Urban et al. 2012; Montesinos et al. 2016). For instance,  about 20\% of the nearby Sun-like stars in the referred spectral range host dusty disks above the current detection limits (Habing et al. 2001; Trilling et al. 2008; G\' asp\' ar et al. 2013; Eiroa et al. 2013). Among the stars with detected discs, approximately 10\% have ages from10 Myr to 1 Gyr (Chen et al. 2006; Bryden et al. 2006; Meyer et al. 2008), with a clear disappearance of  disks among stars older than 300 to 400 Myr (Habing et al. 1999; Wyatt 2008). More recently, Sierchio et al. (2014) have shown that among solar-type stars within the spectral type range F4 to K2, about 13\% of the stars younger than 5 Gyr have dust disks, while stars most older than 5 Gyr do not.

The Wide-field Infrared Survey Explorer, \textit{WISE}, 
(Wright et al. 2010), which made observations centered at wavelengths of 3.4, 4.6, 12 and 22 
$\mu$m (known as the W1, W2, W3 and W4 bands, respectively), offers a unique laboratory to search for 
mid-IR excess in different stellar families. The 3.4-4.6 $\mu$m 
wavelength range is a good diagnostic of the presence of a near-IR 
excess, whereas the 12-22 $\mu$m range is an indicator of the presence of 
cooler dust. Indeed, these latter wavelengths are very sensitive to thermal 
emission from sources at temperatures comparable to the Earth, approximately 
300 K, and to our asteroid belt and interior zodiacal cloud, approximately 
150-250 K. For instance, Lawler $\&$ Gladman (2012) analised the IR 
\textit{WISE} behavior for hundreds of \textit {Kepler} objects, 
including stars with confirmed planets and stars with planet candidates, 
and they identified 8 stars with mid-IR excesses. Morales et al. (2012) performed a analysis for a sample composed of 591 stars with confirmed planets, 
listed in the Extrasolar Planet Encyclopedia (Scheider et al. 2011), 
from which 9 stars revealed excess mid-IR emission. More recently, 
Cotten and Song (2016) presented a large census of infrared excess in main-sequence stars, 
amounting to approximately 1750 nearby and bright stars, most of which
were revealed 
for the first time by \textit{WISE} observations.

The primary aim of this study is to determine the incidence of mid-IR 
excess at wavelengths of 3.4, 4.6, 12, and 22 $\mu$m based on homogeneous 
procedures for analyses of the \textit{WISE} observations for a stellar sample composed of 216 solar sibling, twin and  analog stars selected from the literature. 
Solar siblings refer to stars that were born simultaneously with 
the Sun. By definition, these stars must have a solar chemical composition 
considering that they essentially came from the same gas cloud and, consequently, the age of the Sun. 
However, solar siblings do not need to present physical parameters, such
as effective temperature, mass, luminosity, surface gravity or
rotation, similar to those of the Sun. 
Solar twins refer to stars with high-resolution, high signal-to-noise ratio
(S/N) spectra that are closely identical from the spectrum of
the Sun,
regardless  of  their  origin (e.g.: Cayrel de Strobel 1996; Porto
de  Mello  \&  da  Silva 1997; Melendez \& Ramírez 2007; Ramírez et
al.2011). Solar analogs refer to those stars spectroscopically
similar to the Sun, which are known to have stellar properties close 
to solar values (e.g.: Cayrel de Strobel 1996).

The most straightforward approach for this study is to relate the properties of debris disks around 
the referred stars to identify regions similar in temperature to our 
solar system. We revisit the search for the incidence of debris disks in 216 main-sequence stars of our sample, some of which have infrared excess already reported in the literature, searching for \textit{WISE} infrared excess. For a solid control 
on the reliability of the infrared excess, we applied for each star a homogeneous 
diagnostic consisting of the identification of traces of IR excess in the color--color diagram, 
the determination of the spectral energy distributions and image inspections. 
The remainder of this paper is organized as follows. Section 2 presents the dataset 
used in our study. Section 3 discusses the method used to identify mid-IR 
excess and the criterion used to visually inspect the \textit{WISE} images. Section 4 provides the main results, and the conclusions are presented in Section 5.

\section{The \textit{WISE} data and working stellar sample}

\textit{WISE} is a NASA infrared-wavelength astronomical space telescope 
that was launched in December 2009 for an all sky survey in the mid-IR with very 
high sensitivity (Wright et al. 2010). \textit{WISE} has a greater sensitivity than other 
instruments that detect similar wavelengths. For example, its sensitivity 
corresponds to a factor that is approximately 1000 times greater than that 
obtained by the IRAS satellite (Neugebauer  et al. 1984) in the 12 and 25 $\mu$m bands. 
Compared to the COBE satellite (Boggess et al. 1992), which made observations at 3.3 and 4.7 
$\mu$m, \textit{WISE} is approximately 500,000 times more sensitive in the 3.4 and 4.6 $\mu$m bands.

The present study is focused on searching for \textit{WISE} infrared excess in 
 so-called solar siblings, solar twins and solar analog stars. For this purpose, we have attempted to 
identify the largest set of stars defined as such in the literature. First, we selected 179 solar analogs 
listed by Datson et al. (2015), Chen et al. (2009) and Ramirez et al.(2012); 
117 twins from Ramirez et al. (2014), Datson et al. (2015), Pasquini et al. (2008), Yana Galarza et al. (2016) and 
Meléndez et al. (2006); 6 siblings from Ramirez 
et al. (2014), Liu et al. (2015) and Batista et al. (2014); and 30 siblings 
candidates from Liu et al. (2015). These selections correspond to a total preliminary 
sample of 334 stars, containing positional and photometric information, including J, H, Ks bands, from 2MASS. 
Nevertheless, for our study, we cross-correlated the selected sample by using 2MASS coordinates with
the \textit{WISE} all-sky data catalog, considering only those \textit{WISE} sources located within 
a two arcsecond radius for which the signal-to-noise ratio is 
larger than 3 in all \textit{WISE} bands, excluding stars with 
photometry corresponding to upper limits, as well as those with
saturated fluxes. As a check of the level of saturation, for each
source, we  compared  the \textit{WISE} magnitudes with the saturation
thresholds defined by the \textit{AllWISE} data release (Cutri et
al. 2013), according to which for sources with brightness larger than
approximately 2.0, 1.5, -3.0 and -4.0 mag in W1, W2, W3 and W4,
respectively,  the reliability and completeness of WISE photometric
measurements degrade because there are too few non-saturated pixels
available in the measurement area for reliable source
extraction. Based on these criteria, the final sample consists of 121 solar analogs, 6 solar siblings, 29 solar sibling candidates, and 60 solar twins, amounting to 216 targets, composed of stars with metallicity between -0.50 and 0.45, as shown in Fig. 1. This figure shows that the peak of the distribution is located around the solar values. Based on the definition by Cutri et al. (2013), none of the stars in our final sample have saturation levels higher than acceptable, and therefore, we can consider that the  observed fluxes for our sample are not overestimated. The final sample of 216 stars is listed in Table 2. This table also lists the values of the apparent  
magnitude V, the spectral type, effective temperature, surface gravity, 
metallicity and other parameters associated with the IR 
excess measurements, which will be defined in the next section. Fig. 2 shows the locations of 206 stars 
in the log (g) vs. log (T$_{eff}$) diagram, which shows that
their masses, surface gravities and effective 
temperatures range from approximately 1.0 to 2.0 $M_{\odot}$, 3.24 to 
4.8 dex and 4300 to 6900 K, respectively. These values are compatible 
with stars located in the main sequence stage. From the final sample of 216 stars, the 10 stars from Pasquini et al. (2008) are not represented in this figure because they do not have measured log (g) values.

\subsection{\textit{WISE} data analysis}

The first diagnostic in the search for traces of IR excess was
  the analysis of the distribution of our final stellar working sample
  of 216 stars in the color-color diagram, $J-H$ versus $K-[22]$, derived from 2MASS J, H, K, and {\it WISE} 22 $\mu$m magnitudes, to apply the criteria established by Wu et
  al. (2013). According to these authors, stars with a tendency to show IR excess at 22 $\mu$m  
present $K-[22]$ values greater than $\sim$ $0.22$ (dashed vertical black line in Fig. 3). The referred color-color diagram is then shown in Fig. 3, from which one observes that the large majority of stars in our sample show normal IR behavior, 
presenting essentially photospheric colors, i.e., K-[22] $<$ $\sim$ 0.2 are identified. However, a total of 13 stars show IR excess in the W4 band. These stars are the analogs HD 39060, HD 86087, HD 109573, HD 113766, HD 168746, HD 181296, HD 218396 and HD 224448; the twins HD 63487, HD 145927 and HD 150248; and the siblings HD 21216 and HD 168325. The 2MASS and \textit{WISE} magnitudes used in Fig. 3 were obtained from the All\textit{WISE} Data Release (Cutri et al. 2013).

To verify that the IR excess that emerges from the 
color-color diagram is actually real, we compared the observed and 
model-derived photospheric IR fluxes for each of the 216 stars presented in the diagram using the Virtual Observatory Spectral Analyzer (VOSA, 
Bayo et al. 2008). Indeed, using the VOSA procedure, we calculated the synthetic photometry in a given 
filter set and then performed a $\chi^{2}$ minimization to determine the best 
fit to the data. Kurucz models (Castelli 
et al. 1997) were used for the fit between the observed spectral energy distributions
(SEDs) and the synthetic photometry. In this procedure, we  used  the 
following ranges of stellar parameters as the input parameters: 4000 $\leq$ $T_{eff}$ $\leq$ 7000 K, 0.0 $\leq$ log 
$g$ $\leq$ 5.0 and -2.5 $\leq$ [M/H] $\leq$ 0.5,  covering the 
range of the estimated parameters for our targets.

Measurements of the \textit{WISE} infrared excess for each source were made 
using the excess significance $\chi_{\lambda}$ (Beichman et al. 2006), which can be defined as

\begin{equation}
     \chi_{\lambda} \equiv 
\frac{F^{obs}_{\lambda}-F^{phot}_{\lambda}}{\sqrt{{\sigma_{obs}}^2+{\sigma_{cal}}^2}},
\end{equation}

where $F^{obs}_{\lambda}$ is the observed flux value in the [3.4], [4.6], [12] or [22] 
bands; $F^{phot}_{\lambda}$ is the theoretical photospheric flux value at the same 
wavelength computed from the photospheric modeling; $\sigma_{obs}$ 
represents the errors at $F^{obs}_{\lambda}$; and $\sigma_{cal}$ are the absolute 
calibration uncertainties of the \textit{WISE} estimated at 2.4, 2.8, 4.5 
and 5.7 percent in the [3.4], [4.5], [12] and [22] bands, 
respectively (see the Explanatory Supplement to the \textit{WISE} 
Preliminary Data Release Products by Jarret et al. (2011)). The errors in 
the photospheric theoretical fluxes corresponding to the fit
performed using VOSA are 
considered insignificant and were not used (Ribas et al. 2012). The distribution histograms of $\chi_\lambda$ values for the four \textit{WISE} bands are shown in Fig. 
4. The means and standard deviations from  Gaussian fits (red solid 
curves) are 0.23 and 0.45, 0.72 and 0.94, -0.38 and 0.23, and 0.86 and 
0.53 for the W1, W2, W3 and W4 bands, respectively. The negative
average of $\chi_{\lambda}$ for the W3 band reflects the larger
passband of W3 than the others (Wright et al. 2010; Cotten and Song
2016). Indeed, Wright et al. (2010) describes the inflight discrepancy
found between red and blue sources that implies that the coolest stars
will have an observed W3 flux that is fainter than the theoretical photospheric flux value, that is, a negative significance of excess.

We consider sources with an apparent significant excess as those for which 
$\chi_{\lambda}$ $\geq$ 3.0 (Su et al. 2006) in one or more of 3.4, 4.6, 12, or 22 $\mu$m  bands, which corresponds to
at least 3 $\sigma$ significance of deviation from the expected photospheric
value ($\chi_{\lambda}$ = 0).
By applying this criterion, we find that 14 of the 216 stars present infrared excess in one or more \textit{WISE} 
bands: At  W4,  HD 21216, HD 39060, HD 86087, HD 109573, HD 113766, HD
168325, HD 168746, HD 181296 and HD 218396, and all of these stars
also show traces of infrared excess based on the criteria of Wu et al. (2013); at W3,  HD 39060, HD 109573 and HD 113766; at  W2,  HD 11131, HD 96423, HD 113766 and HD 150248; and  at W1,  HD 6470 and HD 168746.

The stars with an apparent excess from the SEDs at W1, HD 6470
  and HD 168746, and at W2, HD 11131, HD 96423 and  HD 150248, were disregarded from
  our sample of sources considered to have an acceptable excess
  level. Indeed, these stars present flux ratios
  ($F^{obs}_{12}/F^{phot}_{12}$) close to 1, which are very low
  compared to those presented by stars that exhibit significant
  infrared excesses at 22 $ \mu$m, which are approximately 1.8. In the
  W2 band, the stars present an excess ratio that is more important, but according to the \textit{WISE} Team (Cutri et al. 2012), there is an overestimation in the brightness of 
the 4.6 $\mu$m band, and the bias can reach nearly 1 mag. Stars with apparent excess in bands W1 and W2 exhibit saturated pixels fractions ranging from 8\% to about 50\%, significantly higher than the saturated level of stars with detections in W3 and W4 which presents a range level of saturated pixels of 0.0 to 0.8\%. Moreover, the overestimation of W2 magnitudes applies at most to saturated objects. The overestimation by up to 1 mag applies to objects with W2 $\sim$ 3 mag, and it is smaller for stars with $3.5 < W2 < 6.5$ magnitudes. For instance, the 5 stars in our sample showing an apparent excess in W1 (HD 6470, HD 168746) and W2 (HD 11131, HD 96423, HD 150248) have W2 magnitudes between 4.8 and 7.4.

For the stars HD 150248, HD 63487, HD 145927  and HD 22448, their locations in the 
color-color diagram (see Fig. 3) also suggest a slight IR excess in the 22 
$\mu$m band. Despite this result, the estimated values of
  $\chi_{22}$ for HD 150248, HD 63487, HD 145927  and HD 22448 of 0.373, 1.090, 2.148 and  2.023, respectively, do not show
confidence levels higher than 3$\sigma$, which represents the acceptable level of IR excess in our study. The SEDs for the referred stars whose IR excess were disregarded are shown in Fig. 9, presented in the online material.

\subsection{\textit{WISE} image inspection}

Because different factors such as artifacts, background galaxies and other 
fundamental problems can contaminate the IR excesses observed in stars, we have applied a third diagnostic to define which stars with IR excess traces revealed from the color-color diagram and from the SEDs have a reliable \textit{WISE} IR excess.  
In this sense, we inspected the {\it WISE} images for all the stars, which were 
downloaded from the IRSA\footnote{http://irsa.ipac.caltech.edu/applications/wise/}  archive, for 
the 1'.4 $\times$ 1'.4 regions surrounding each star in all \textit{WISE} bands. The inspection is  based on the following criteria:

\begin{itemize}
\item PSF (Point Spread Function): Check whether the photocenter of the band 
in which the excess is found appears to be a bona fide point source or an 
extended PSF.
\item Offsets: Check whether the photocenter position changes in the 
band in which excess is found compared to its position in all \textit{WISE} bands.
\item Condition confusion flag (ccf): Check whether the photometry and/or 
position of a source may be contaminated or biased due to 
image artifact (Cutri et al. 2013). The available artifact flags are diffraction spikes (orange dots), scattered light halos (green squares) and optical  ghosts (pink diamonds), as described by IRSA.
\end{itemize}

Figure 5 shows some examples of stars with excesses in the 22 $\mu$m band as well as fundamental problems. A visual inspection shows that 
HD 168325 is a spurious detection in which no 
object is evident in the W3 band (middle left panel) and W4 band (middle right panel). This source suggests dust emission 
when viewed through the SEDs, but the visual analysis of the WISE image 
shows a lack of a point source in the W4 band. 

Another star that presents fundamental problems in the image is HD 
109573. Although it is a point source and has no indications of 
contamination by the background, the excess may be due to contamination by a diffraction spike from a nearby bright star on the same 
image in 12 the $\mu$m band (top left panel) and by a scattered light halo 
surrounding a nearby bright source in the 22 $\mu$m band (top right panel). 
Although HD 168746 (bottom panel) can be associated with a point source, 
this star displays a substantial contribution from the background in the 
W4 band (bottom right panel), which may be related to the fact that this star is located near a 
stellar formation region. In this context, the objects with problems in the images, HD 168325, HD 109573 and HD 168746, were rejected from our final list of  
excess candidates. Their SEDs are show in Fig.9, presented in the online material.

For the remaining stars, HD 21216, HD 39060, HD 86087, HD 113766, HD 181296 and HD 
218396, the image examination indicates that such sources 
are clear detections of single isolated PSF sources without source 
confusion, and the centroid position is preserved in each image. The four \textit{WISE} bands (from left 
to right, W1, W2, W3, and W4) for these stars are shown in Fig. 6.

\subsection{Solar twins from M67 open cluster}

An additional step in our search for \textit{WISE} IR excess among solar twin stars was dedicated to the open cluster M67, one of the most important laboratory for studying stellar evolution (Burstein et al. 1986; Carraro et al. 1996). Commonly mentioned as a solar age cluster, with chemical composition very close to the solar values, M67 hosts 10 potential solar twins identified by Pasquini et al. (2008). These stars are  MMJ5484, 
MMJ6055, MMJ6384,  S770, S779, S785, S945, S966, S1041 and S1462, following the nomenclature used by the referred authors. All these 
stars present an apparent excess in the color-color diagram presented
in Fig. 3, but one should consider such behavior with caution because
\textit{WISE} data show an upper limit for the magnitude K-[22], with a
S/N ratio lower than 3 in the referred \textit{WISE}
band. The values of $\chi_\lambda$, in the various {\it WISE} bands, for the M67 sample of stars are also listed in Tab. 2. By applying as a diagnostic the analysis in the \textit{WISE} bands W1, W2 and W3 based on their SEDs, we find an IR excess in the W3 band, namely, in 12 $\mu$m, for the star S966. Nevertheless, the significance of the excess at 12 $\mu$m for S966 is relatively low ($\chi_{12}=3.8$). No traces of excess were observed from the SEDs of the remaining nine stars, MMJ5484, 
MMJ6055, MMJ6384, S770, S779, S785, S945, S1041 and S1462, which show essentially photospheric colors. The SEDs for the stars from M67 also are shown in the online material.
 Unluckily, a visual image inspection shows that 
S966  is a spurious detection. The visual analysis of the WISE images shows a lack of a point source in the W3 band, with a complex background characterizing the  
image in the referred band. Therefore, the apparent IR excess of the solar twin S966,  emerging from the  \textit{WISE} W3 band, is indeed artificial. Therefore, we conclude that none of the solar twins stars in the M67 cluster have a detectable IR excess in the {\it WISE} bands.

\section{Results and Discussion}
\label {Results}

Because IR excess reveals the presence of a circumstellar debris 
disk, we fitted the observed reliable excess for the stars HD 21216,
HD 39060, HD 86087, HD 113766, HD 181296 and HD 218396 with a
black body function (blue dashed lines in Fig. 7) to determine the color temperature, which is
defined here as the disk temperature $T_{d}$. In the debris disk modeling,  it is assumed that an optically
thin dust is in thermal equilibrium with the stellar radiation
field. Based on such a condition, the temperature of a dust grain with a defined chemical composition and size will depend only on the radial distance to the central star. The corresponding SEDs and the referred fits are presented in Fig. 7.

We estimated three fundamental disk properties,  using equations 2 (Beichman et al. 2005), 3 ({Backman \& Paresce 1993}) and 4 (Liu et al. 2014): the luminosity fraction $f_{d} = L_{IR}/L_{*}$, which is
defined as the ratio of infrared luminosity from the dust to the 
stellar luminosity; the orbital distance of the dust or disk radius $R_{d}$;
and the disk mass $M_{d}$. In these equations, $T_{*}$ is the effective 
temperature of the star, $L_{*}$ is the stellar luminosity, F$^{d}_{\lambda}$= F$^{obs}_{\lambda}$ - F$^{phot}_{\lambda}$ and $\frac{F^{d}_{\lambda}}{F^{phot}_{\lambda}}$
are the dust flux  and fractional excess at 12 or 22 $\mu$m, respectively.

\begin{equation}
     f_{d} = 
\frac{L_{IR}}{L_{*}}=kT_{d}^{4}\frac{\left(\exp^{h\nu/kT_{d}}-1\right)}{h\nu 
T_{*}^{3}}\frac{F^{d}_{\lambda}}{F^{phot}_{\lambda}}
\end{equation}

\begin{equation}
     R_{d} 
=\left(\frac{278}{T_{d}}\right)^{2}\left(\frac{L_{*}}{L_{\odot}}\right)^{0.5}
\end{equation}

\begin{equation}
     M_{d} = f_{d}(R_{d}/9.12)M_{\oplus}
\end{equation} 

From this modeling, our small sample of solar analog stars with 
\textit{WISE} IR excesses shows warm circumstellar material with disk temperatures 
$T_{d}$ within the range of 150 to 270 K, luminosity fractions $f_{d}$ 
from about $0.85\times10^{-4}$ to  $199\times10^{-4}$, disk radii $R_{d}$ between 2.54 and 19.43 AU and disk masses from $1.11\times10^{-4}$ to $4.23\times10^{-2}$ $M_{\oplus}$, which also 
agree with the overall results found in the literature (e.g.: Chen et al. 2006; Cruz-Saenz de Miera et al. 2013; Chen et al. 2014). See table 1 for derived disk properties.

The stars identified to have reliable \textit{WISE} IR excesses, i.e., HD 21216, HD 39060, HD 86087, HD 113766, HD 181296 and HD 218396, 
were previously studied by other authors by also using  \textit{WISE} data. 

For HD 86087 and HD 113766, Chen et al. (2014) found a clear IR excess
from \textit{WISE} as well as from Spitzer/IRS observations.  For HD 86087, we computed a \textit{WISE} IR excess significance $\chi_{22}$ of 9.89 and fractional excess of 1.55. We also estimated a dust temperature of 175 K, luminosity 
fraction of approximately 1.77 $\times$ $10^{-4}$, orbital distance of 19.43 AU 
and disk mass of $M_{d}=7.3 \times 10^{-3}$ $M_{\oplus}$. These characteristics  
suggest that this star is surrounded by warm circumstellar material with a
temperature similar to our asteroid belt but located at a distance from the 
Sun greater than 7 times the distance of our belt. Based on the Spitzer spectrum 
in the range 5.5 - 35 $\mu$m, Chen et al. (2014) identified two dust
components with dust temperatures of 381 K and 91 K, luminosity
fractions of approximately 3.7 and 9.5 $\times$ $10^{-5}$, orbital distances of 3.9 and 84.8 AU 
and disk masses of M$_{d}$ = 2.7 $\times$ 10$^{-4}$ M$_{moon}$ and 1.6 $\times$ 10$^{-4}$ M$_{moon}$.

For HD 113766, we computed \textit{WISE} IR excess significance in two
bands, with $\chi_{12}$ = 20.0 and $\chi_{22}$ = 17.0 and fractional
excesses of 11.42 and 63.11 for W3 and W4, respectively. The thermal emission from this object 
was well modeled with a single black body component, with a estimated dust temperature of 270 K, luminosity fraction of 199 $\times$ $10^{-4}$, orbital distance of 4.4 AU, and disk mass of  $M_{d}=4.23 \times 10^{-2}$ $M_{\oplus}$. The estimated dust temperature and disk radius from the present analyses are 
in agreement with the results obtained by Chen et al. (2006). Indeed,
this is an expected result considering that the wavelengths observed
by WISE (4.6 - 22 $\mu$m) are within the range of Spitzer observations (5.5 - 35 $\mu$m).

For HD 39060, Morales et al. (2012) identified warm dust emission at
22 $\mu$m, estimating a dust temperature of approximately 199 K and a fractional excess of 22.8. 
For the stars , HD 21216, HD 181296 and HD 218396, Patel et al. (2014) identified significant levels of \textit{WISE} 
IR excess in the 12 and 22 $\mu$m bands and estimated disk properties. The dust temperatures, 
disk radii and luminosity fractions are  167 K, 3.7 AU and 4.1$\times$ $10^{-4}$, respectively;
for HD 21216; 177 K, 11.0 AU and 2.5$\times$ $10^{-4}$ for HD 181296 and  225 K, 3.3 AU and 0.6 $\times$
$10^{-4}$, for HD 218396. 
For these four stars, we estimated dust temperatures and disk radii
that are in agreement with those reported by these authors, as shown 
in Table 1. The disk masses listed in Table 1 (10$^{-4}$ - 10$^{-2}$ M$_{\oplus}$), estimated from the observed \textit{WISE} IR 
excesses at 12 and/or 22 $\mu$m, are reported for the first time for the considered stars.
In general, for these stars, our results confirm and reinforce the 
indication of the presence of the previously detected circumstellar material with grain temperatures comparable to those of the grains in our asteroid belt and the interior 
zodiacal zone (Lawler \& Gladman 2012; Morales et al. 2012).

\subsection{On the dearth of \textit{WISE} mid-IR excess among solar twin stars}

Inspired by the reported incidence of debris disks around Sun-like stars, revealed by IR excess detections among F , G and K stars (Rieke et al. 2005; Bryden et al. 2006; Chen et al. 2006; Su et al. 2006; Wyatt et al. 2007; Meyer et al. 2008; Urban et al. 2012), we have searched in detail for traces of WISE excess emission among solar twin stars. Also, it is clear from different studies underlined in section 1, that the likelihood of a detectable debris disk depends strongly on stellar age, with a higher percentage of young stars harboring disks than older ones. For comparison purposes, as underlined in Sec. 1, approximately 10\% of Sun-like stars,  with ages from 10 Myr to 1 Gyr, have IR excesses reported in the literature (Chen et al. 2006; Bryden et al. 2006; Meyer et al. 2008). Nevertheless, the incidence of IR excess become sparse among stars older than 300 to 400 Myr (Habing et al. 1999; Wyatt 2008).

Our study shows that among the solar twin stars here analyzed, none have a statistically significant WISE IR excess compared to the predicted stellar photospheric flux. Indeed, the null detection of debris disks around solar twins stars can reflect different root-causes, including that dust disks around the referred stars become so optically thin to be undetectable, at their age, cold disks, which are detectable only at longer wavelengths, as well as disk disappearance due to stellar ages. At this point, the analyses of  the age distribution for our sample of solar twin stars it is mandatory. Fig. 8 shows the age distribution for our sample of 60 solar twin stars, with ages taken from Yana Galarza et al. (2016), Ramirez et al. (2014), and Melendez et al. (2006); for stars from Datson et al. (2015) and Pasquini et al. (2008), age were taken from Holmberg et al. (2007) and \"Onehag et al. (2011), respectively. From the range of ages shown in Fig. 8, one observes that the large majority of stars are older than the expected age interval (300 to 400 Myr) for disks disappearance (Habing et al. 1999, Wyatt 2008).  In this context, the dearth of mid-IR excess among solar twin stars is rather in agreement with the scenario expected for main-sequence Sun-like stars in the same age range, pointing for a possible debris disk disappearance.

\section{Conclusions}
\label{Conclus}

In this study, we analyzed the \textit{WISE} emission flux properties of a 
sample of 216 main sequence stars, consisting of 121 solar analogs, 6 solar siblings, 29 solar sibling candidates, and 60 solar twins, in the search for infrared emission excess in four WISE bands (W1-3.4 $\mu$m, W2-4.6 $\mu$m, W3-12 $\mu$m, and W4-22 $\mu$m). Nine stars present excesses in the 12 and/or 22$\mu$m bands at the
$3\sigma$ level of confidence, but three were rejected after visual 
inspection of the WISE images because of fundamental problems 
(contamination  by artifacts, no evident source and/or background emission). \\


We confirm {\it WISE} excess in 12 and/or 22$\mu$m in the solar analogs HD 39060, HD 86087, HD 113766, HD 181296, and HD 218396, corresponding to 
4.1$\%$ of the analyzed sample, and in the solar sibling candidate HD 21216, 1 out of 29 solar siblings candidates. The estimations of the dust properties for these stars with IR excesses 
are consistent with those given in the literature. For HD 113766,
we confirm a \textit{WISE} excess with a dust temperature of 270$\,K$,
which is compatible with other studies. These values strengthen the fact that these stars present warm 
circumstellar material with temperatures that are similar to the asteroid belt and interior zodiacal zone. The fractional dust luminosity ranges from about 0.85 $\times10^{-4}$ to  199$\times10^{-4}$, with HD 113766 presenting the highest fractional luminosity in the sample. The orbital distances of the dust disks estimated in the present work range from 2.54 to 19.43 AU. Star HD 21216  show dust disk 
within the solar asteroid belt region, whereas the disks of HD 39060, HD 86087, HD 113766, HD 181296  and HD 218396  are outside the asteroid belt, reflecting probable differences in planetary systems. 
The present analyses reinforce that for the definition of IR excess, a visual inspection of the WISE 
images is mandatory to check the reliability of the mid-IR excess in the WISE bands because 
such excesses can be contaminated by artifacts or background emissions. For 
instance, the apparent IR excess associated with the stars S996,  HD 109573, HD 168325 and HD 168746 are due to the presence of artifacts.\\


Finally, some relevant points should be highlighted. No stars having
\textit{WISE} photospheric flux excess at the W1 (3.4 $\mu$m) and W2 (4.6
$\mu$m) bands were identified, indicating that, in principle, no detectable hot dust is generated in the present stellar
sample. The detection of a mid-IR excess revealed by \textit{WISE} W3
(12 $\mu$m) and W4 (22 $\mu$m) suggests the presence of circumstellar warm
material from primary dust generation. In this context, the
identification of debris disks around stars similar to the Sun is
extremely important because the circumstellar dust also represents a fundamental factor to show how special the Sun 
and the solar system are compared to other stars and other planetary
systems. Interestingly, among the 60 solar twins  analyzed here, no stars present \textit{WISE} IR excess at a 3$\sigma$ level 
of confidence. Such a finding may point to a very interesting
scenario: stars with physical parameters similar to the Sun, as is the
case of the solar twins, can in fact be very different from the Sun once the star and its circumstellar environment are considered.
However, a null-detection excess does not necessarily indicate the absence of dust around a star.  In this context, additional studies for the search of IR excess in other wavebands, including those by Spitzer and Herschel, are mandatory to confirm the present results about the dearth of debris disks around stars with physical parameters similar to the Sun.

\acknowledgments   Research activity of the Observational Astronomy Board of the Federal University of Rio Grande do Norte (UFRN) is supported by continuous grants from CNPq and FAPERN Brazilian agencies. We also acknowledge financial support from INCT INEspa\c{c}o/CNPq/MCT. ADC acknowledges a CAPES/PNPD fellowship. 
ICL acknowledges a CNPq/PDE fellowship. JEL and DFS acknowledge graduate fellowships from CAPES. This work is based on data products from the Wide-field Infrared Survey Explorer, a joint project of the University of California, Los Angeles, and the Jet Propulsion Laboratory/California Institute of Technology, supported by the National Aeronautics and Space Administration. This study has used the NASA’s Astrophysics Data System (ADS) Abstract Service, the SIMBAD database, operated at CDS, Strasbourg, France, and data products from the Two Micron All-Sky Survey (2MASS), a joint project of the University of Massachusetts and the Infrared Processing and Analysis Center, supported by the National Aeronautics and Space Administration and the National Science Foundation. This study has used the VOSA support, developed under the Spanish Virtual Observatory project funded by the Spanish MICINN through grant AyA2011-24052. Finally, a special thanks to the anonymous referee for providing very helpful comments and suggestions that improved largely this publication.

\newpage

\begin{sidewaystable}
\label{TableXRay}
\caption{\textit{WISE} flux and disc properties. }
\center
\scriptsize
\begin{tabular}{lccrrccrcccccc}
\hline \hline
~~~~Star & L$_{*}$  & Age  & J-H   & K- [22] & $\frac{F^{d}_{12}}{F^{phot}_{12}}$ & SNR$_{12}$  & $\frac{F^{d}_{22}}{F^{phot}_{22}}$ & SNR$_{22}$  & T$_{d}$ & f$_{d}$          & R$_{d}$ & M$_{d}$&Ref \\
     & L$_{\odot}$ & (Myr)& (mag) & (mag)   &    &   &     &   & (K)   & ($\times$ 10$^{-4}$) & (AU)  &  (M$_{\oplus}$)&\\

\hline
\multicolumn{14}{c}{Analogs} \\  
\hline

 HD 39060   & 9.1	 &$\sim{12}$	&0.125	& 3.512 & 1.11	&	165.8&	22.43 	&	69.3&150	&25.50	&10.36	& 3.00 $\times$ 10$^{-2}$& 1,6	\\
 HD 86087   & 59.3 & $\sim{50}$	&-0.063	& 1.180 &  -0.10	&	74.6	& 	1.55&	43.5 &175	&	1.77&19.43	& 7.33 $\times$ 10$^{-3}$&1,4	\\
 HD 113766  & 17.3 &	$\sim{16}$&0.131	& 4.737 &  11.42	&107.8	&63.11	& 87.7	&	270	&199	&4.40	& 4.23 $\times$ 10$^{-2}$&1,5	\\
 HD 181296  & 22.0 & $\sim{12}$	&-0.052	& 1.747 & -0.02 		&74.4	& 3.05	&	56.1 &150	&2.11	&16.11	&6.01 $\times$ 10$^{-3}$&2,6	 \\
 HD 218396  & 7.8 &$\sim{60}$	&0.103	& 0.387 & -0.03 	&69.8	&0.46	&30.6 	&	250	&0.85	&3.45	& 1.11 $\times$ 10$^{-4}$& 1,8	 \\

\hline 
\multicolumn{14}{c}{Siblings candidates}\\  
\hline
  HD 21216  &1.73 & $\sim{3.2}$ & 0.198  & 0.670	& -0.01 &  61.4	& 0.74	&	15.8 & 200 	&	1.94	&2.54	&1.37 $\times$ 10$^{-4}$&3,7	 \\
\hline                                                                        


\end{tabular}

\begin{tablenotes}
	Notes.  The reference to the stellar luminosity and age are shown in the last column: (1) Chen et al. (2014), (2) Wyatt et al. (2007), (3) McDonald et al. (2012), (4) Chen et al. (2006), (5) Mamajek et al. (2002), (6) Zuckman et al. (2001), (7) Liu et al. (2015) and (8) Marois et al. (2008). For calculate the luminosity fraction we use $\lambda$=12 $\mu$m (\textit{WISE} W3 band)  for HD 113766 and  $\lambda$=22 $\mu$m (\textit{WISE} W4 band) for HD 39060, HD 86087, HD 113766, HD 181296 and HD 218396 in the equation 2, where $\nu$= c/$\lambda$.
	
\end{tablenotes} 
\end{sidewaystable}

\clearpage
\begin{figure}[ht]
\epsscale{0.95}
\plotone{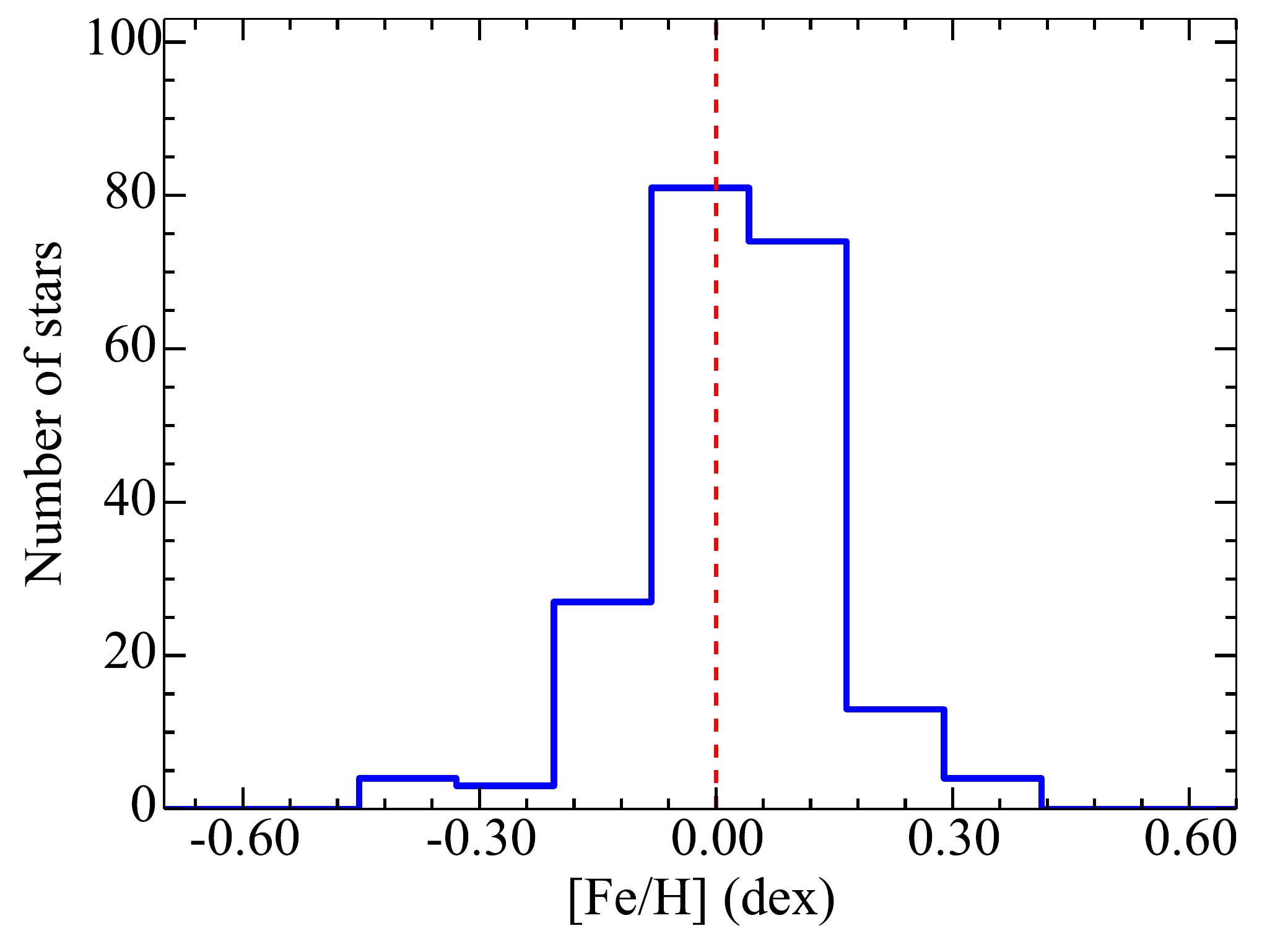}
\caption{Distribution of metallicity for the 206 Sun-like stars analyzed in this work. The 
canonical solar value is presented by the red dashed line.}
\label{}
\end{figure}

\begin{figure}[ht]
\epsscale{1.0}
\plotone{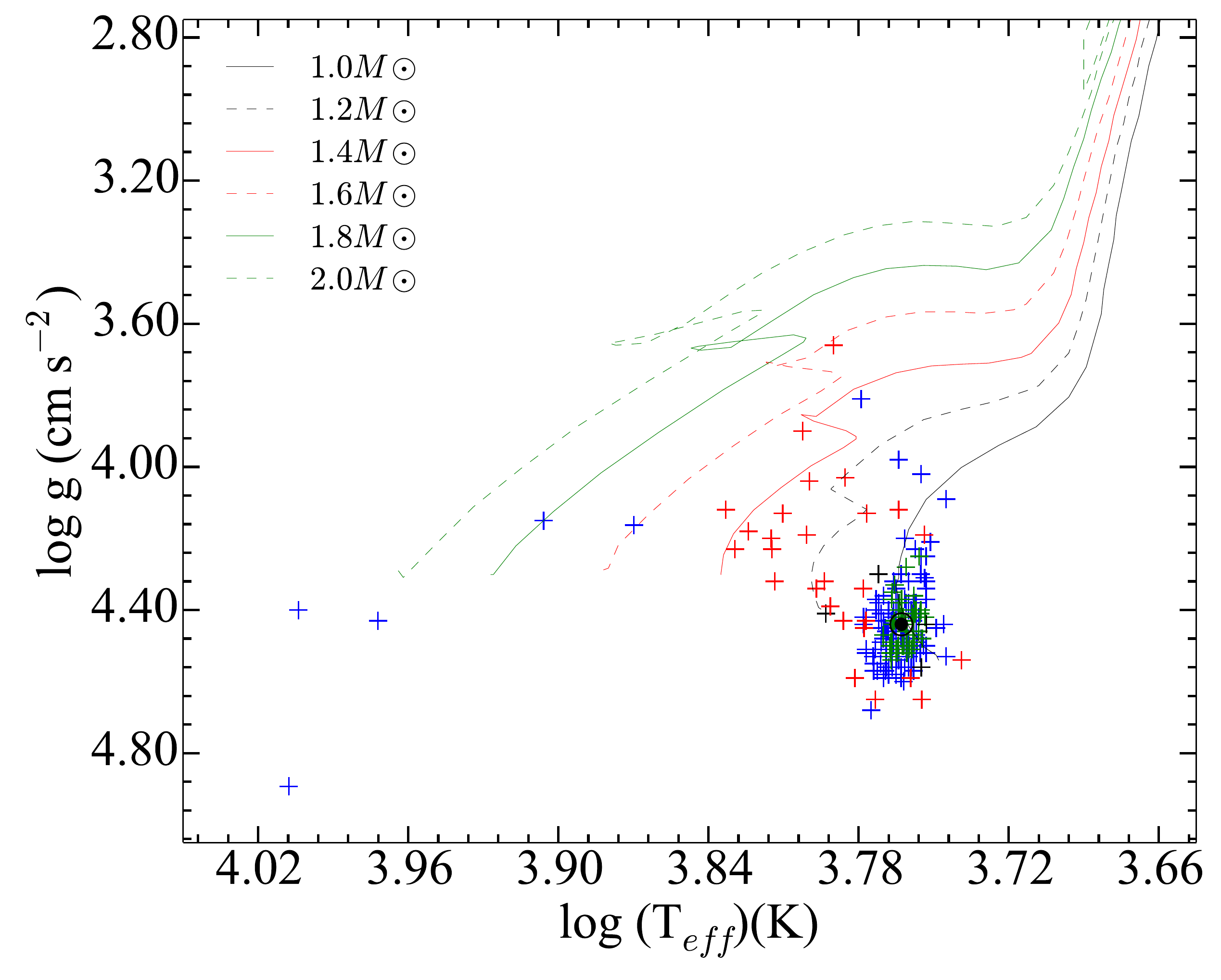}
\caption{Distribution of the 206 Sun-like stars from our sample in a log (g) vs. log (T$_{eff}$) diagram. The solid and dashed lines represent the evolutionary tracks for stars with [Fe/H] = 0 and masses ranging from 1.0 to 2.0 M$_\odot$ from Girardi et al. (2000). Stars classified as solar analogs, twins, siblings and sibling candidates are represented by blue, 
green, black and red symbols, respectively. The position of the Sun is represented by its usual symbol.}
\label{}
\end{figure}

\begin{figure}[t]
\epsscale{1.0}
\plotone{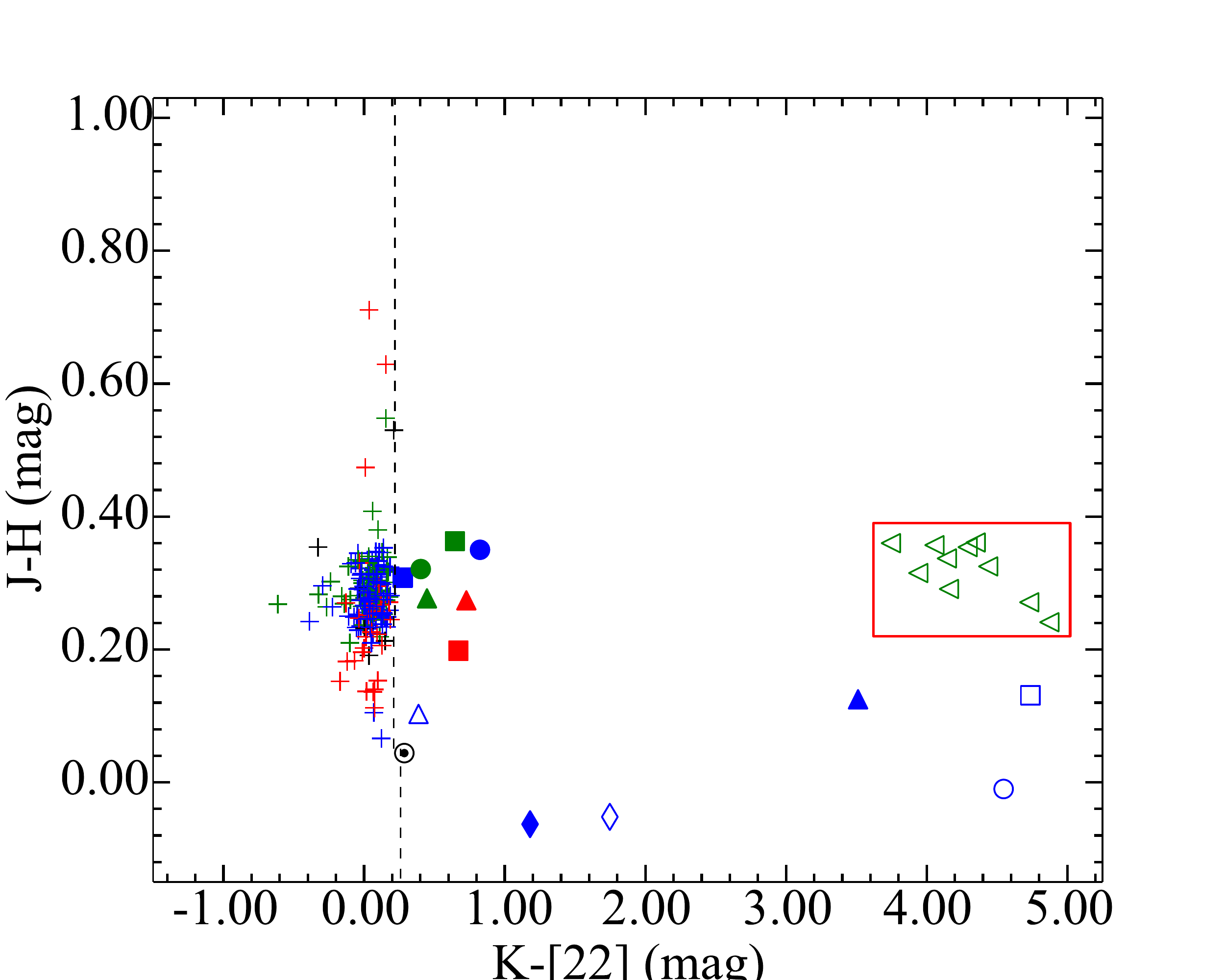}
\caption{Color-color diagram $J-H$ vs. $K-[22]$ for the 216 Sun-like 
stars that comprise our sample. The colors of the symbols are the same as in Fig. 2.  The different symbols represent different stars: red filled square, HD 21216; blue filled triangle, HD 39060; green filled triangle, HD 63487; blue filled diamond, HD 86087; blue circle, HD 109573; blue square, HD 113766; green filled square, HD 145927; green filled circle, HD 150248; red filled triangle HD 168325;  blue filled circle, HD 168746; blue diamond, HD 181296 and blue triangle, HD 218396 and blue filled square, HD 224448. The black dashed line shows the criterion to define \textit{WISE} 22 $\mu$m 
excess from Wu et al. (2013), according to which stellar excess candidate should populate the region with $K-[22]$ larger than approximately 0.2. Triangles in the red box represent stars from M67 and they represent upper limits in $K-[22]$.}
\label{hrcorcor}
\end{figure}

\begin{figure*}[ht]
\centering
\epsscale{1.0}
\plotone{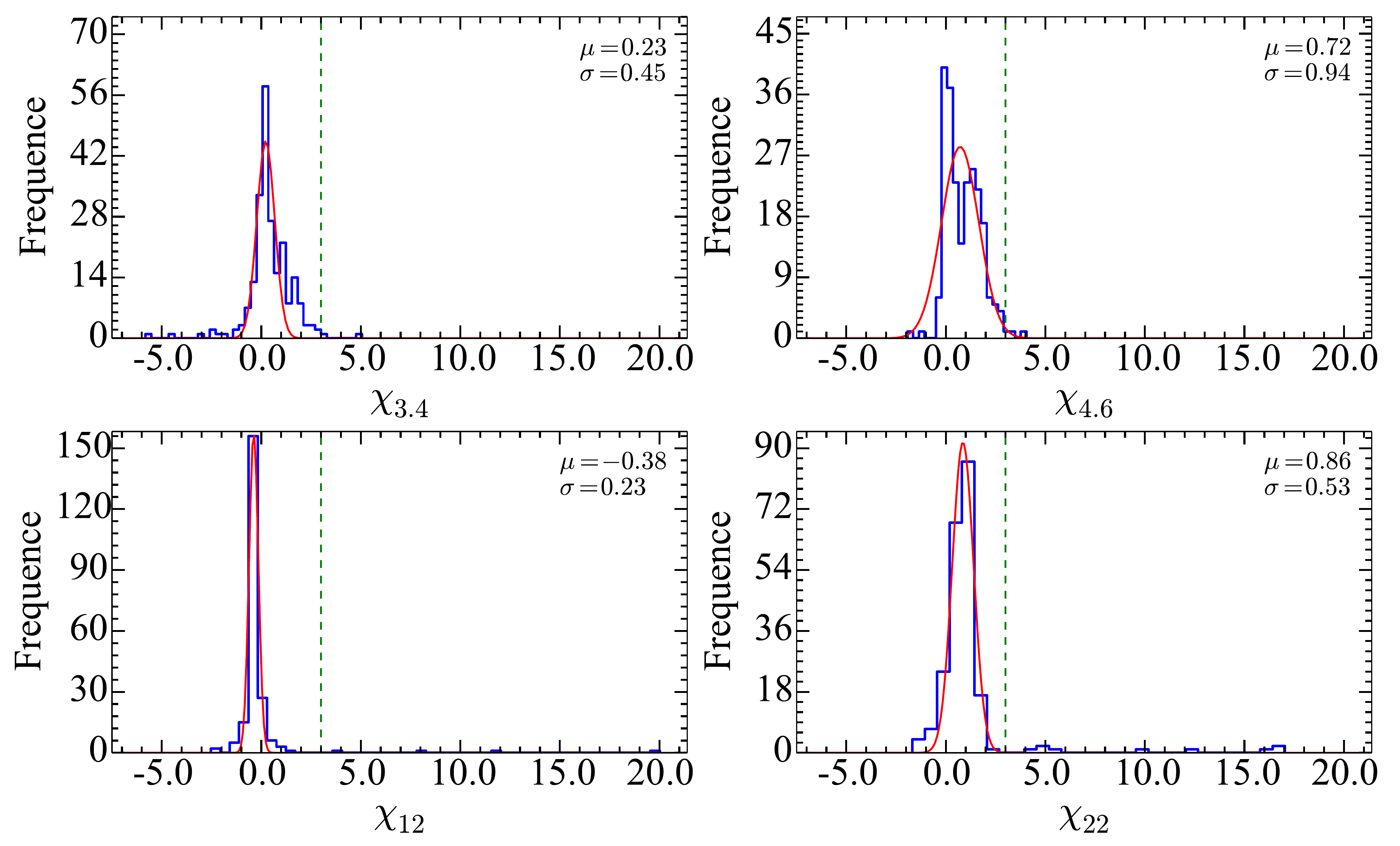}
\caption{Distribution of $\chi_{\lambda}$  values at 3.4 $\mu$m (top left 
panel), 4.6 $\mu$m (top right panel), 12 $\mu$m (bottom left panel) and 22 
$\mu$m (bottom right panel) for stars with S/N $\geq$ 3 at the 
respective wavelength. The green dashed line indicates an excess significance $\chi_{\lambda}$ = 3.0.}
\end{figure*}

\begin{figure}[ht]
\centering
\epsscale{0.65}
\plotone{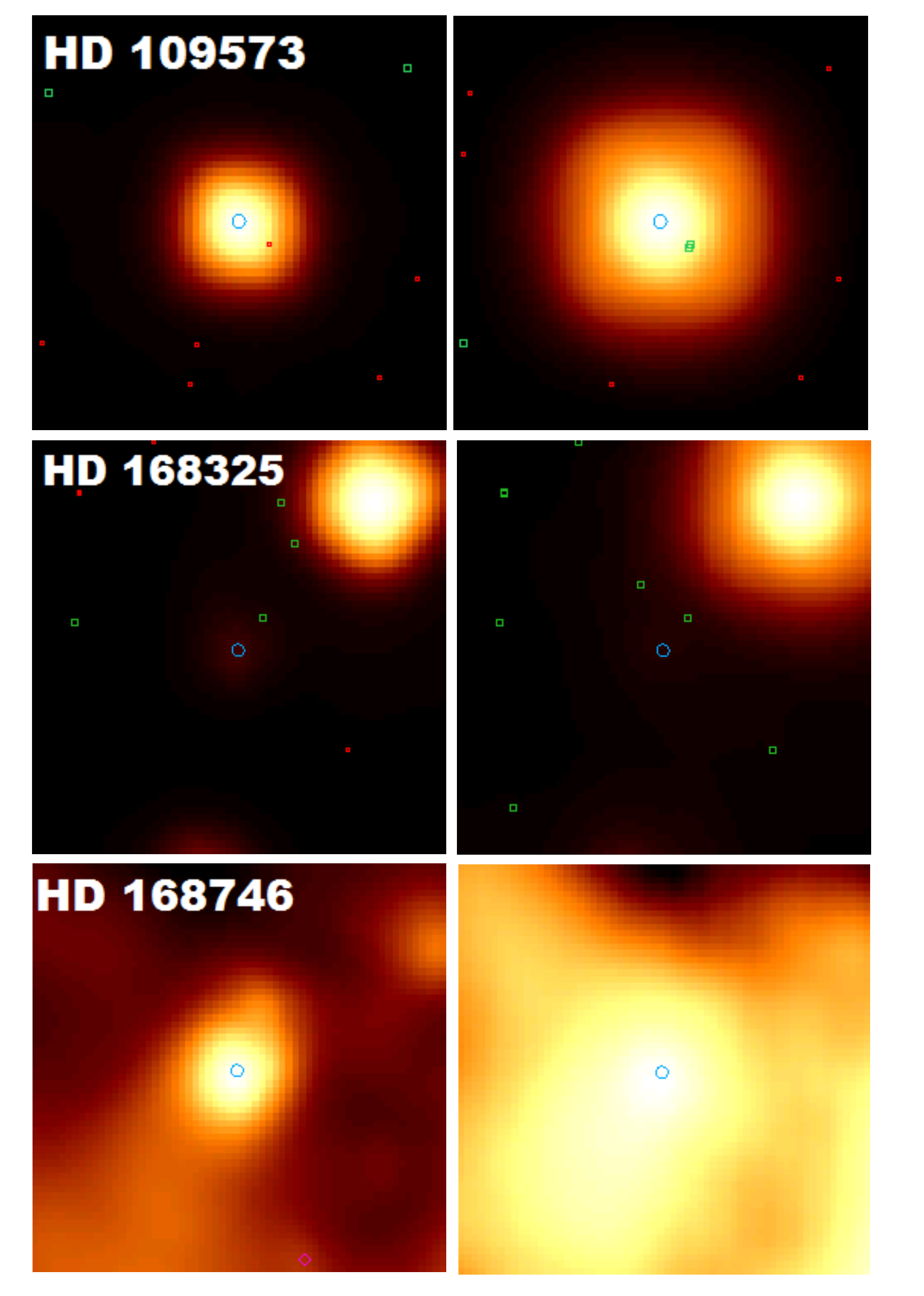}
\caption{\textit{WISE} images (from left to right W3 and W4) in min–max log scales for stars 
with IR excesses in the 12 and/or 22 $\mu$m bands but presenting fundamental 
problems in the \textit{WISE} images: (top  
panel) artifact in the W3 and W4 bands for HD 109573; (middle panel) object  absent in the W3 and W4 band for HD 168325. (bottom  panel) point source without a fundamental problem in the W3 band and contaminated by 
background emission in the W4 band for HD 168746.}
\end{figure}

\begin{figure*}[ht]
\centering
\epsscale{0.55}
\plotone{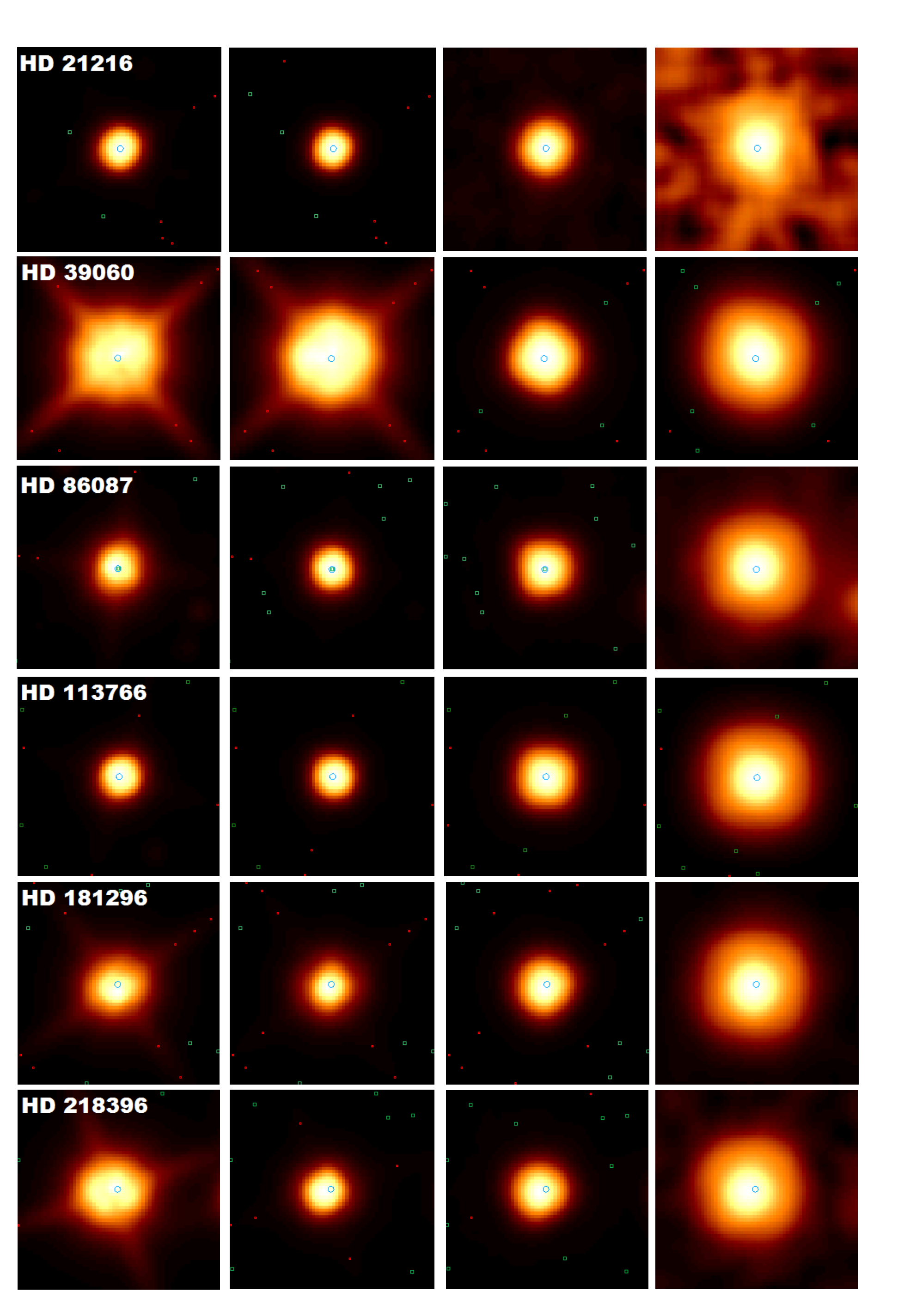}
\caption{\textit{WISE} images (from left to right: W1, W2, W3, and W4) in min–max log scales 
for 6 stars with infrared excesses in the 22$\mu$m band that survived the 
visual image inspection. These stars are examples of clear point sources in all WISE bands. No contamination by artifacts has been found in bands with observed IR excess.}
\end{figure*}

\begin{figure*}[ht]
\centering
\epsscale{1.0}
\plotone{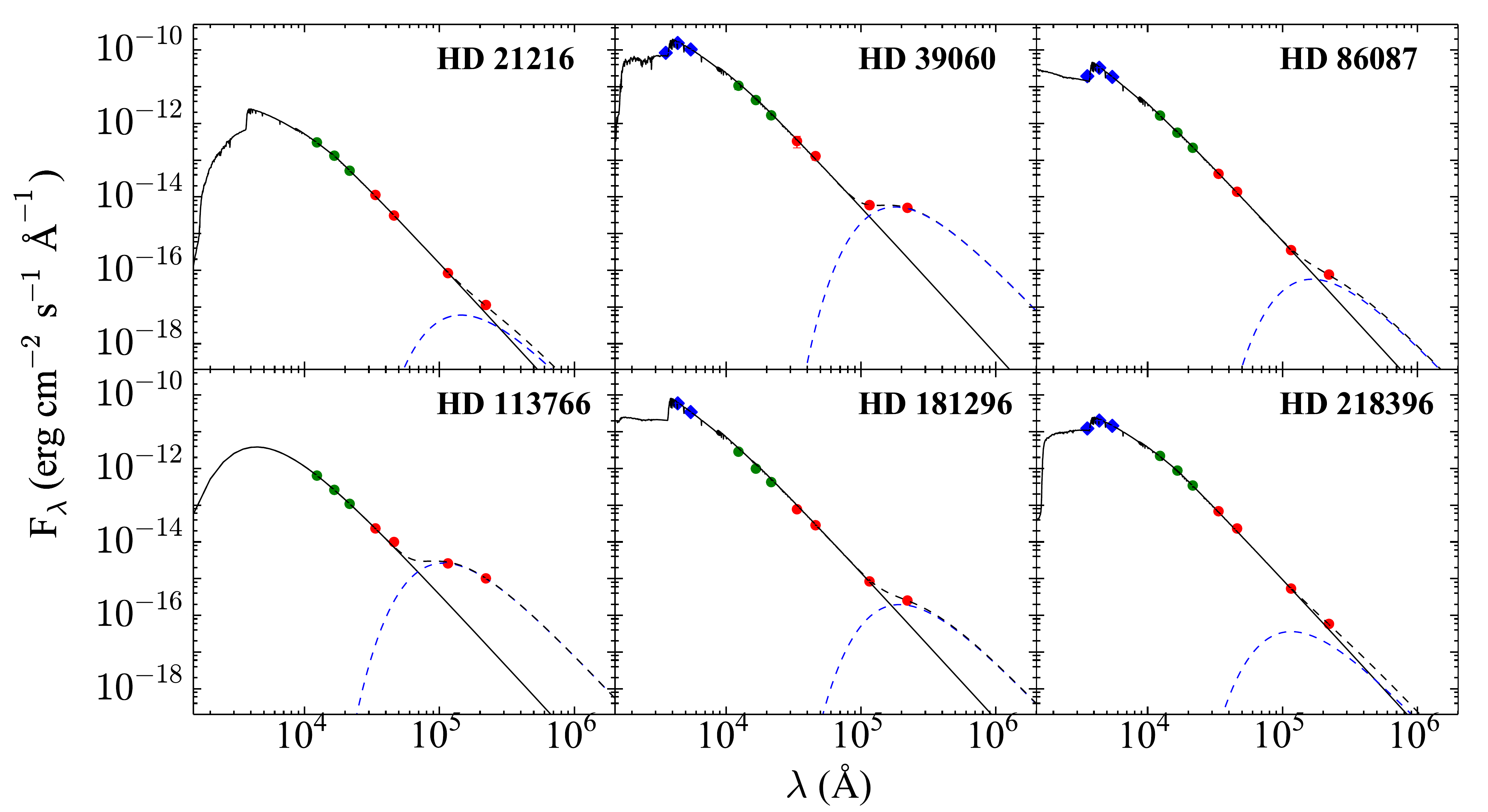}
\caption{SEDs for stars with WISE excesses at the 3$\sigma$ level of confidence (HD 21216, HD 39060, HD 86087, HD 113766, HD 181296, and HD 218396). 
The UBV fluxes (Mermilliod 2006) are plotted as blue squares; 2MASS J, H, and K fluxes (Cutri et al. 2013) as green circles; and the WISE bands W1, W2, W3, and W4 (Cutri et al. 2013) as red circles. The black solid line represents the stellar Kurucz model (Castelli et al. 1997). 
For HD 113766, the stellar Kurucz model does not have enough points to construct a 
fit, and therefore, we fit the photometry to this star using a 
black body model (Allard et al. 2012). The blue dashed line shows the 
best fit using a single black body model for the WISE bands with IR 
excess, while the black dotted line indicates the sum of the two components.}
\end{figure*}

\begin{figure*}[ht]
\centering
\epsscale{0.95}
\plotone{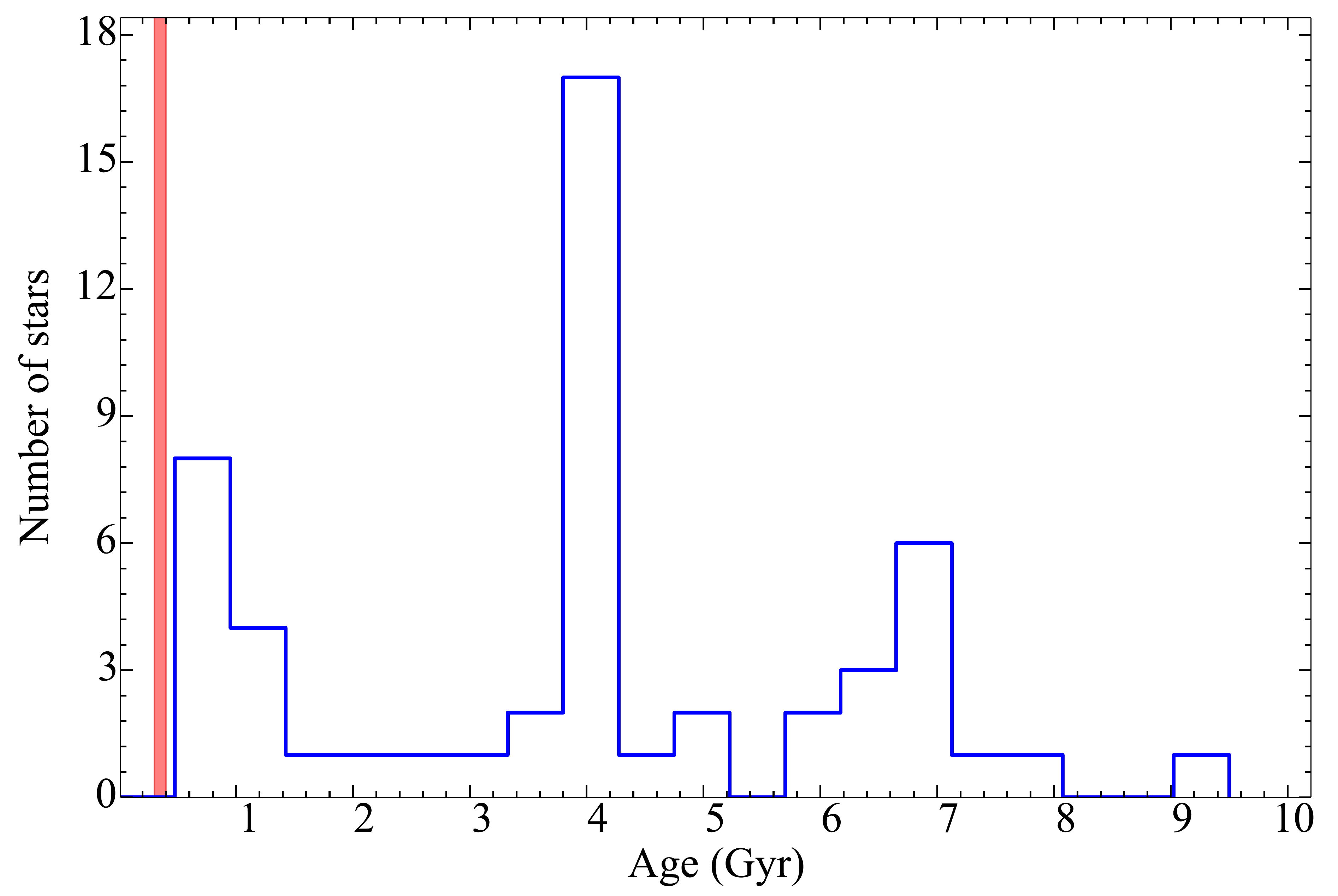}
\caption{Age distribution for the sample of 60 solar twin stars with {\it WISE} measurements. Ages were taken from Yana Galarza et al. (2016), Ramirez et al. (2014), and Melendez et al. (2006); for stars from Datson et al. (2015) and Pasquini et al. (2008), age were taken from Holmberg et al. (2007) and \"Onehag et al. (2011), respectively. The shaded region indicates the expected age interval of  300 to 400 Myr for disks disappearance (Habing et al. 1999).}
\end{figure*}

\clearpage
\section{Online material}

\begin{figure*}[ht]
\centering
\epsscale{0.35}
\plotone{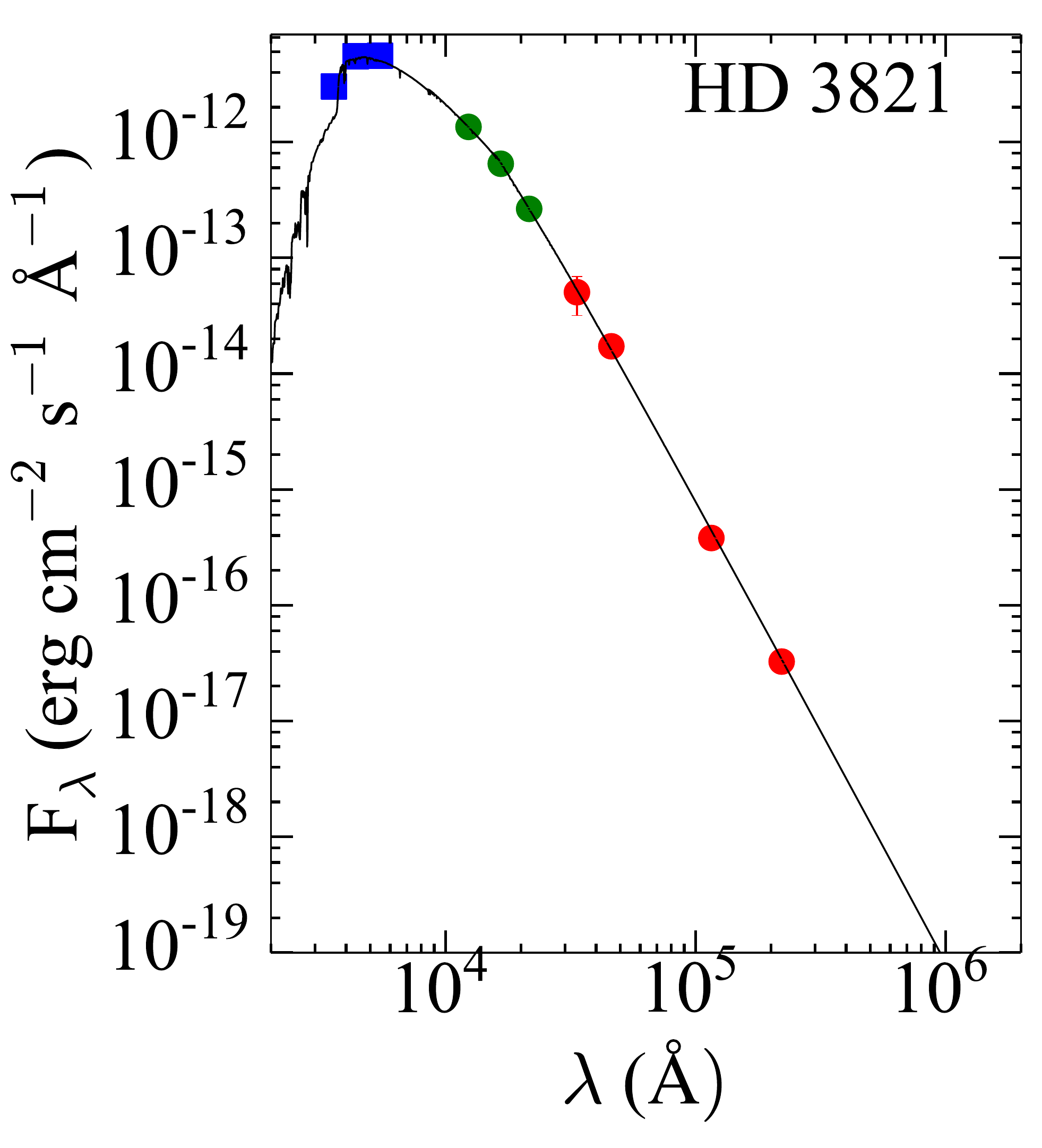}
\plotone{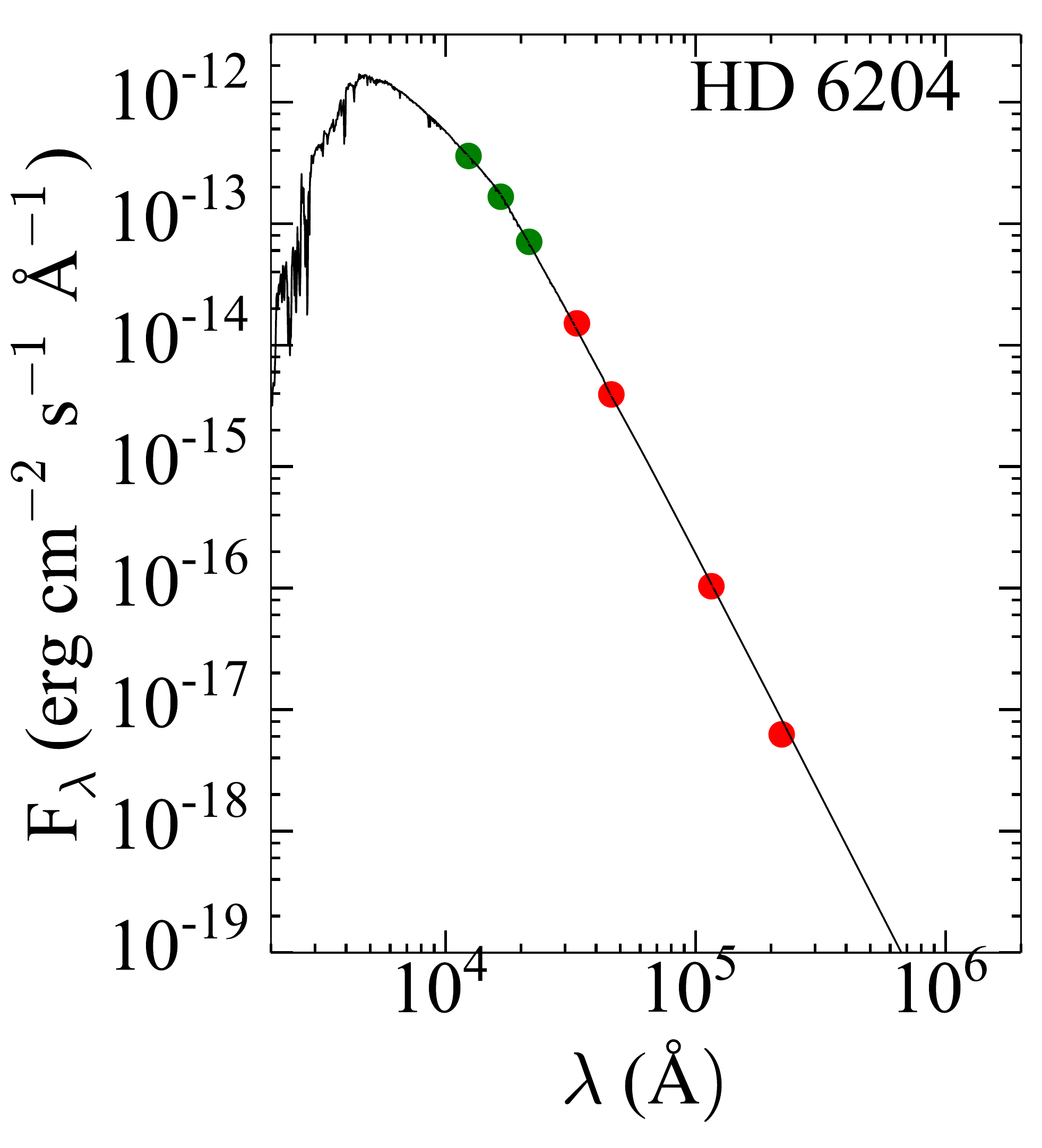}
\plotone{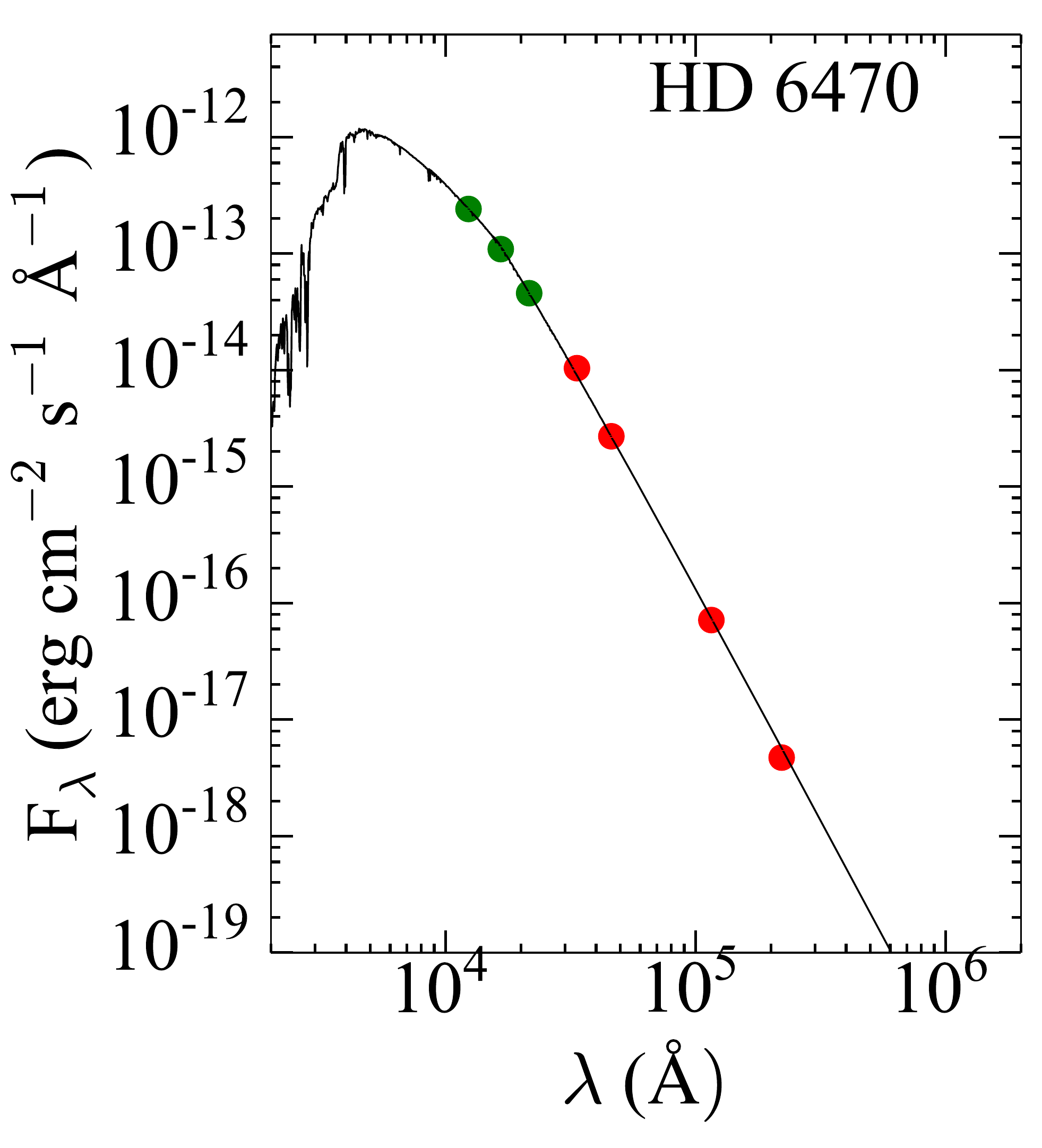}
\plotone{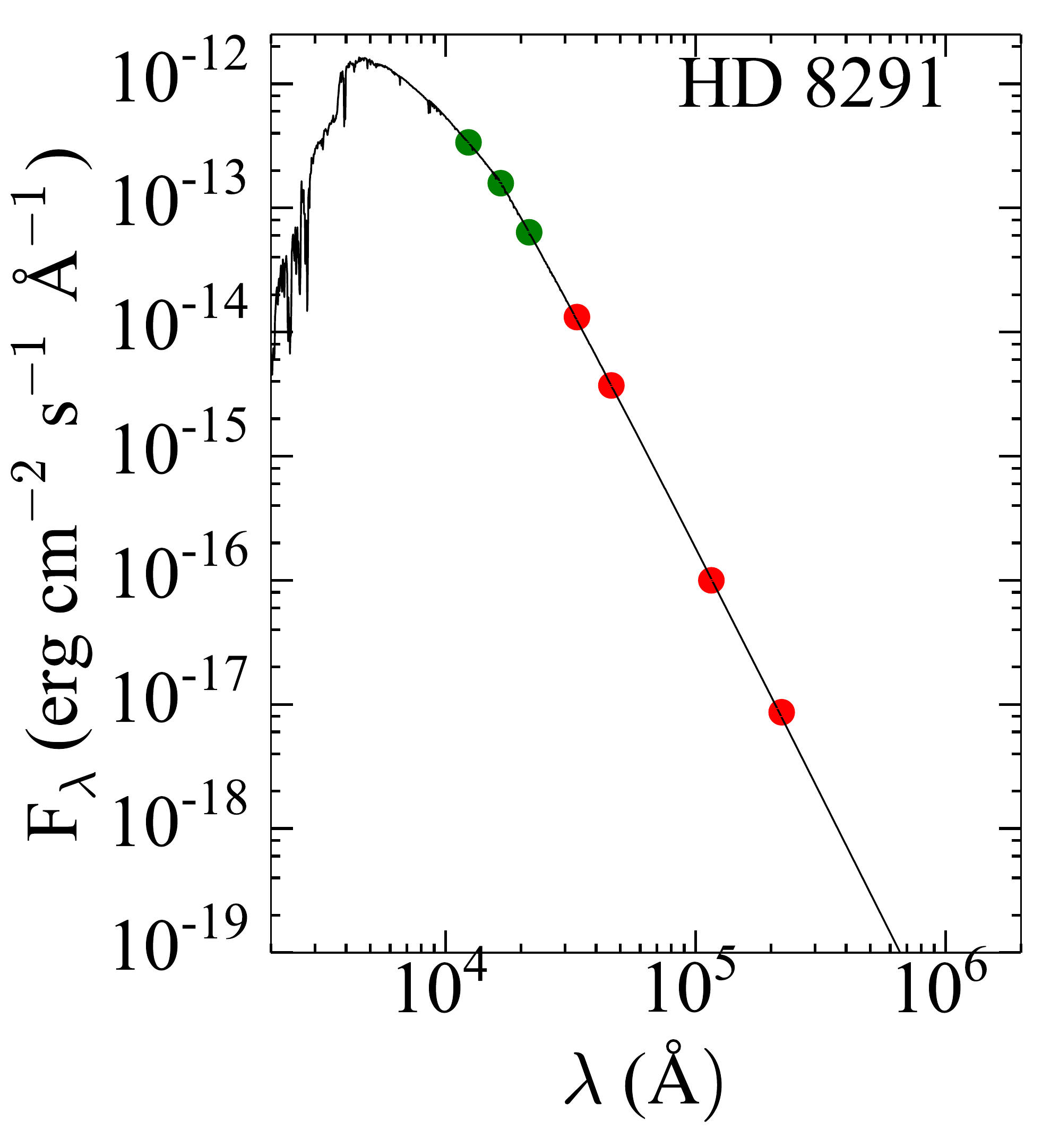}
\caption{SEDs for twin and sibling stars without WISE excesses. 
The  2MASS J, H, and K fluxes (Cutri et al. 2013) as green circles; and the WISE bands W1, W2, W3, and W4 (Cutri et al. 2013) as red circles. When available, the UBV fluxes (Mermilliod 2006) and SDSS ugriz fluxes (Ahn C. P. et al. 2012) are plotted as blue squares and magenta circle, respectively. Red open triangles presents the WISE upper limits. The black solid line represents the stellar Kurucz model (Castelli et al. 1997).}
\end{figure*}

\begin{figure*}[ht]
\centering
\epsscale{0.35}
\plotone{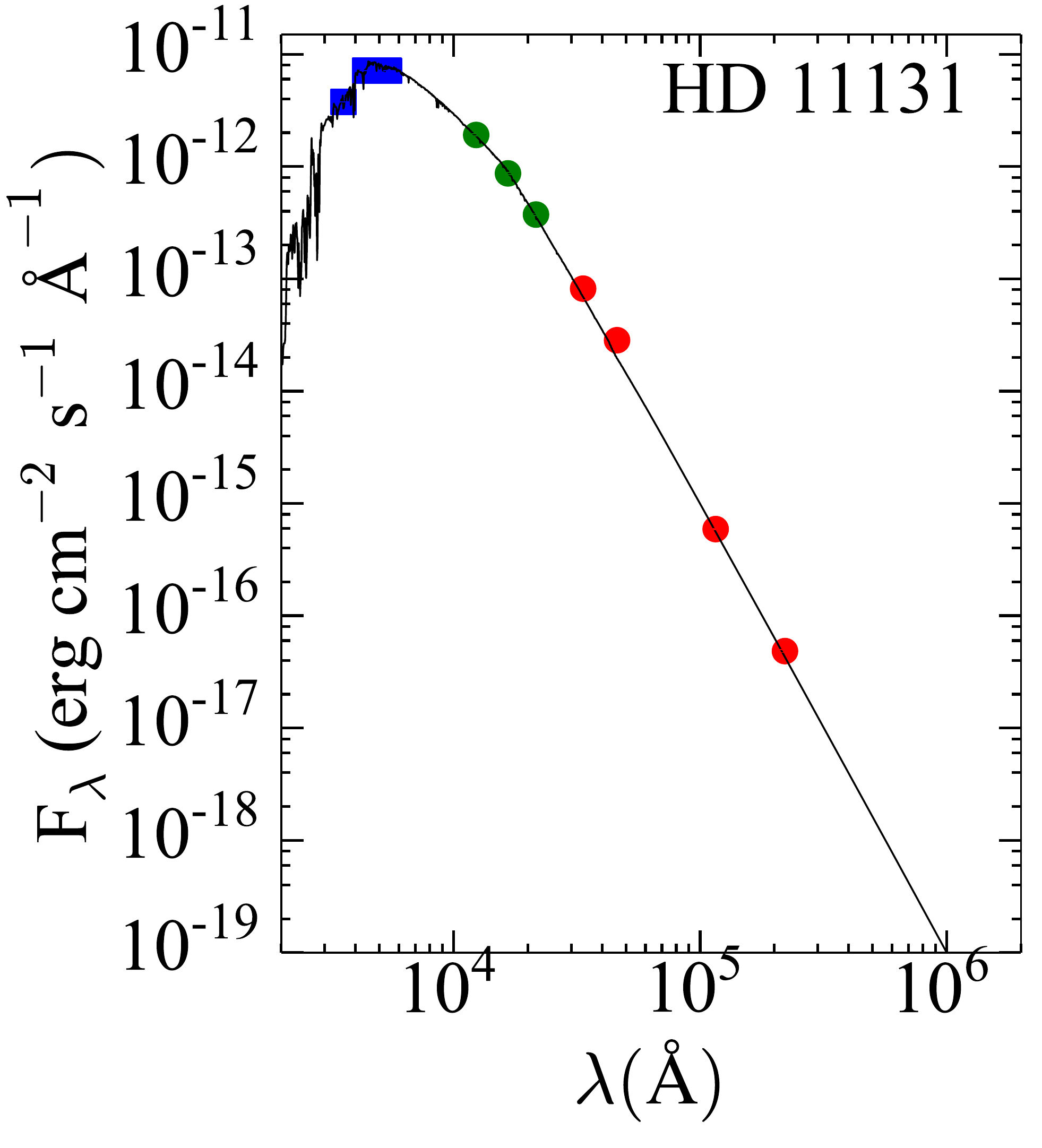}
\plotone{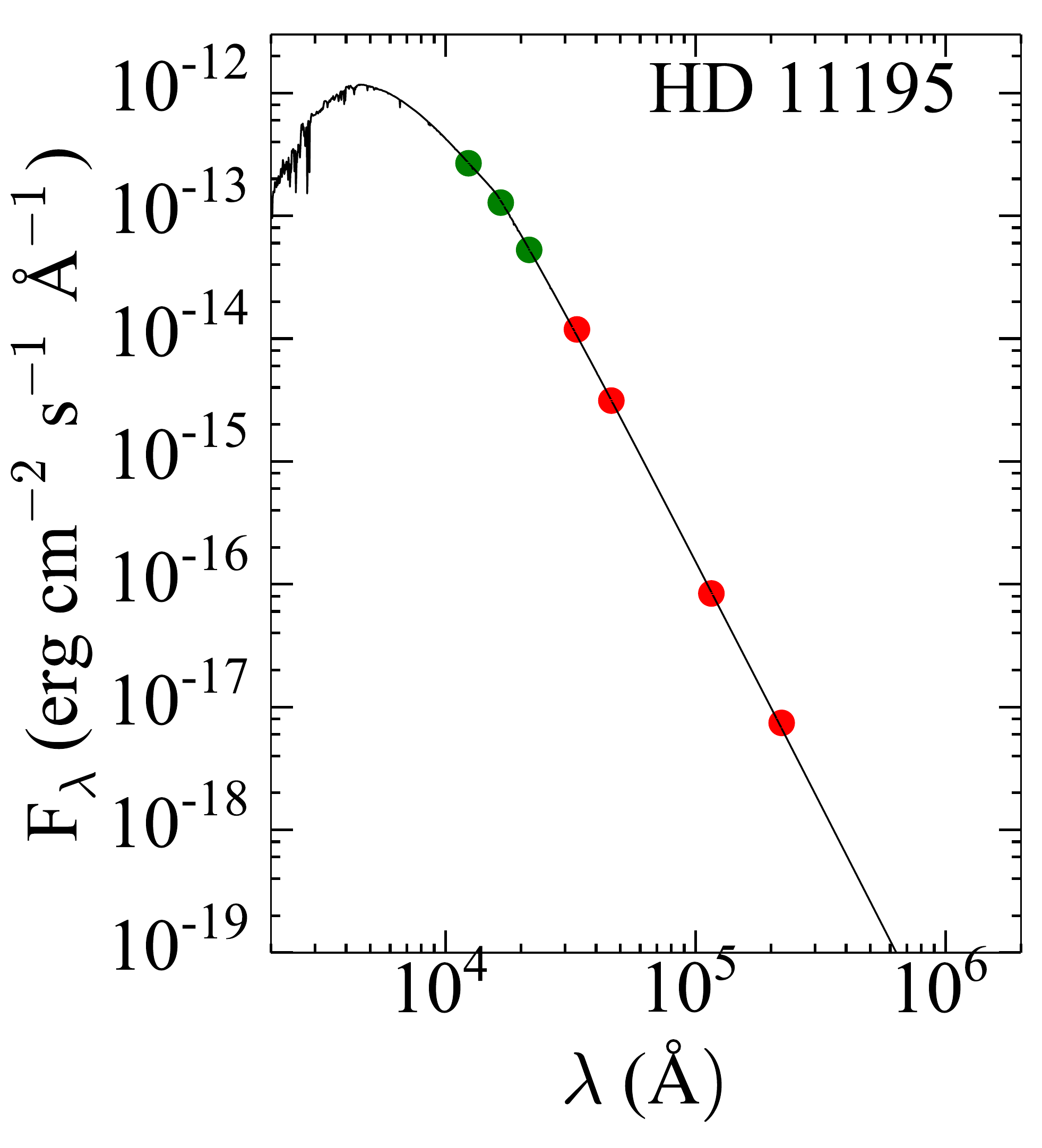}
\plotone{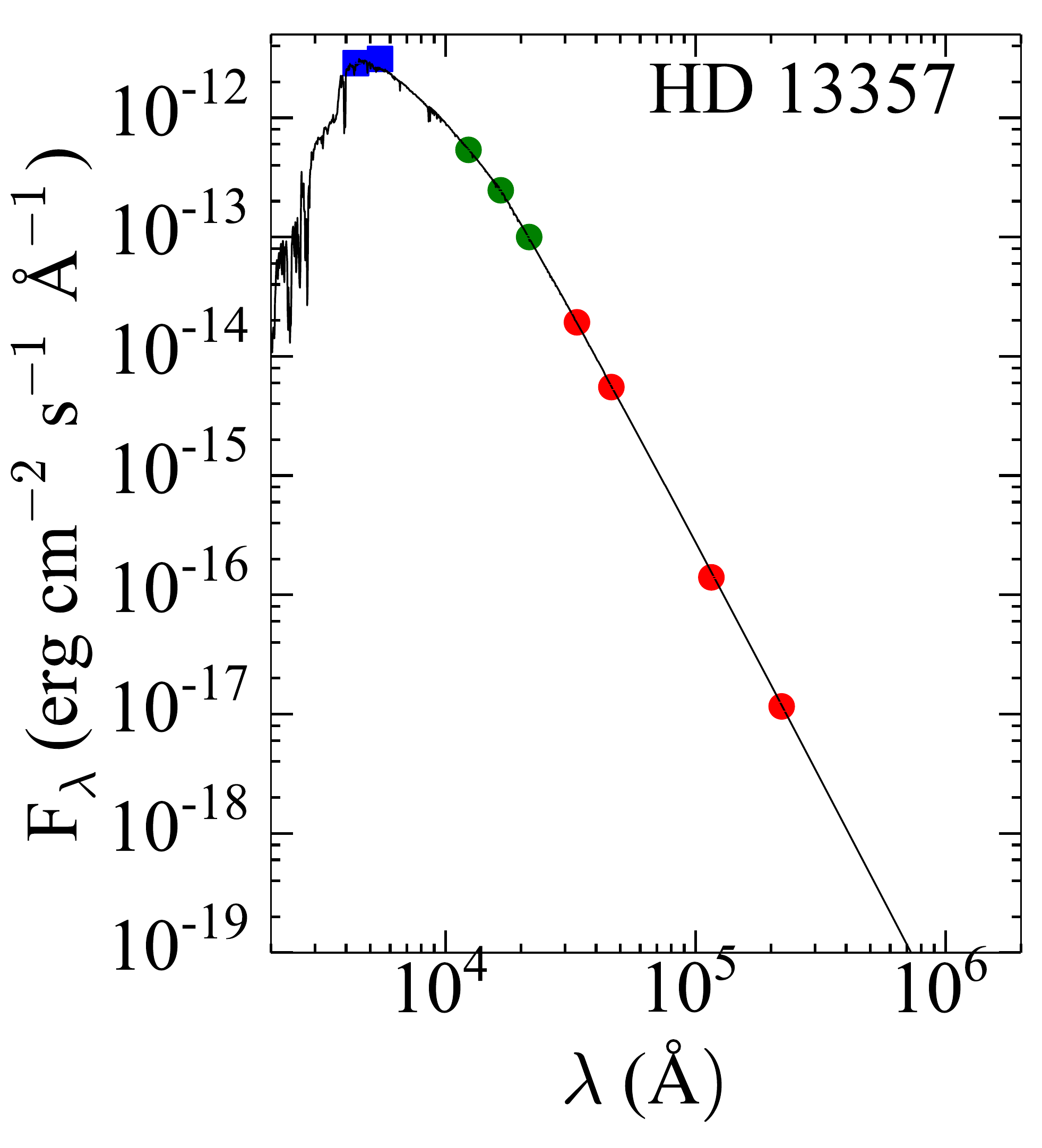}
\plotone{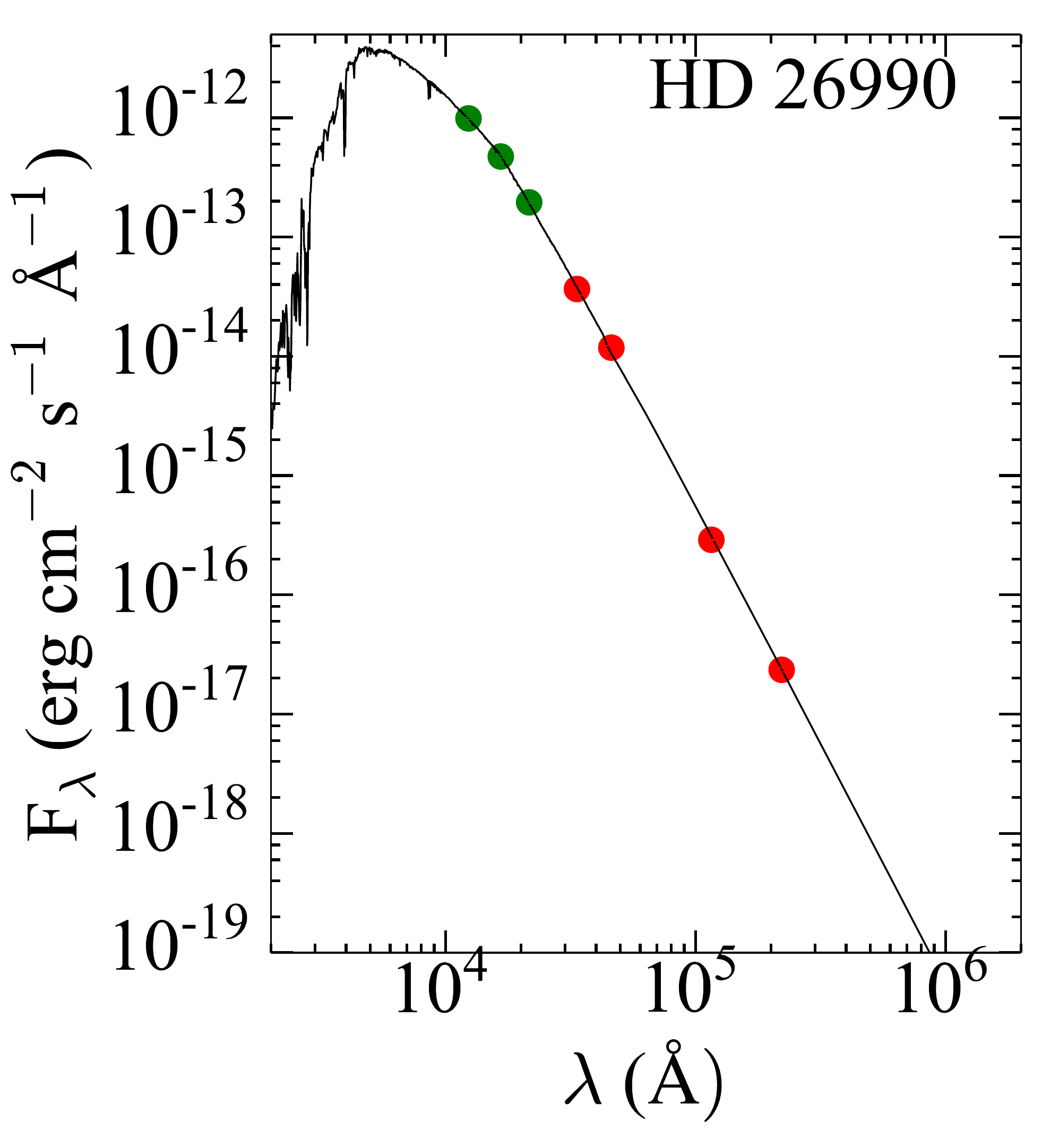}
\plotone{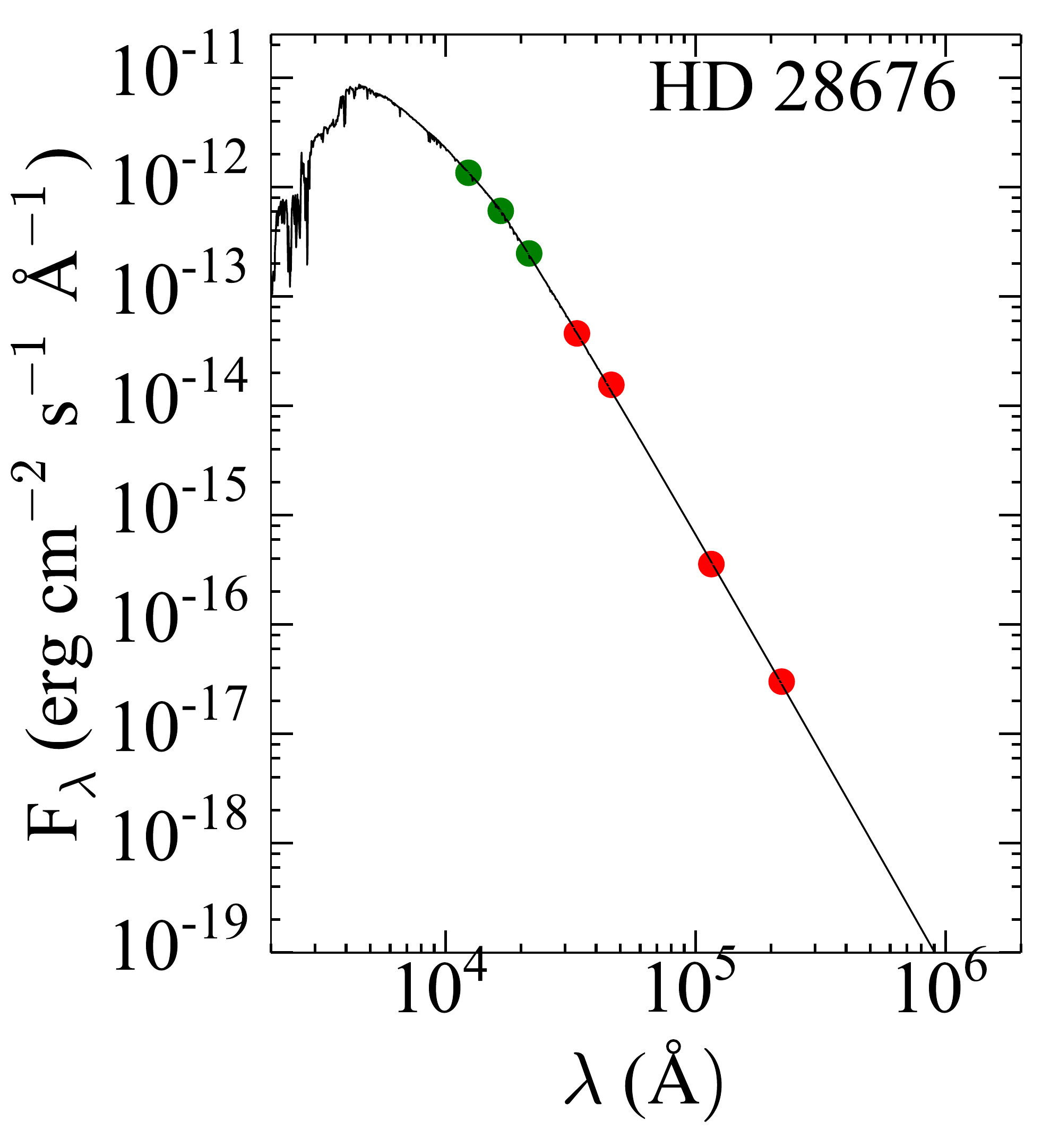}
\plotone{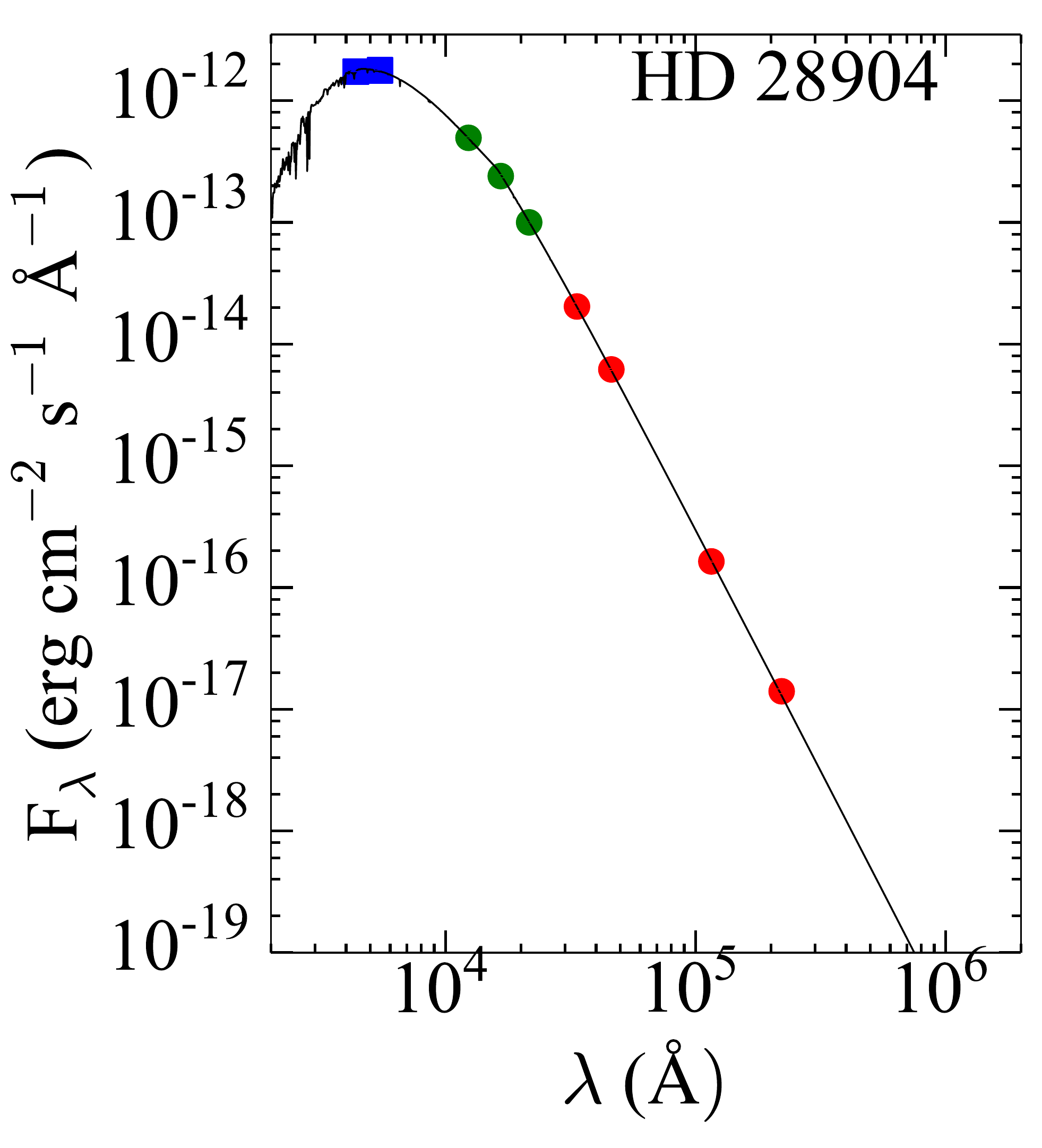}
\caption{Cont. Fig. 9}
\end{figure*}

\begin{figure*}[ht]
\centering
\epsscale{0.35}
\plotone{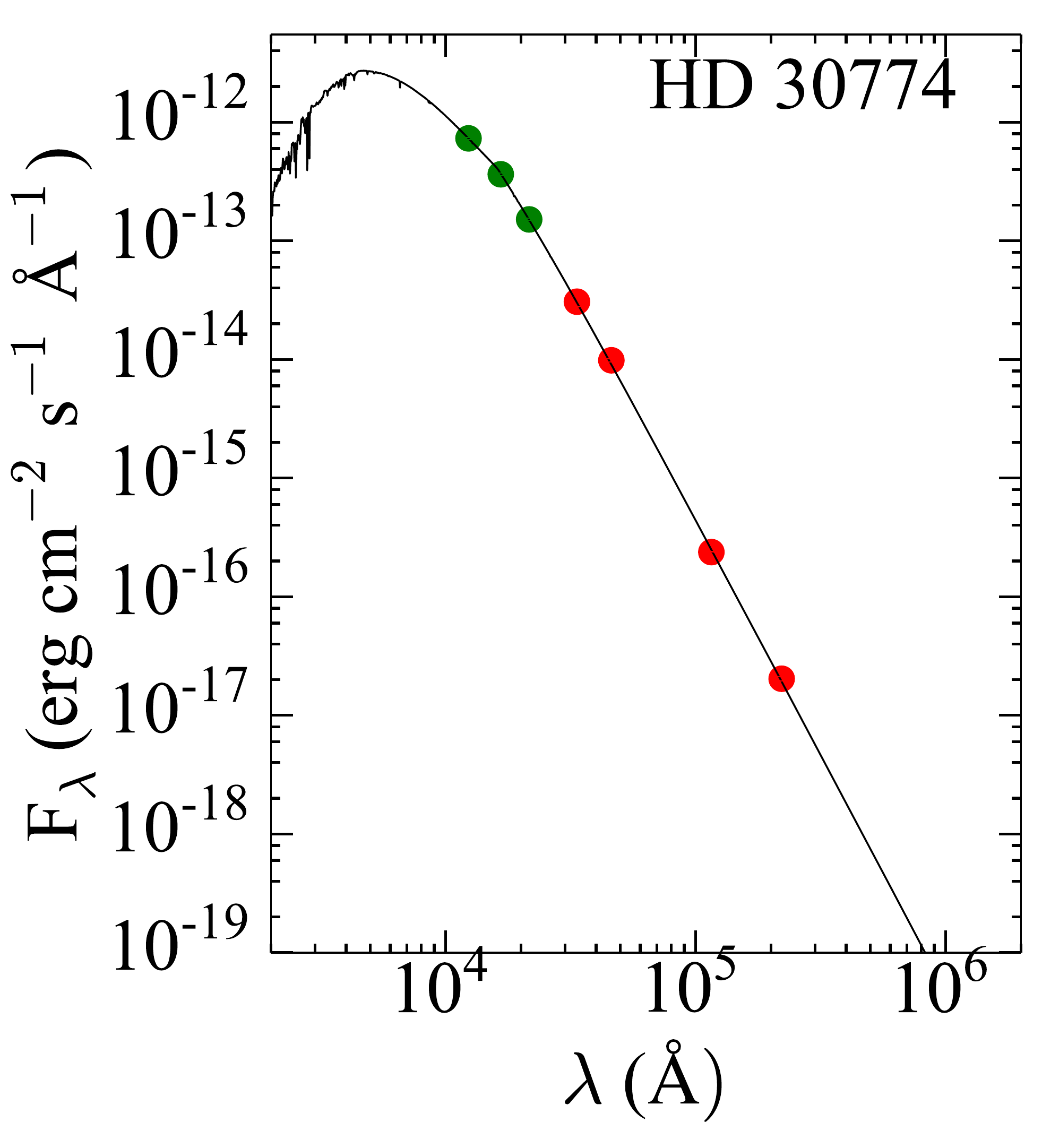}
\plotone{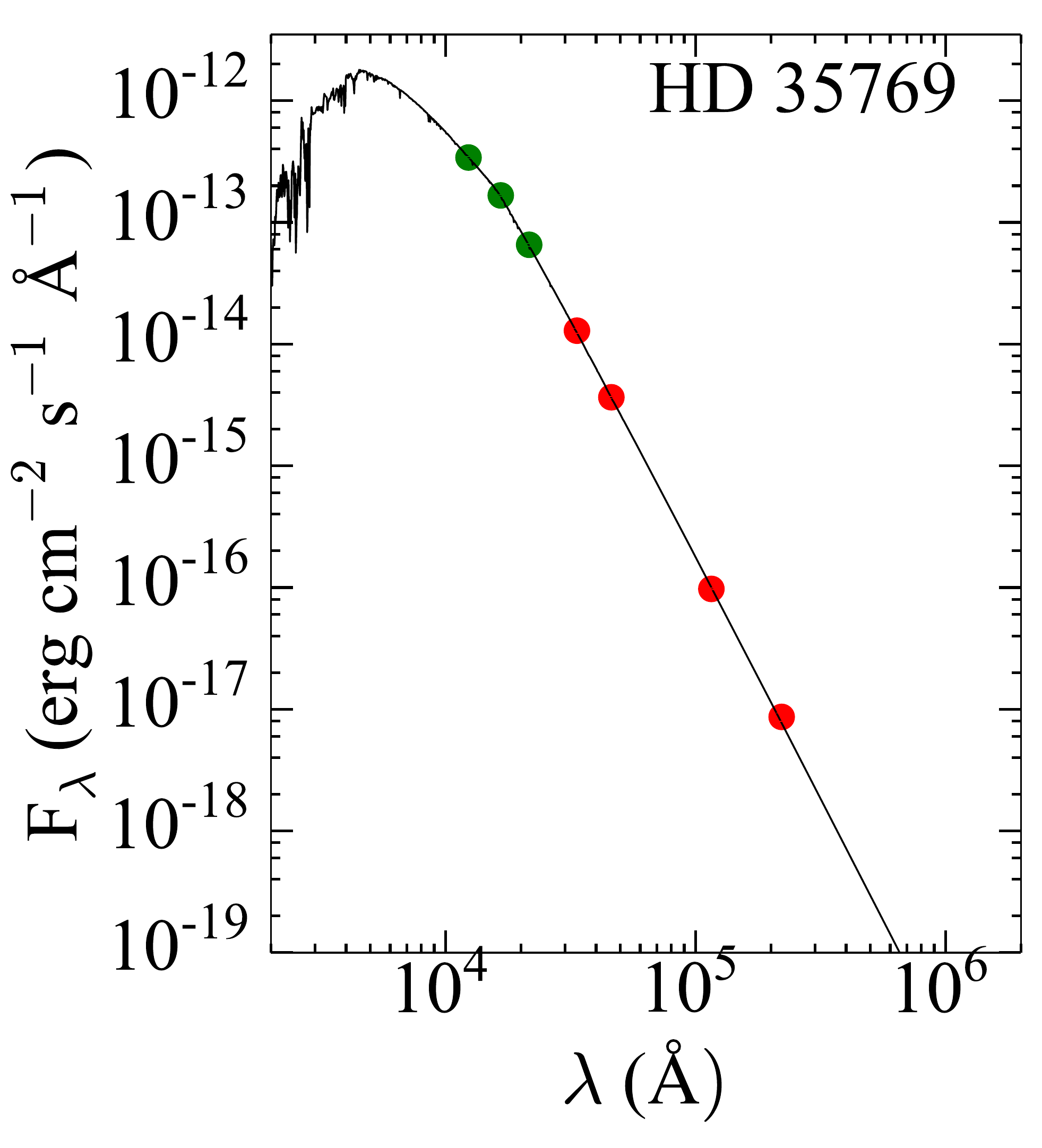}
\plotone{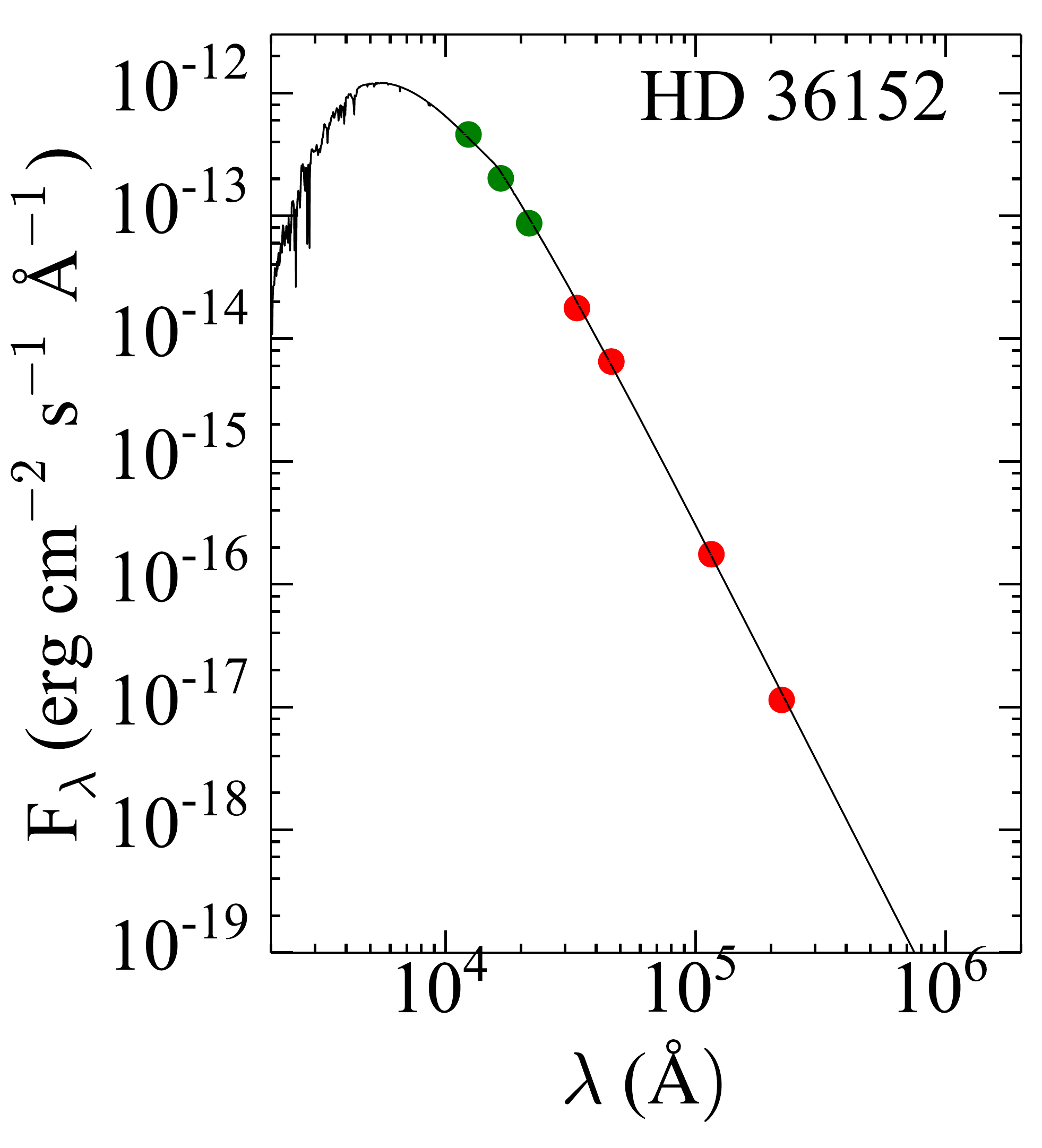}
\plotone{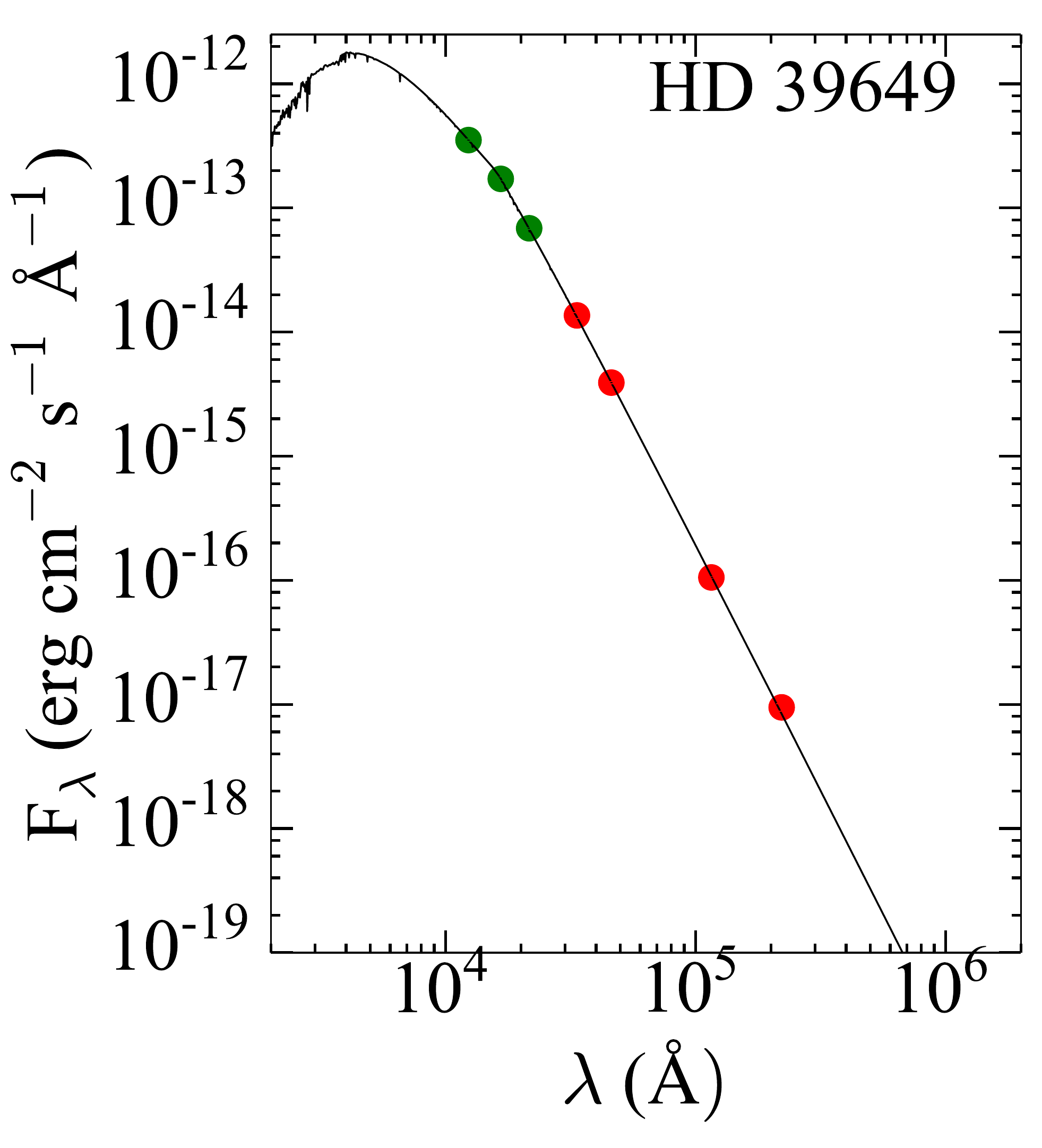}
\plotone{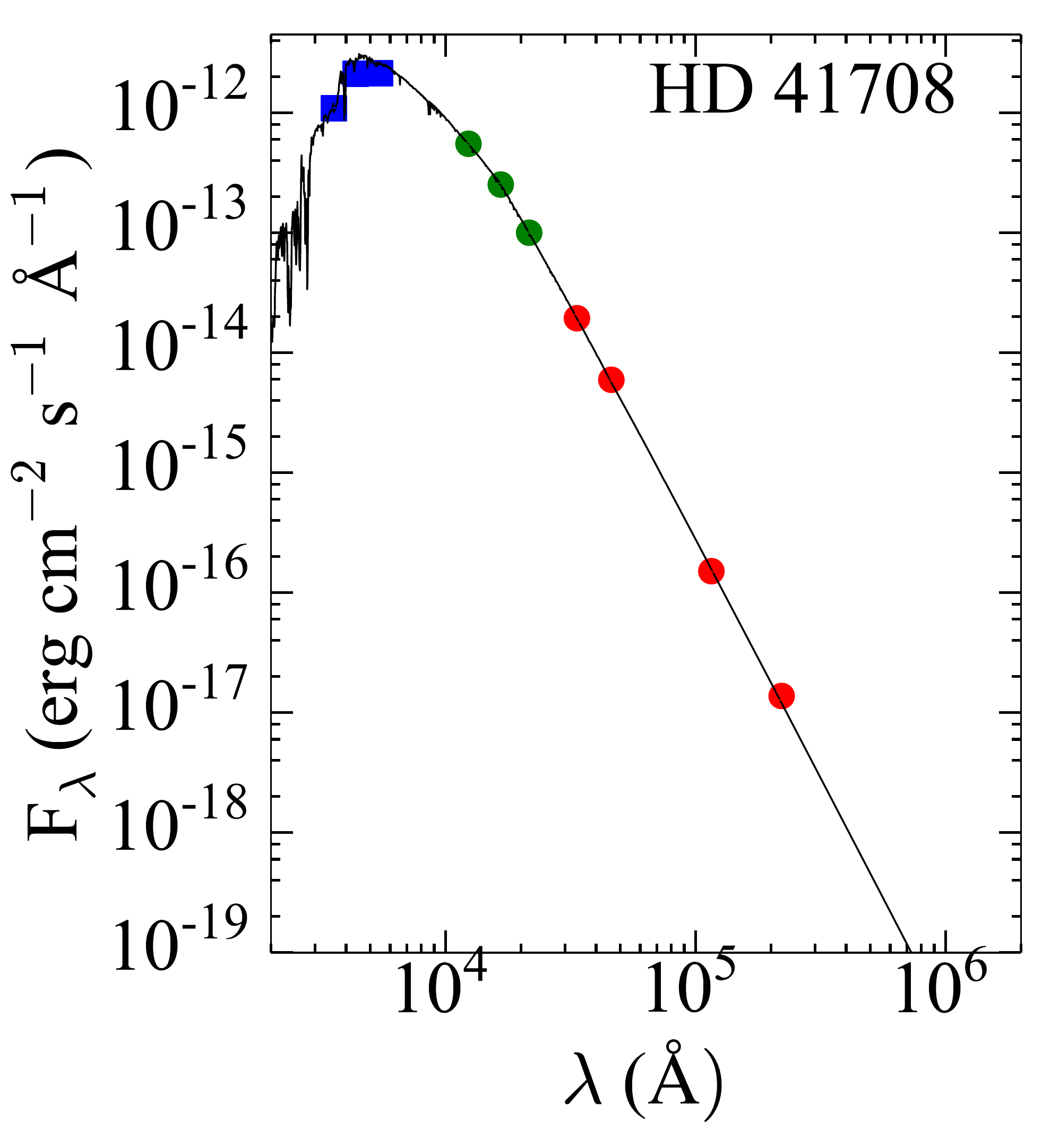}
\plotone{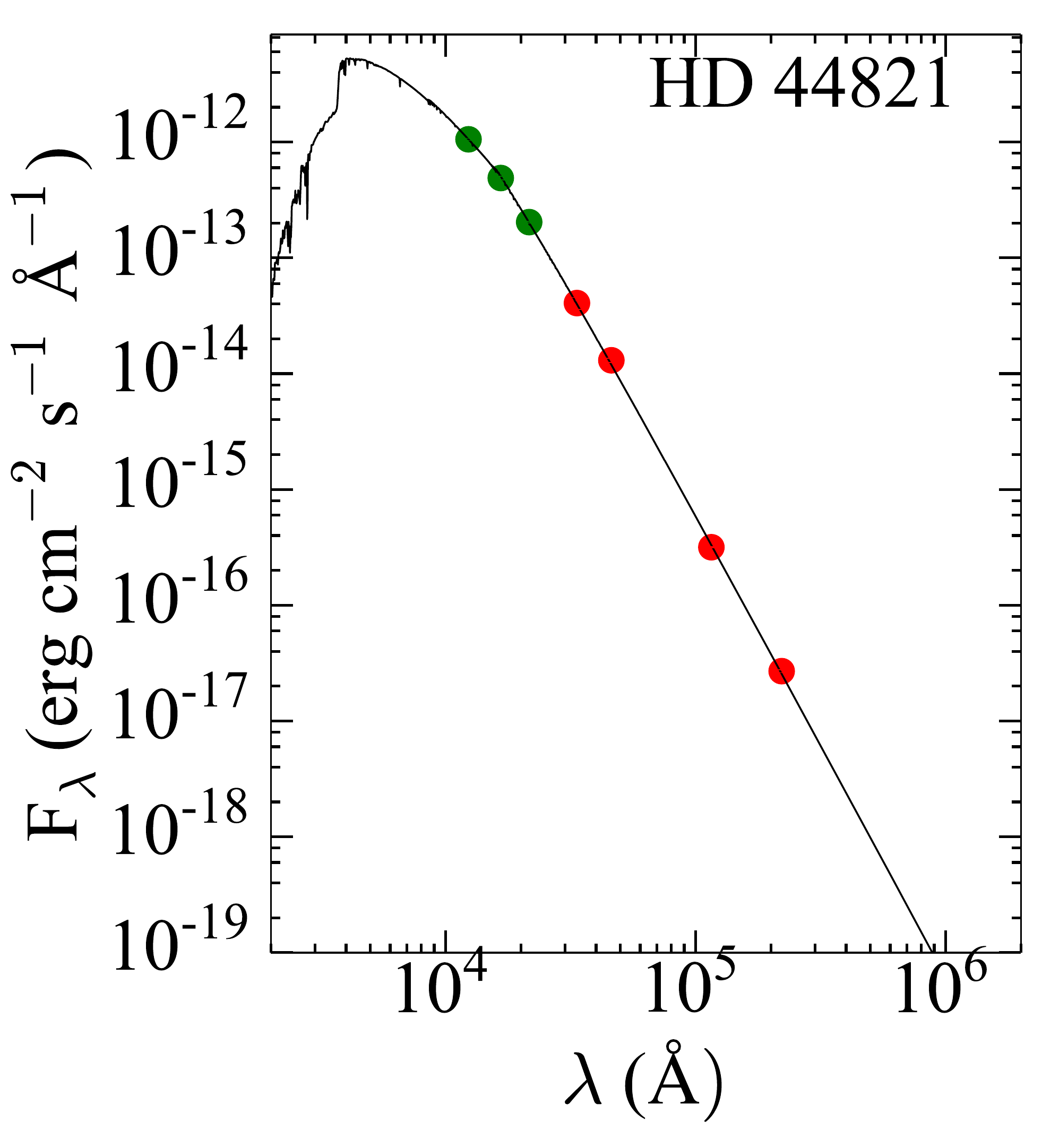}
\caption{Cont. Fig. 9}
\end{figure*}

\begin{figure*}[ht]
\centering
\epsscale{0.35}
\plotone{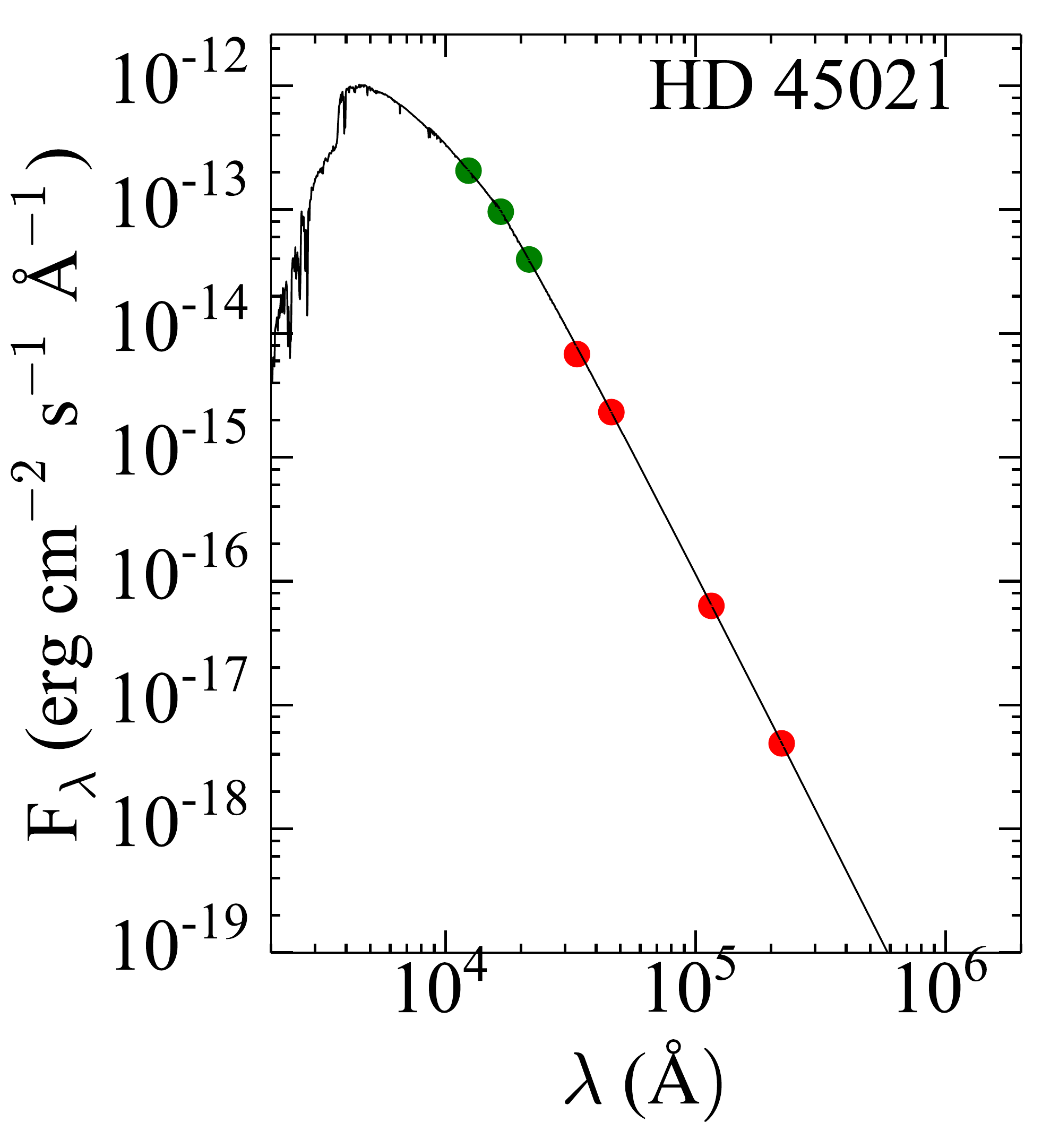}
\plotone{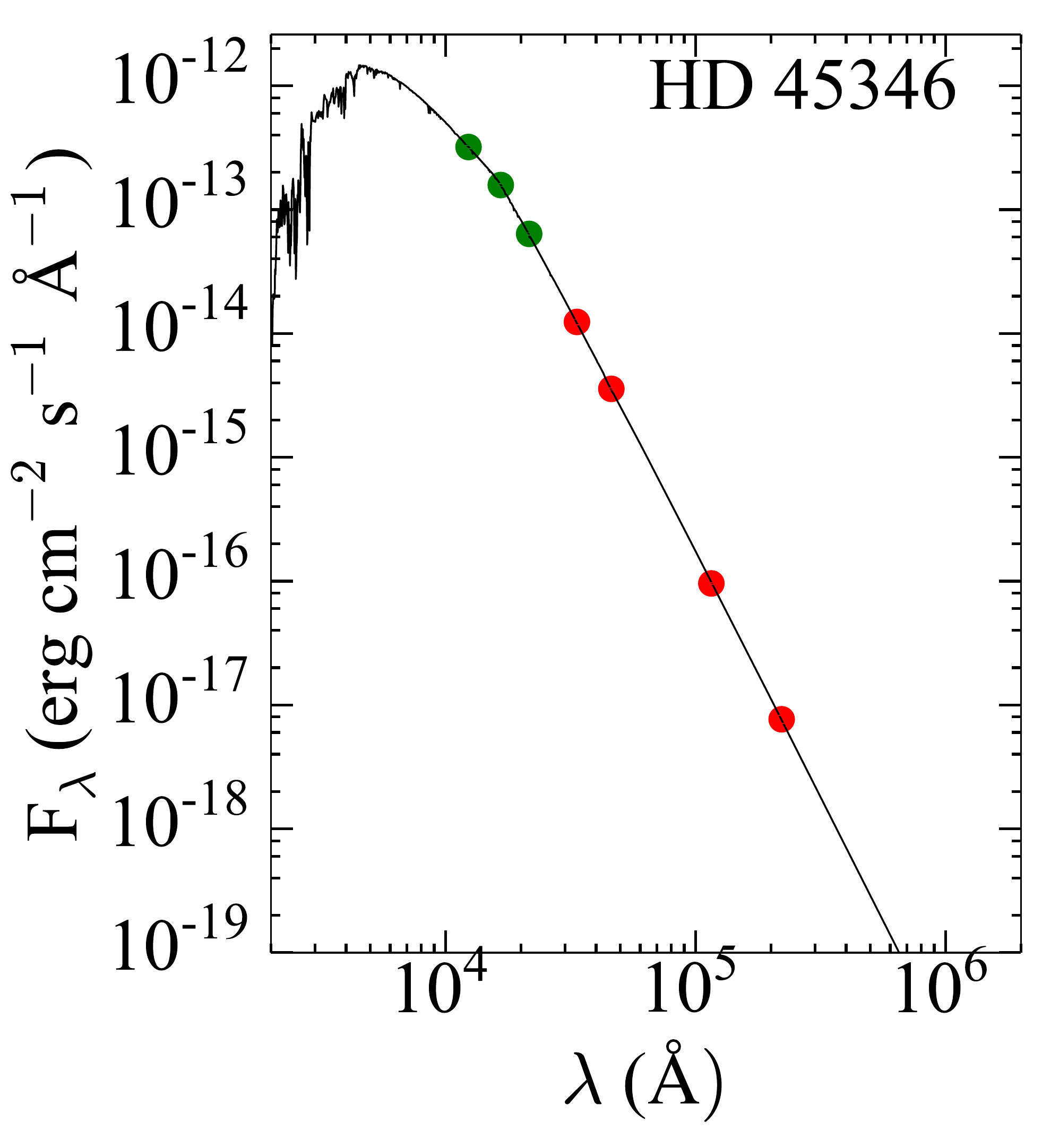}
\plotone{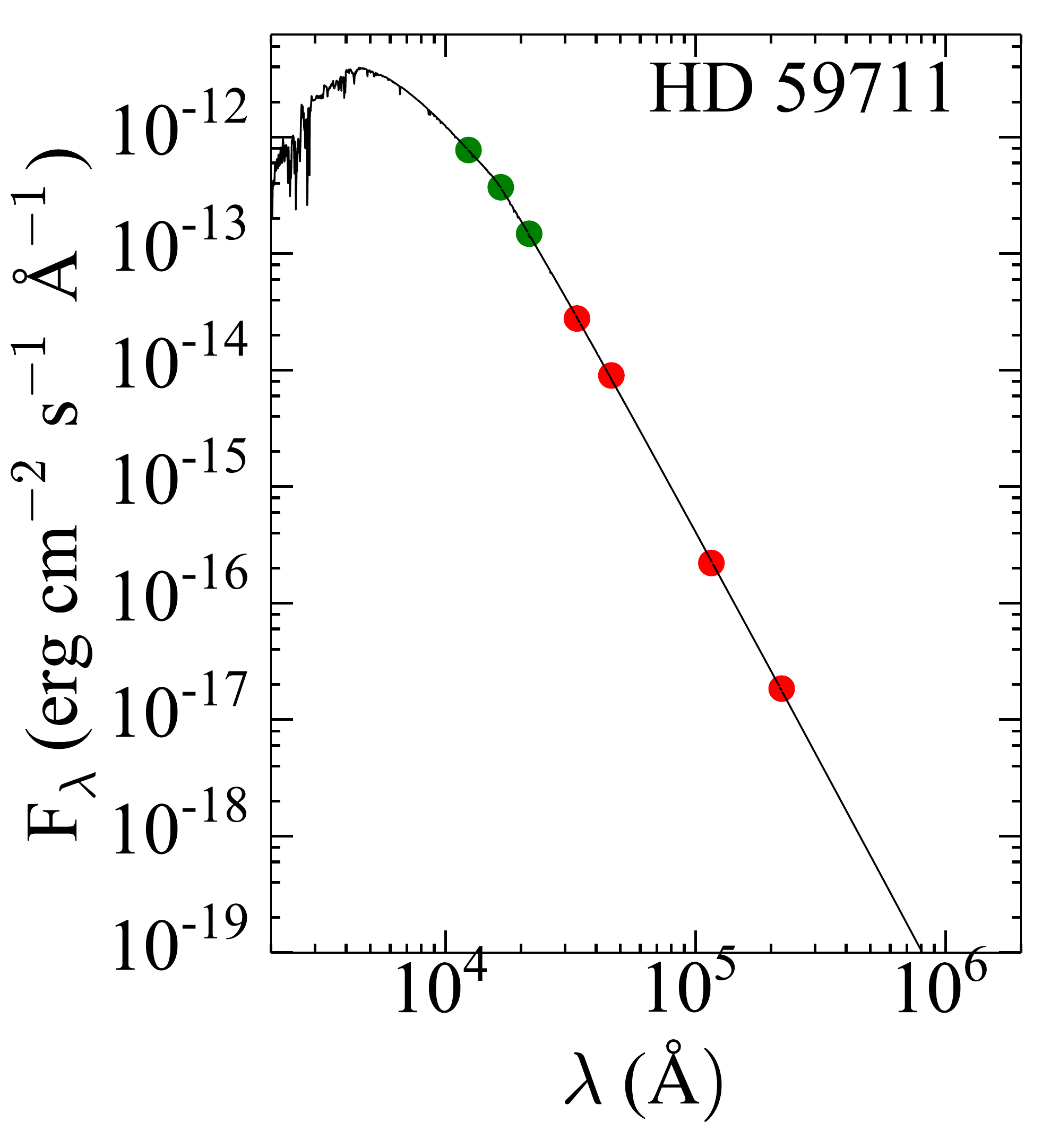}
\plotone{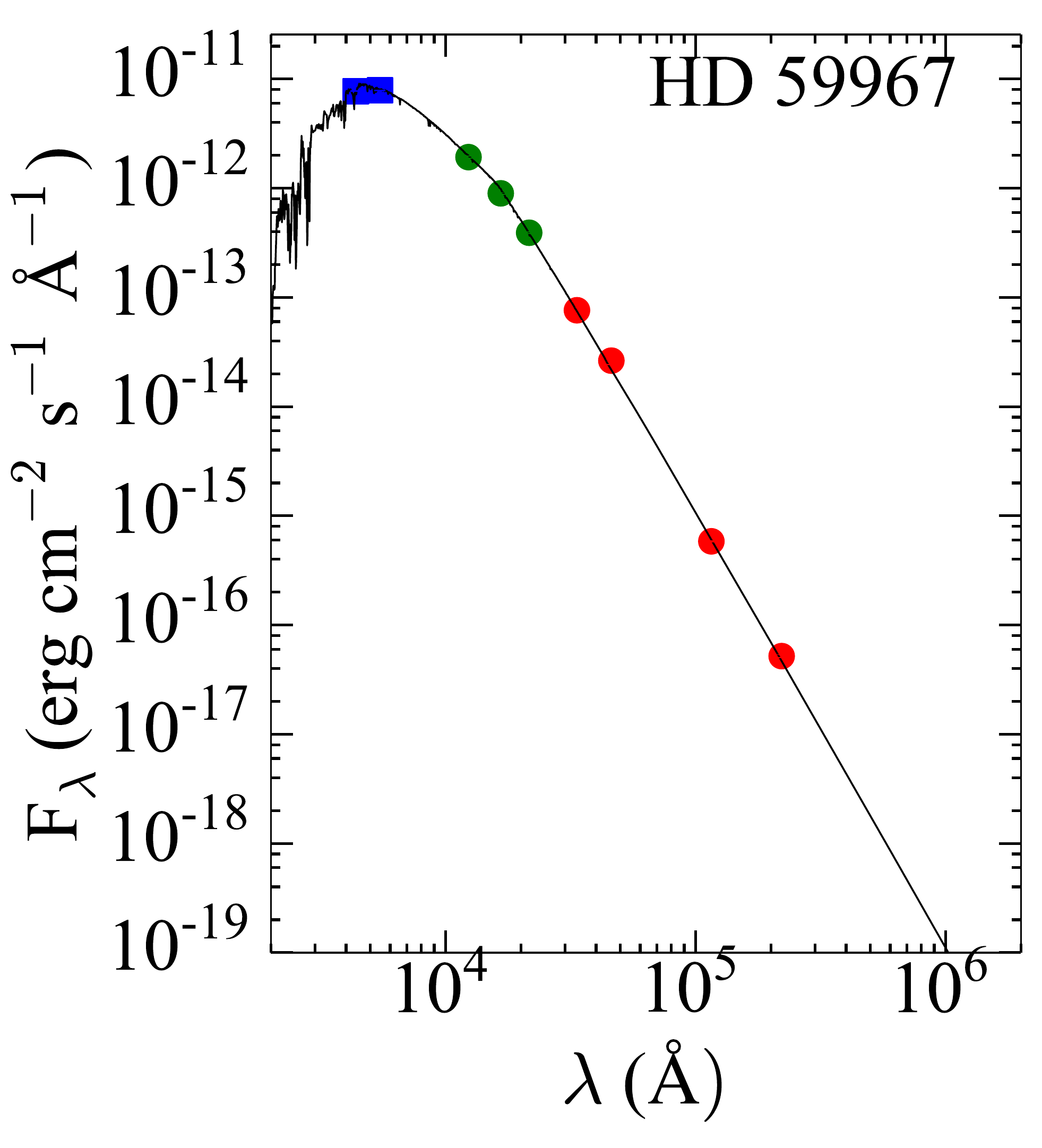}
\plotone{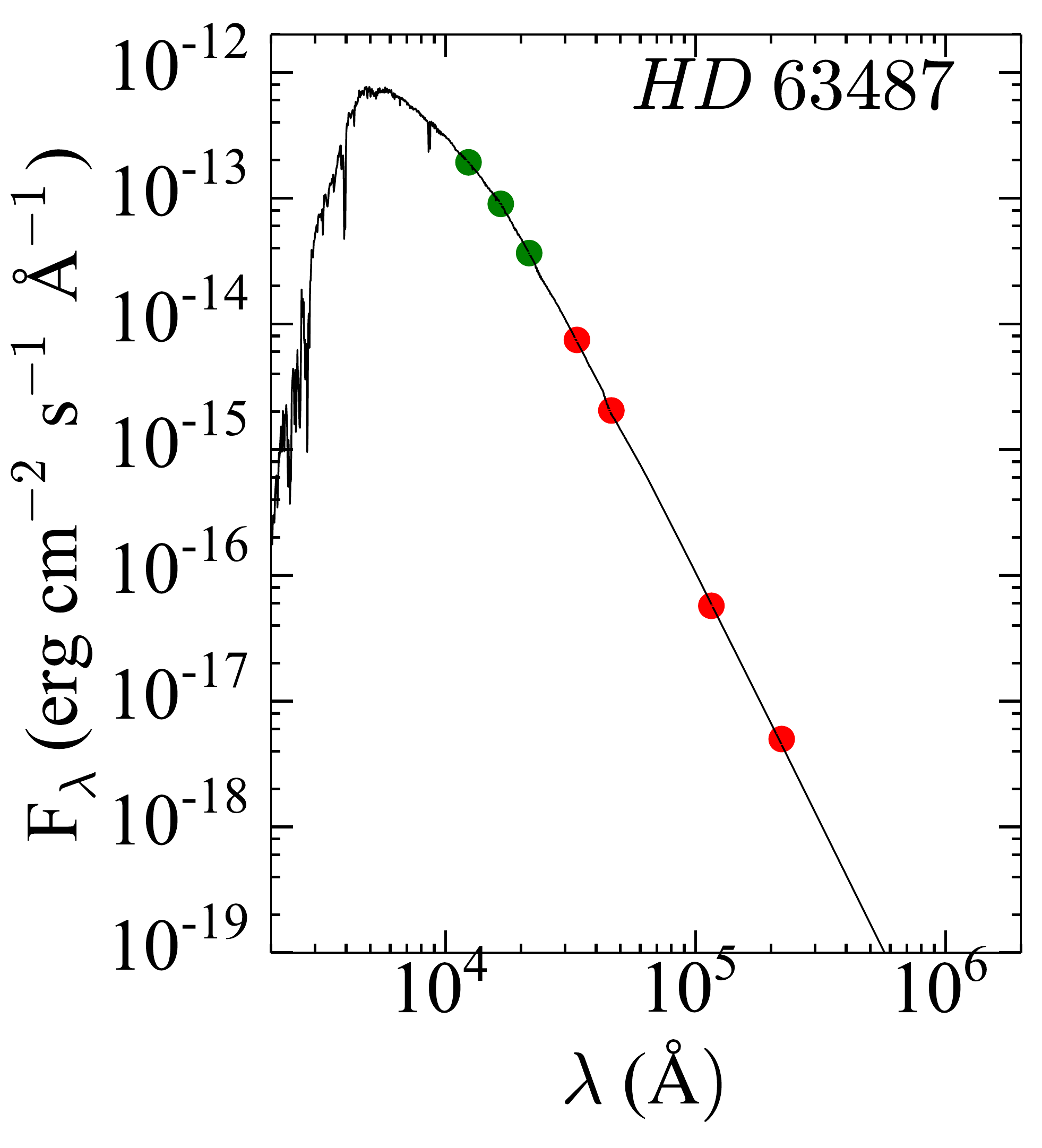}
\plotone{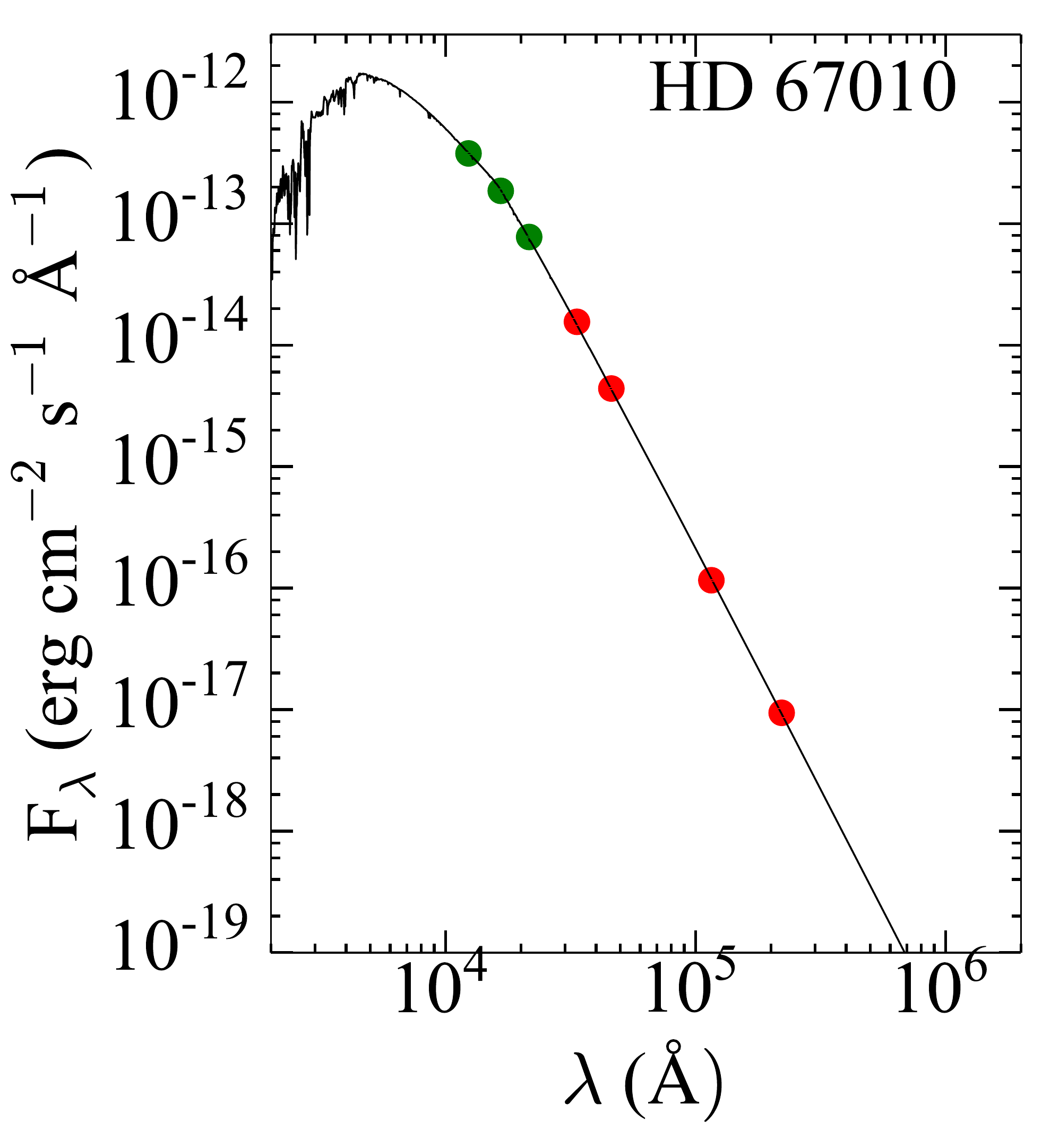}
\caption{Cont. Fig. 9}
\end{figure*}

\begin{figure*}[ht]
\centering
\epsscale{0.35}
\plotone{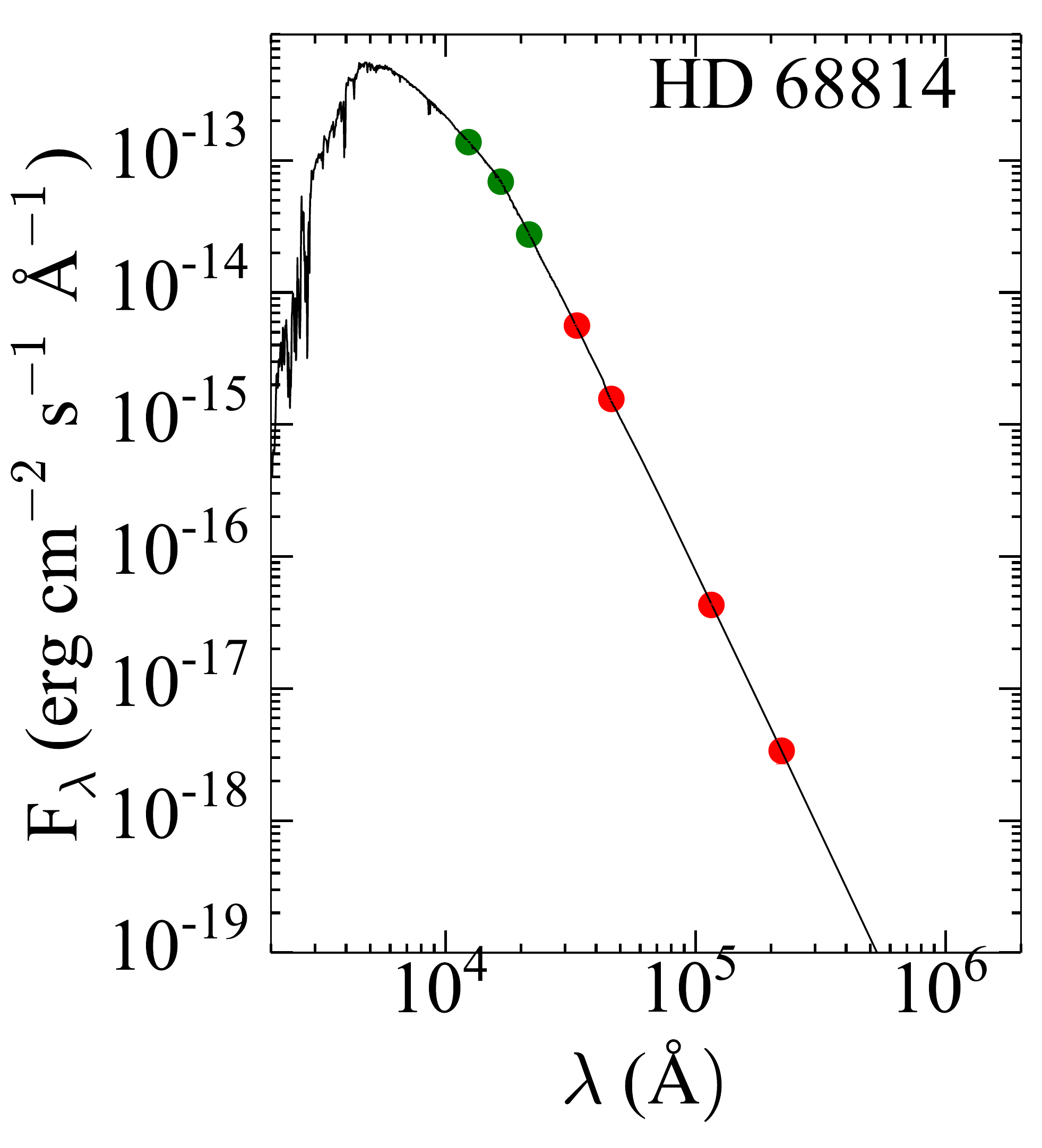}
\plotone{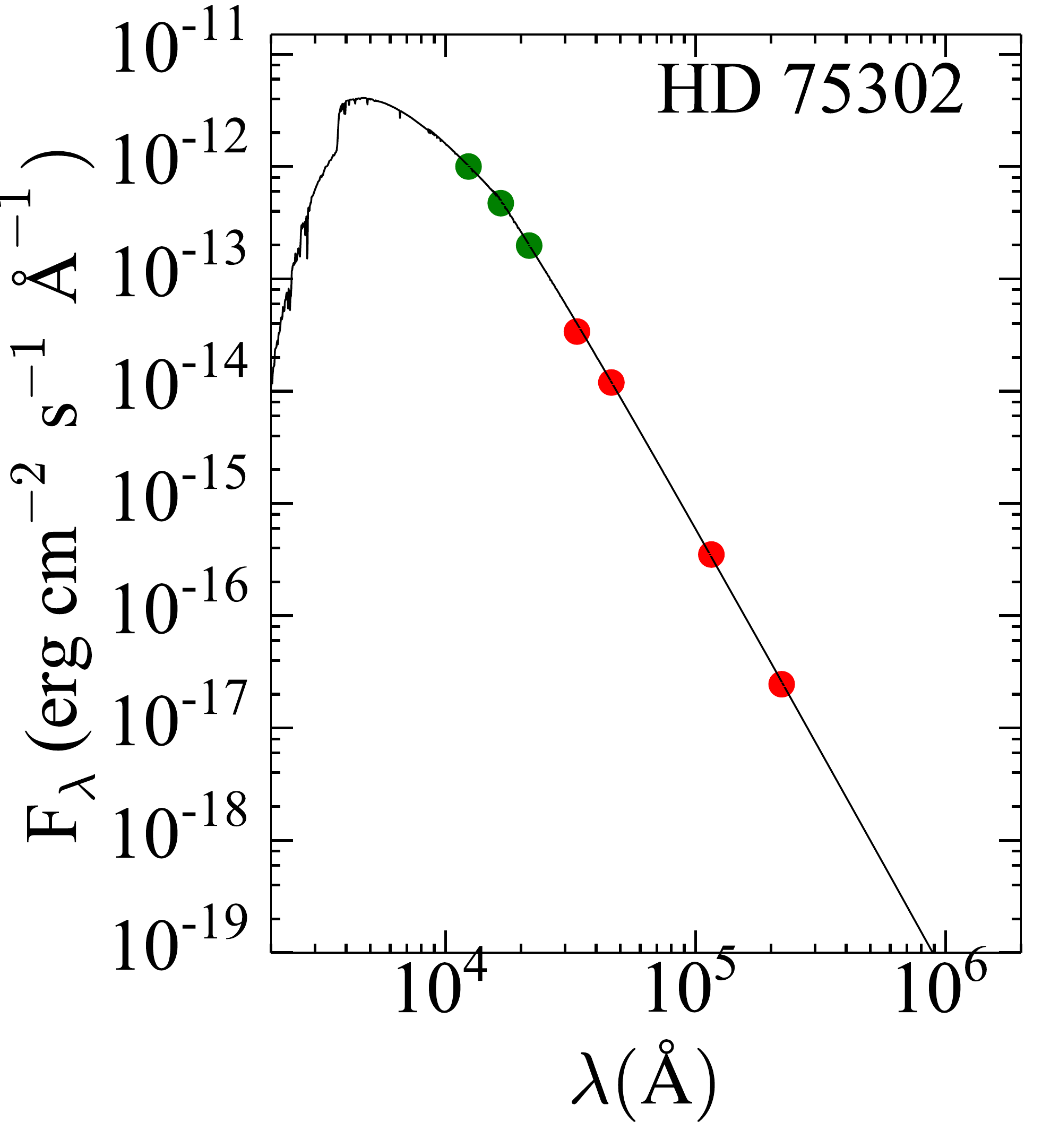}
\plotone{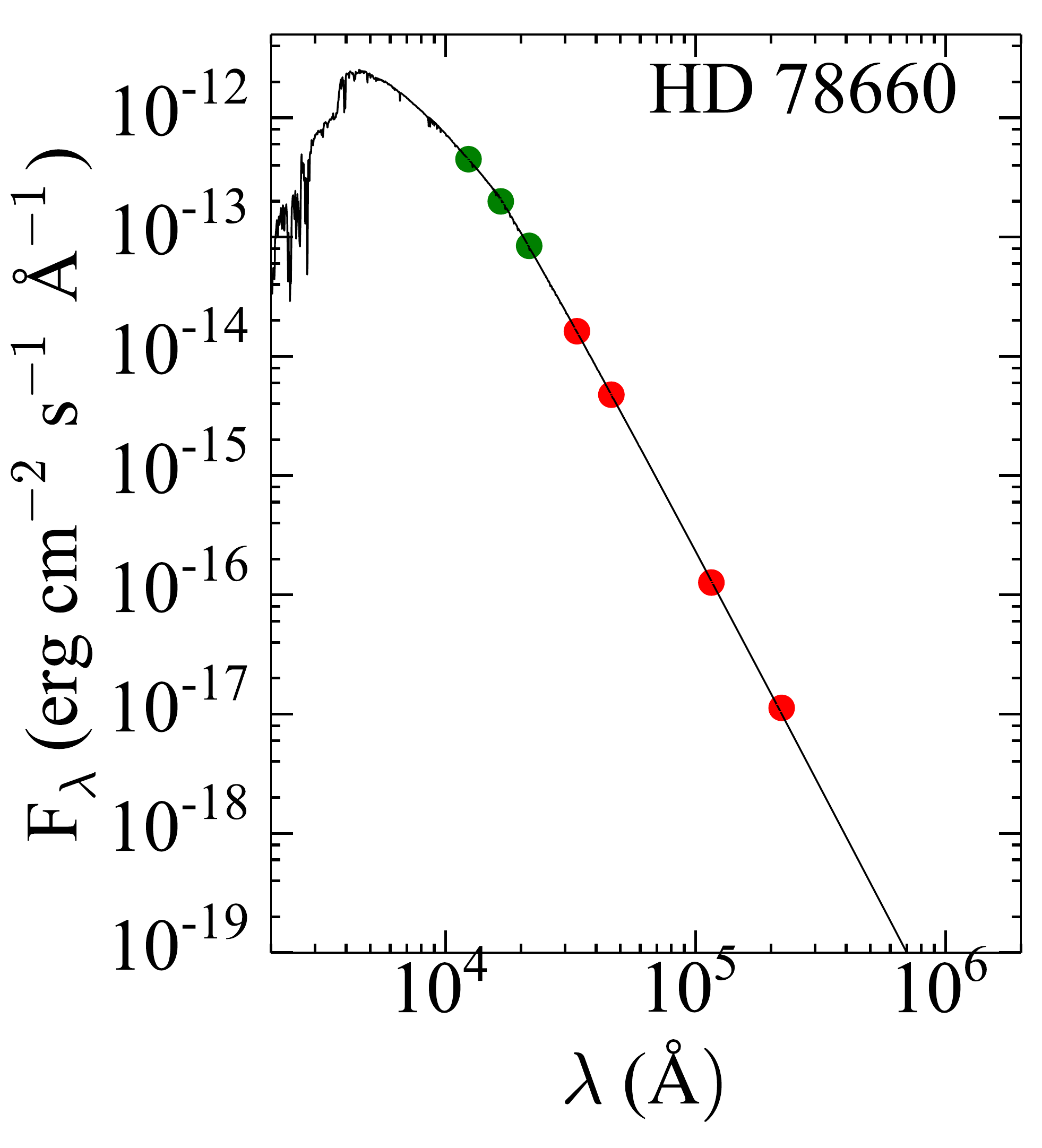}
\plotone{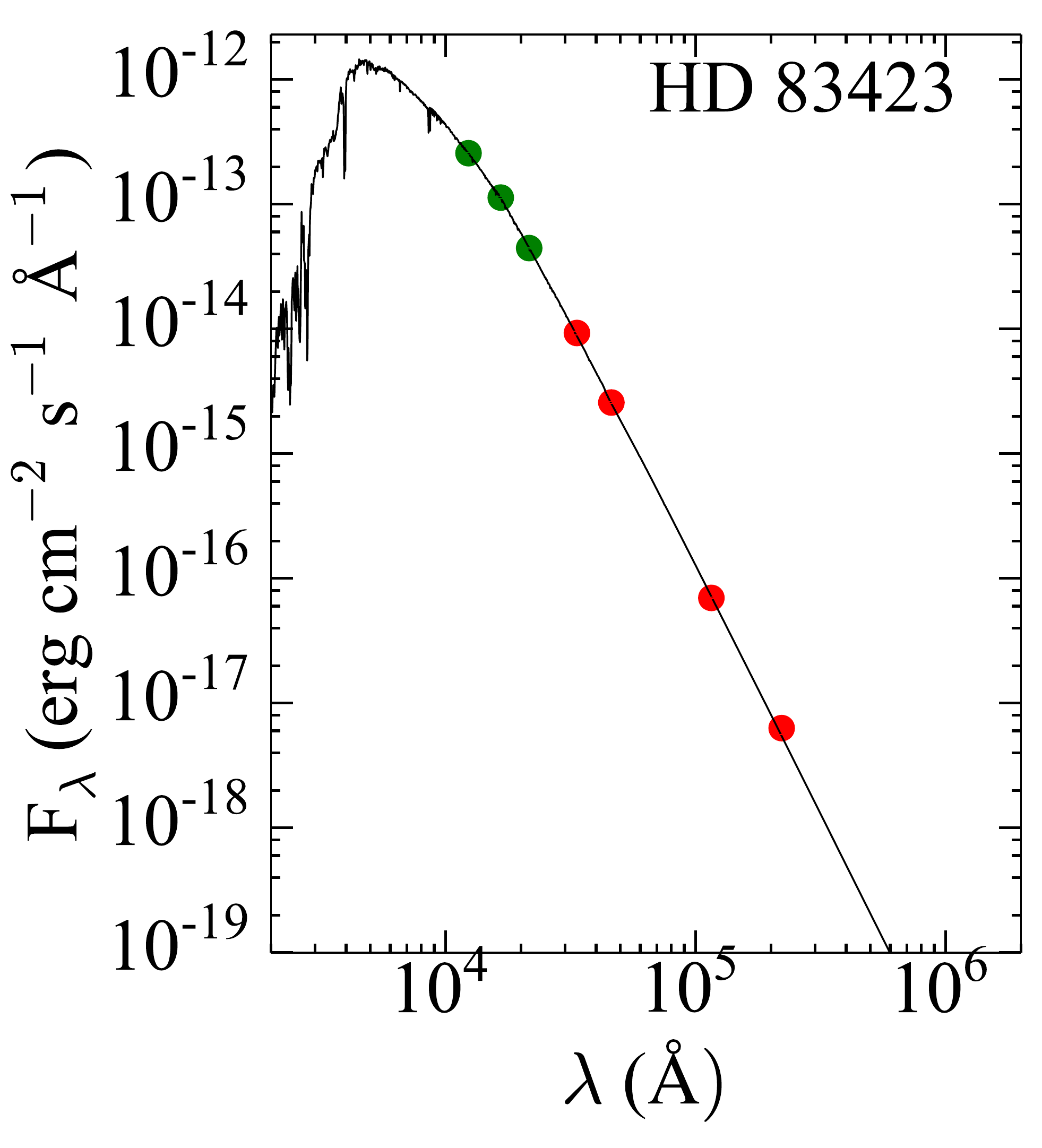}
\plotone{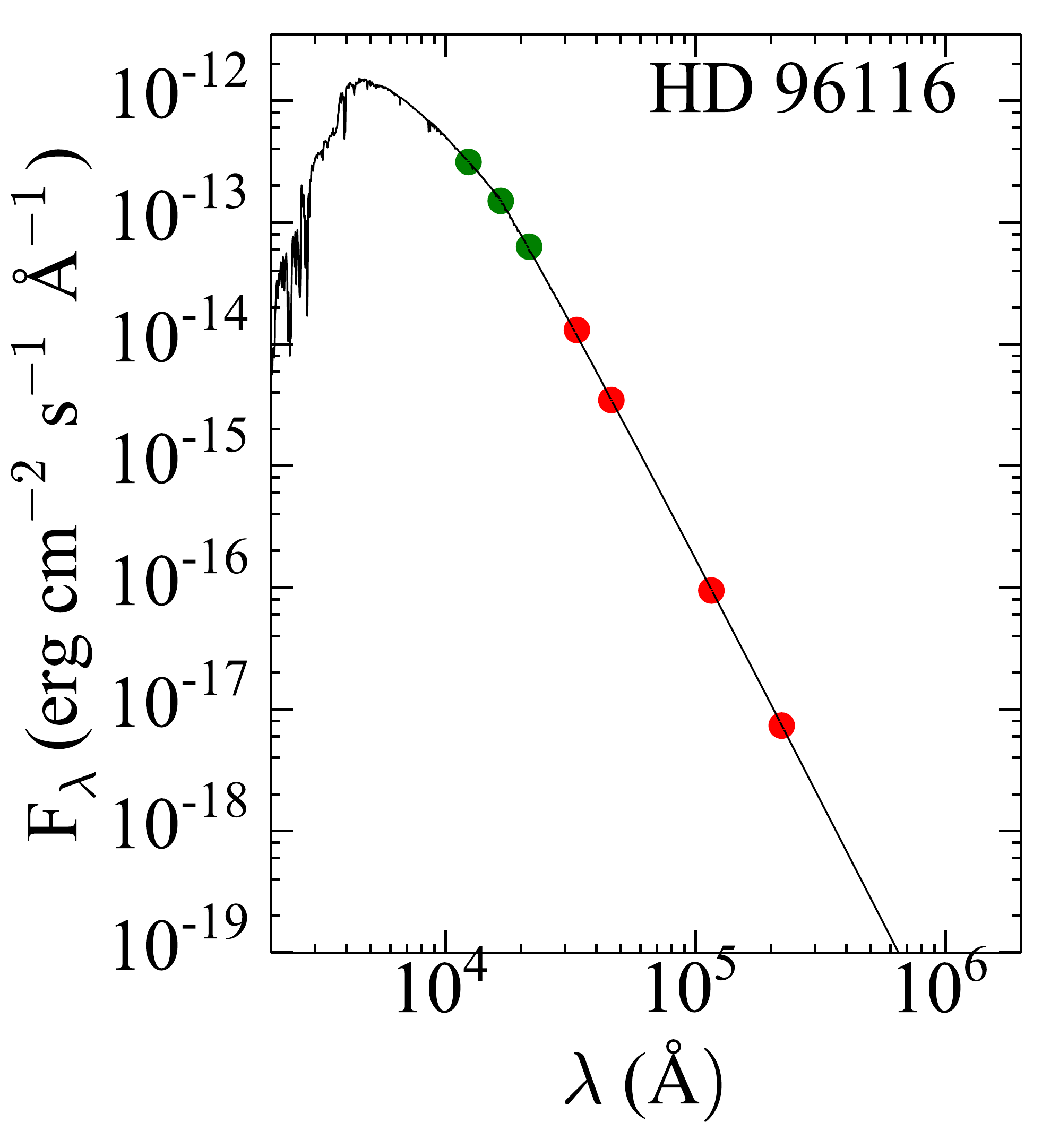}
\plotone{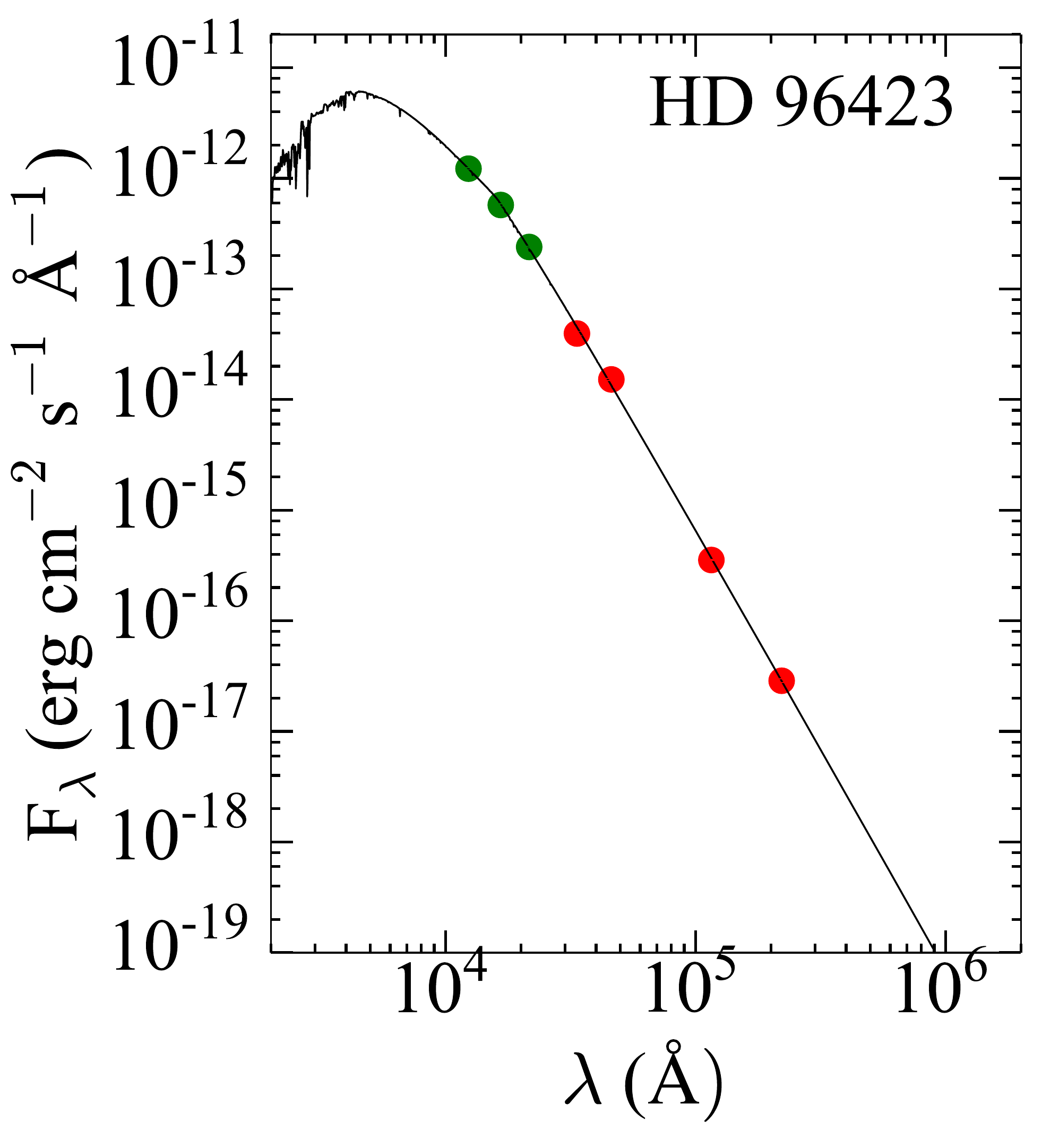}
\caption{Cont. Fig. 9}
\end{figure*}

\begin{figure*}[ht]
\centering
\epsscale{0.35}
\plotone{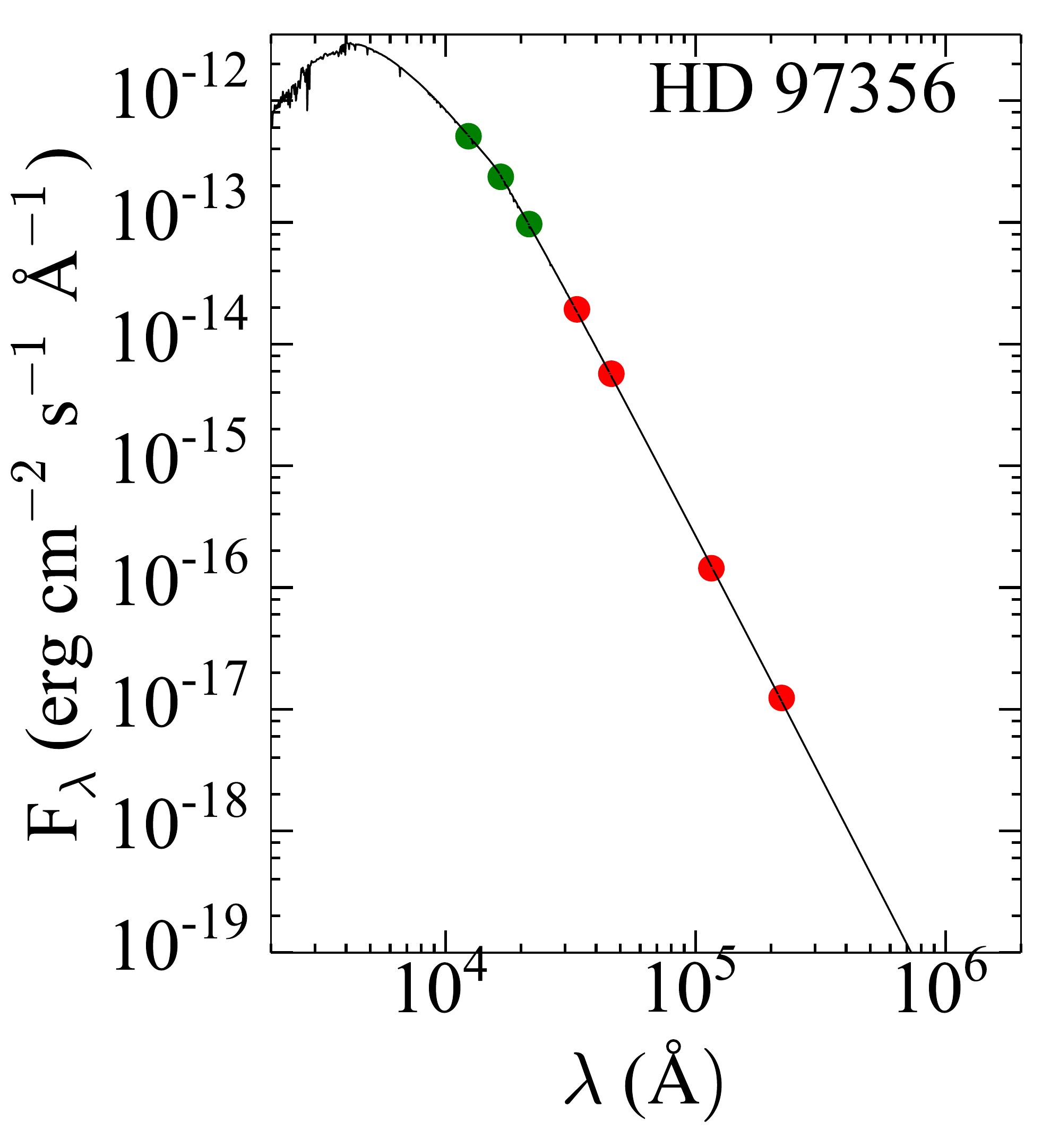}
\plotone{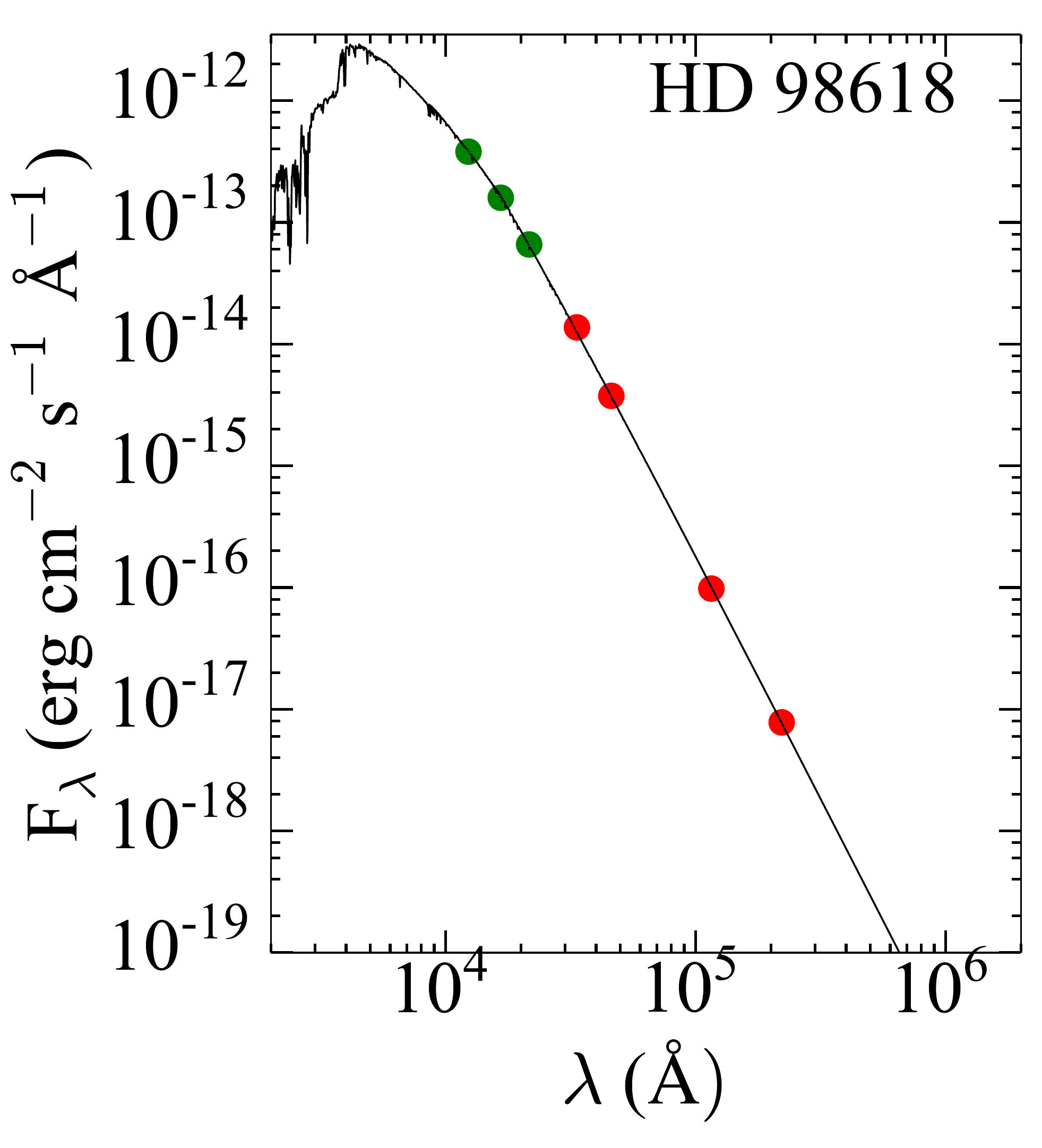}
\plotone{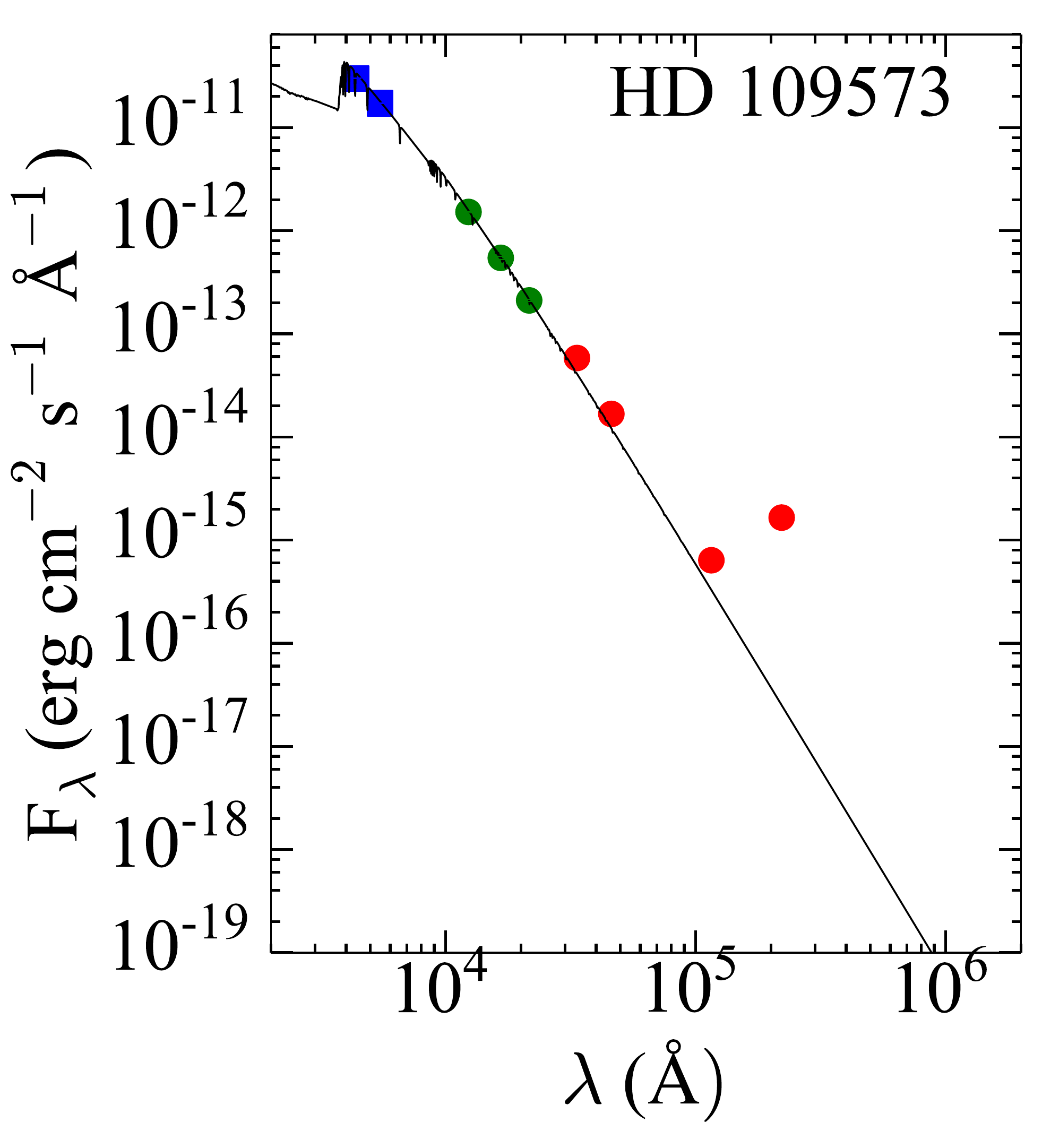}
\plotone{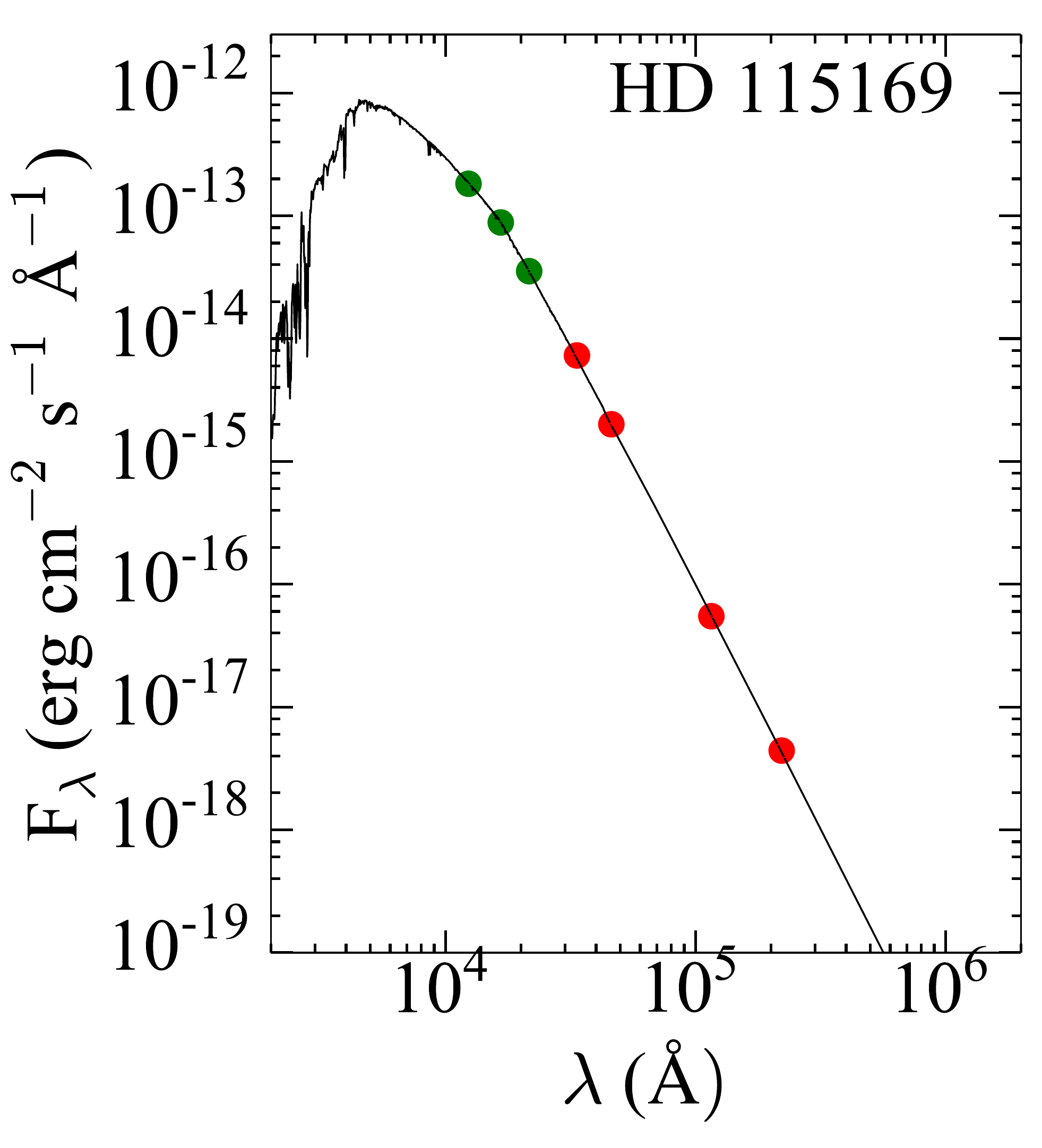}
\plotone{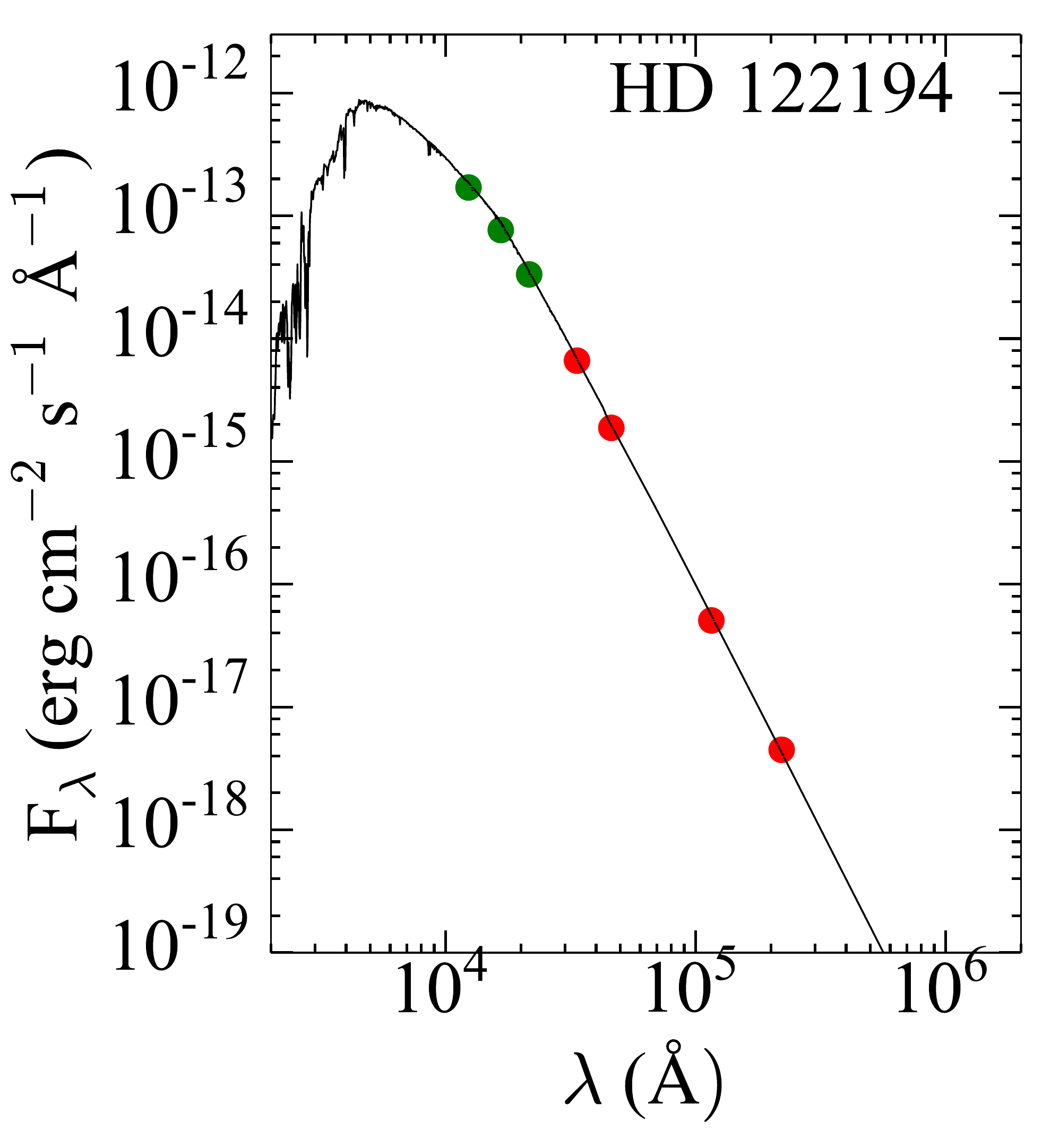}
\plotone{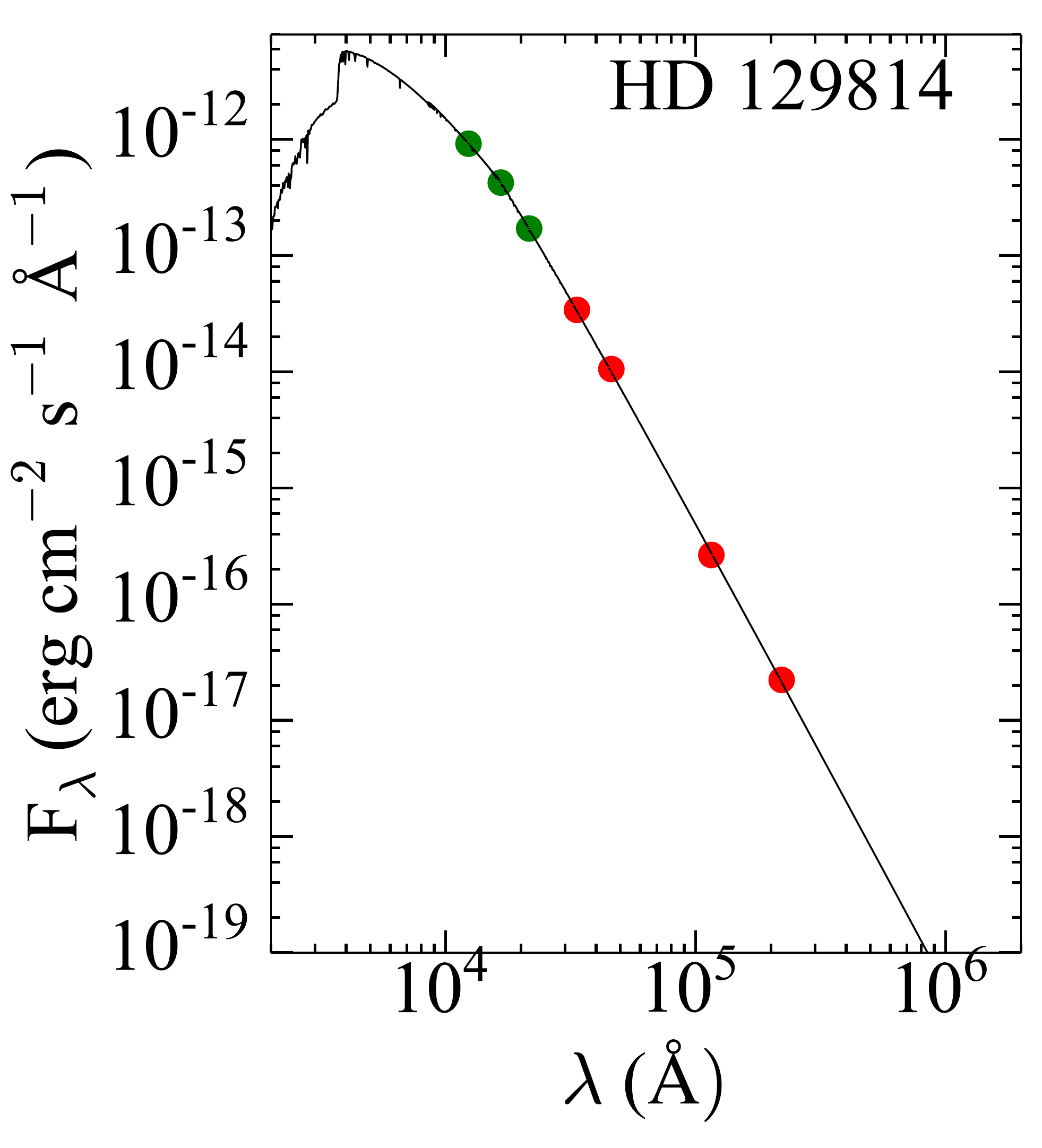}
\caption{Cont. Fig. 9}
\end{figure*}

\begin{figure*}[ht]
\centering
\epsscale{0.35}

\plotone{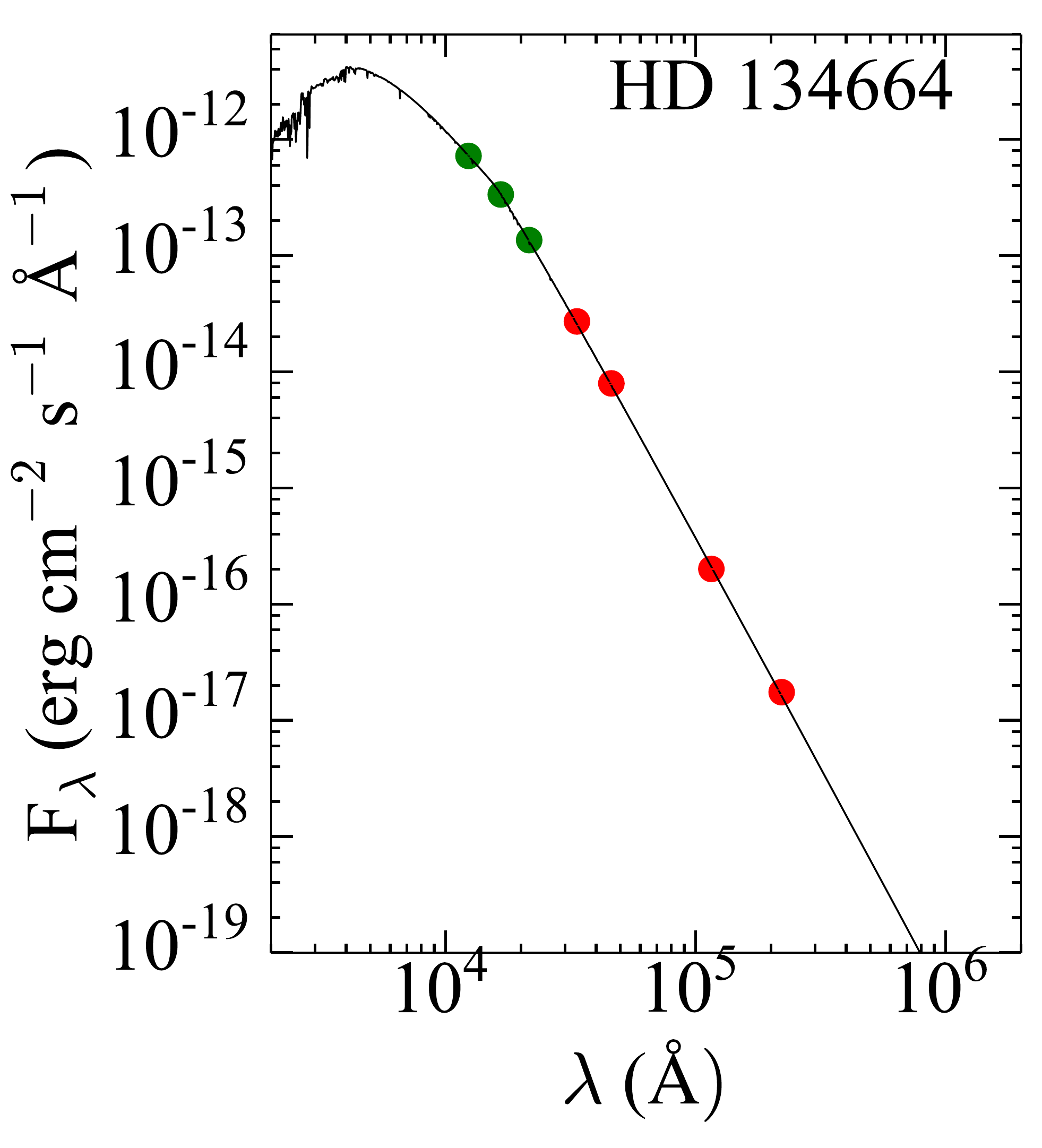}
\plotone{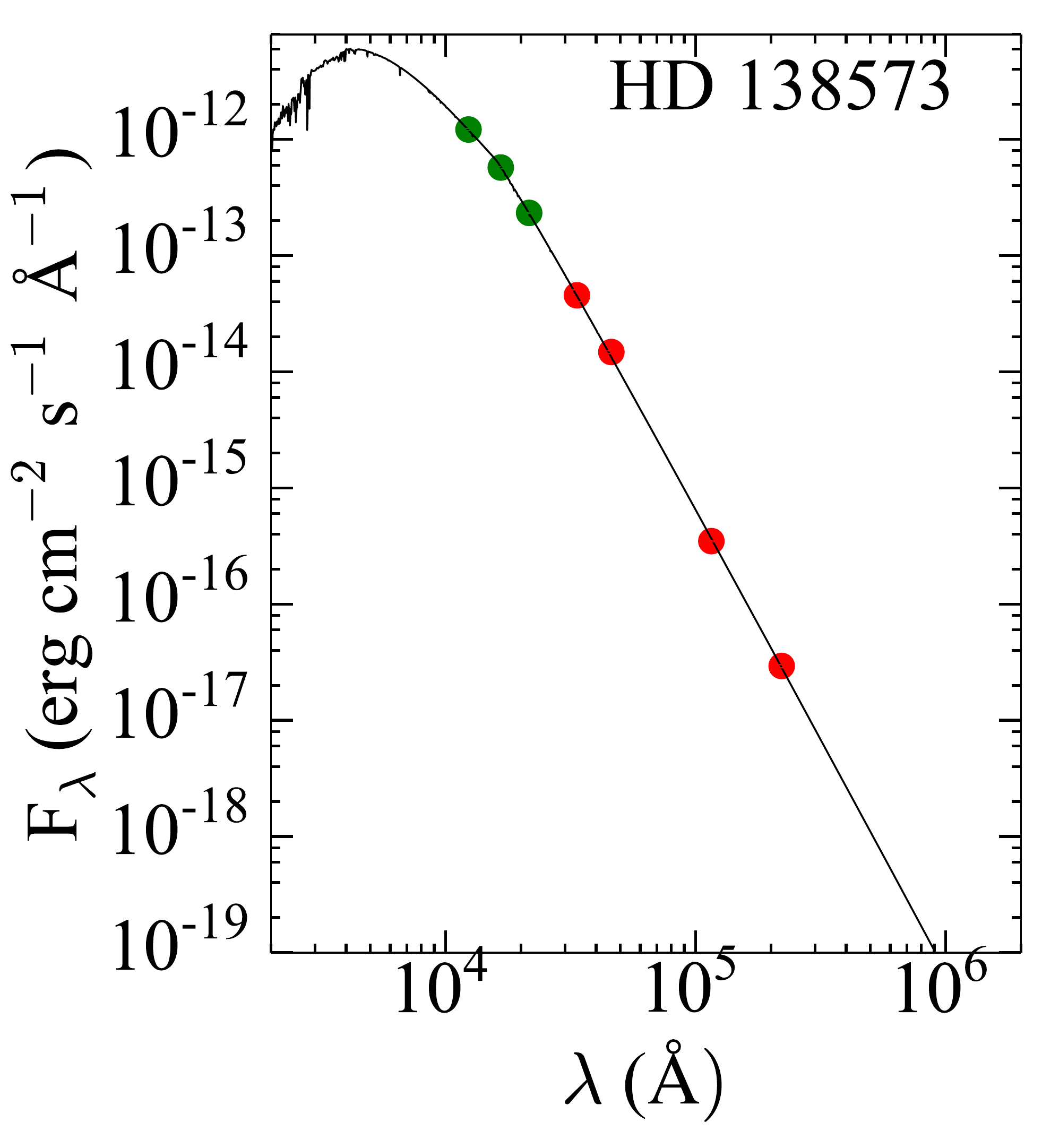}
\plotone{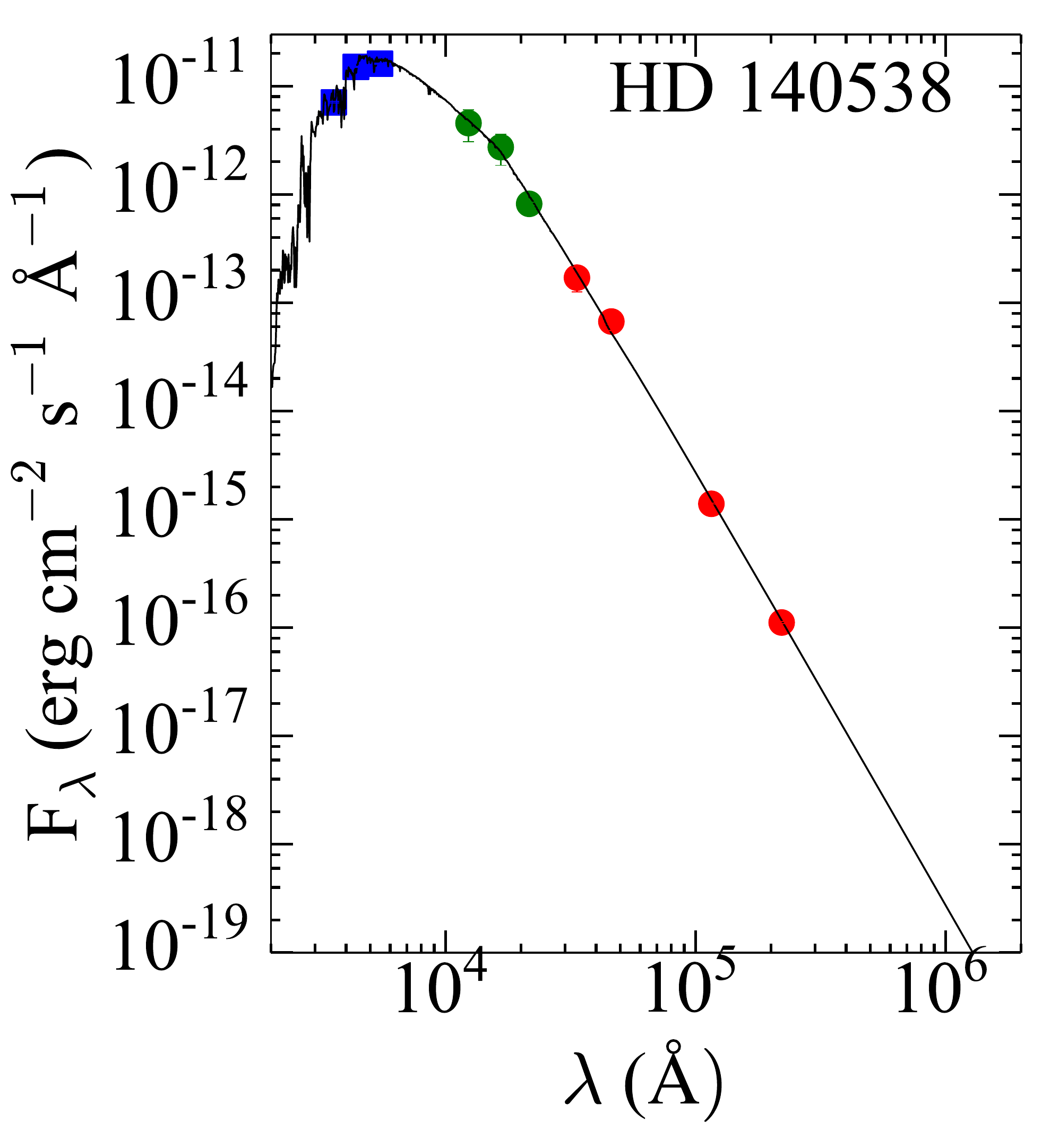}
\plotone{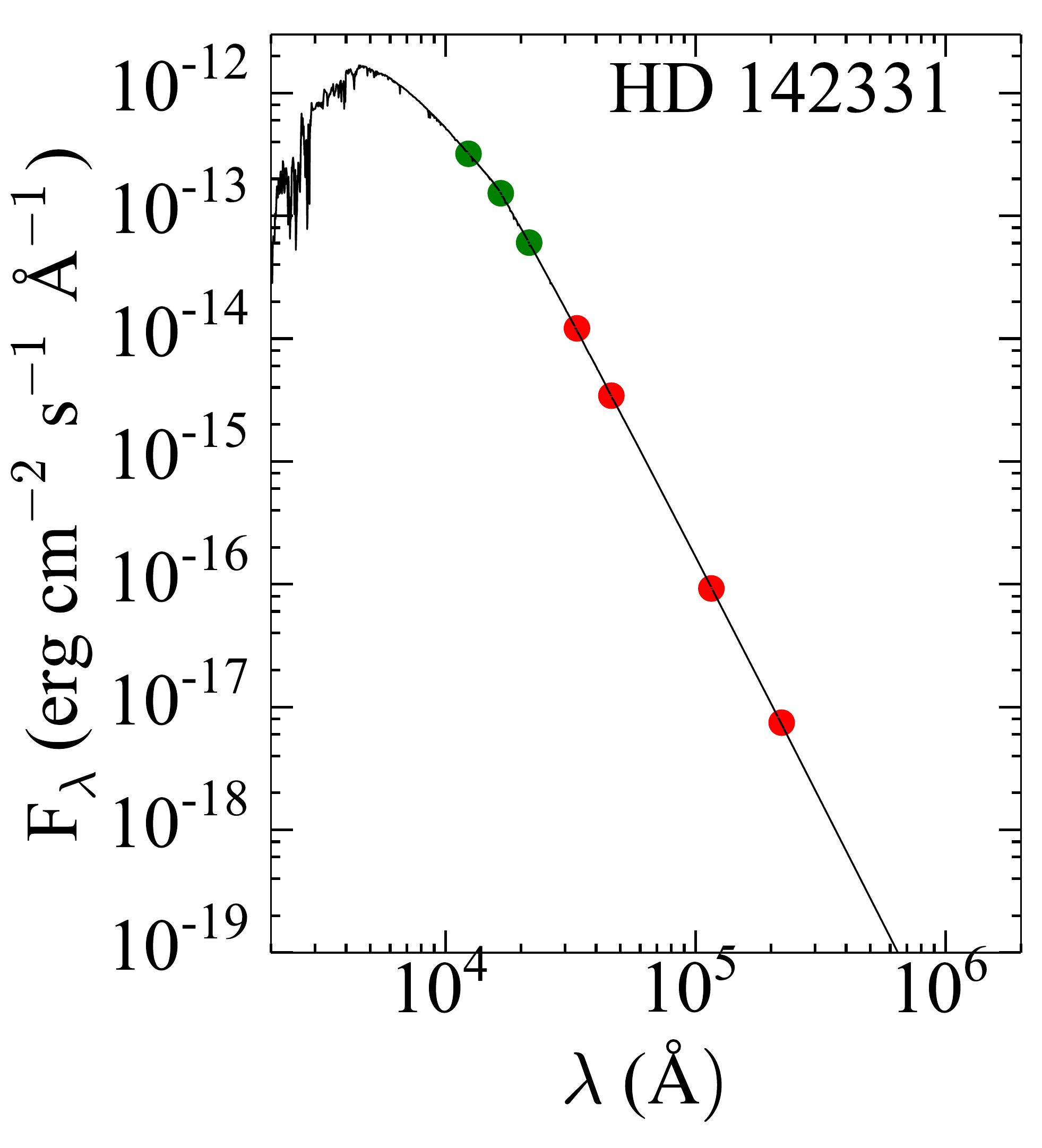}
\plotone{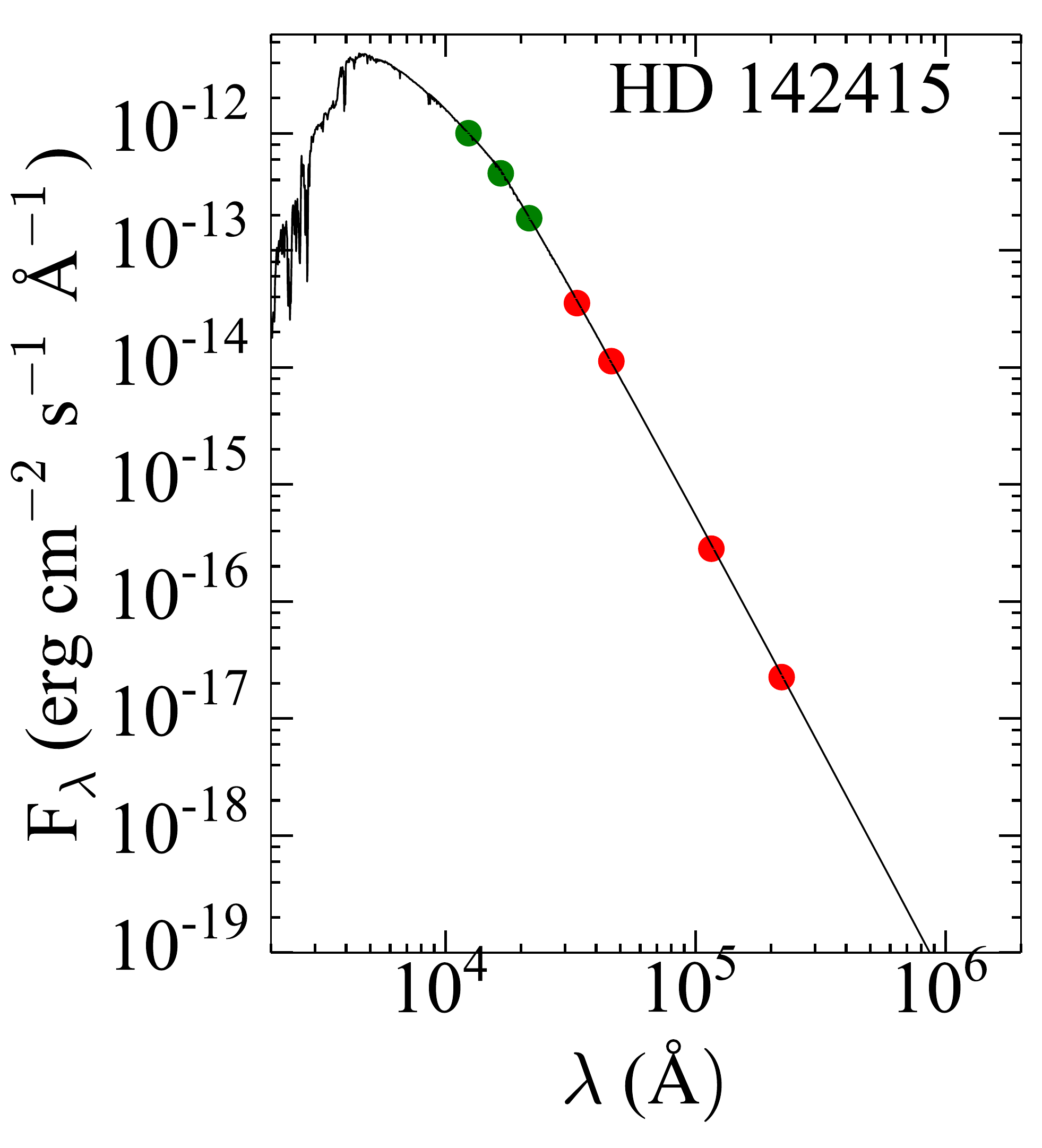}
\plotone{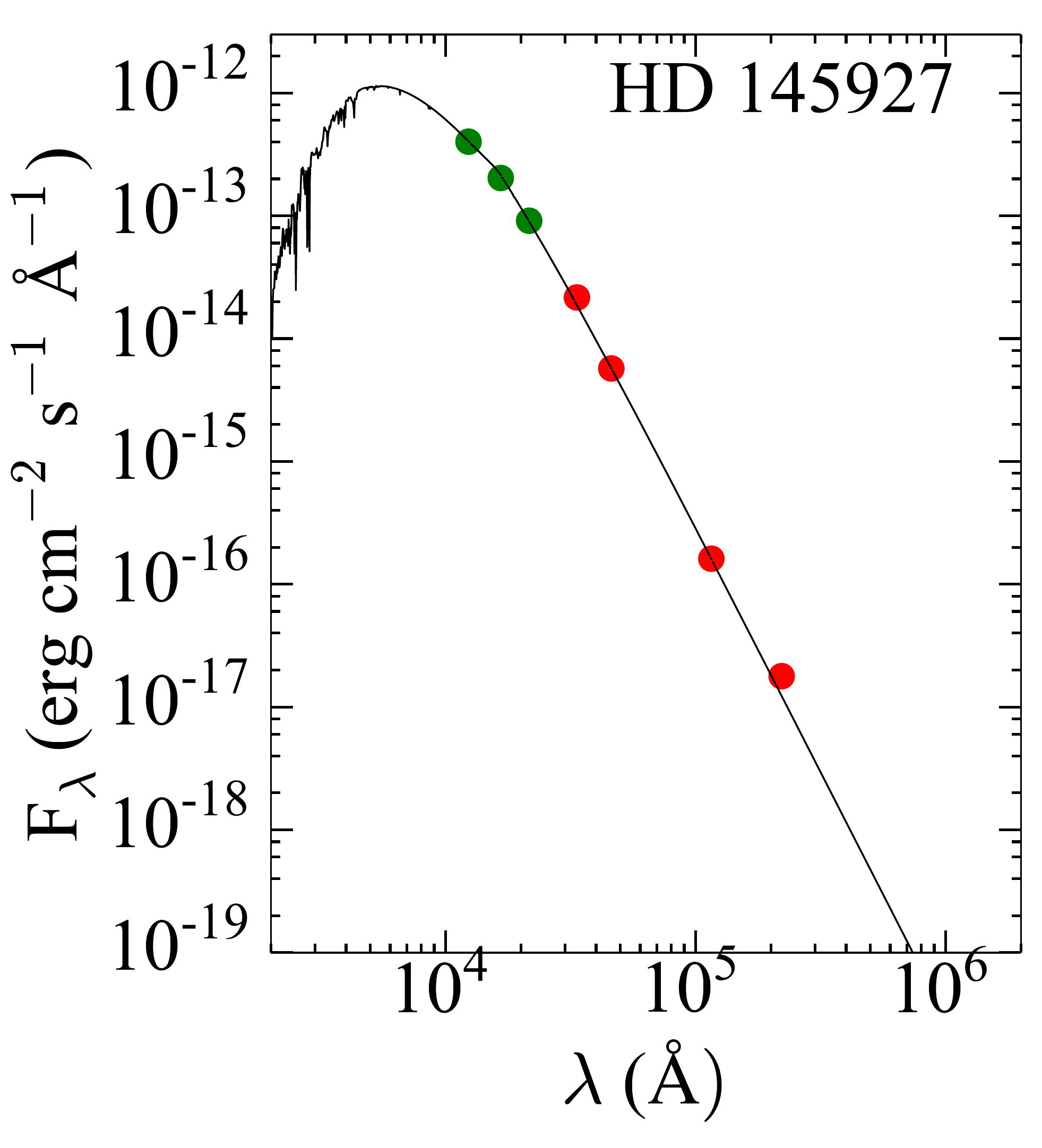}
\caption{Cont. Fig. 9}
\end{figure*}

\begin{figure*}[ht]
\centering
\epsscale{0.35}
\plotone{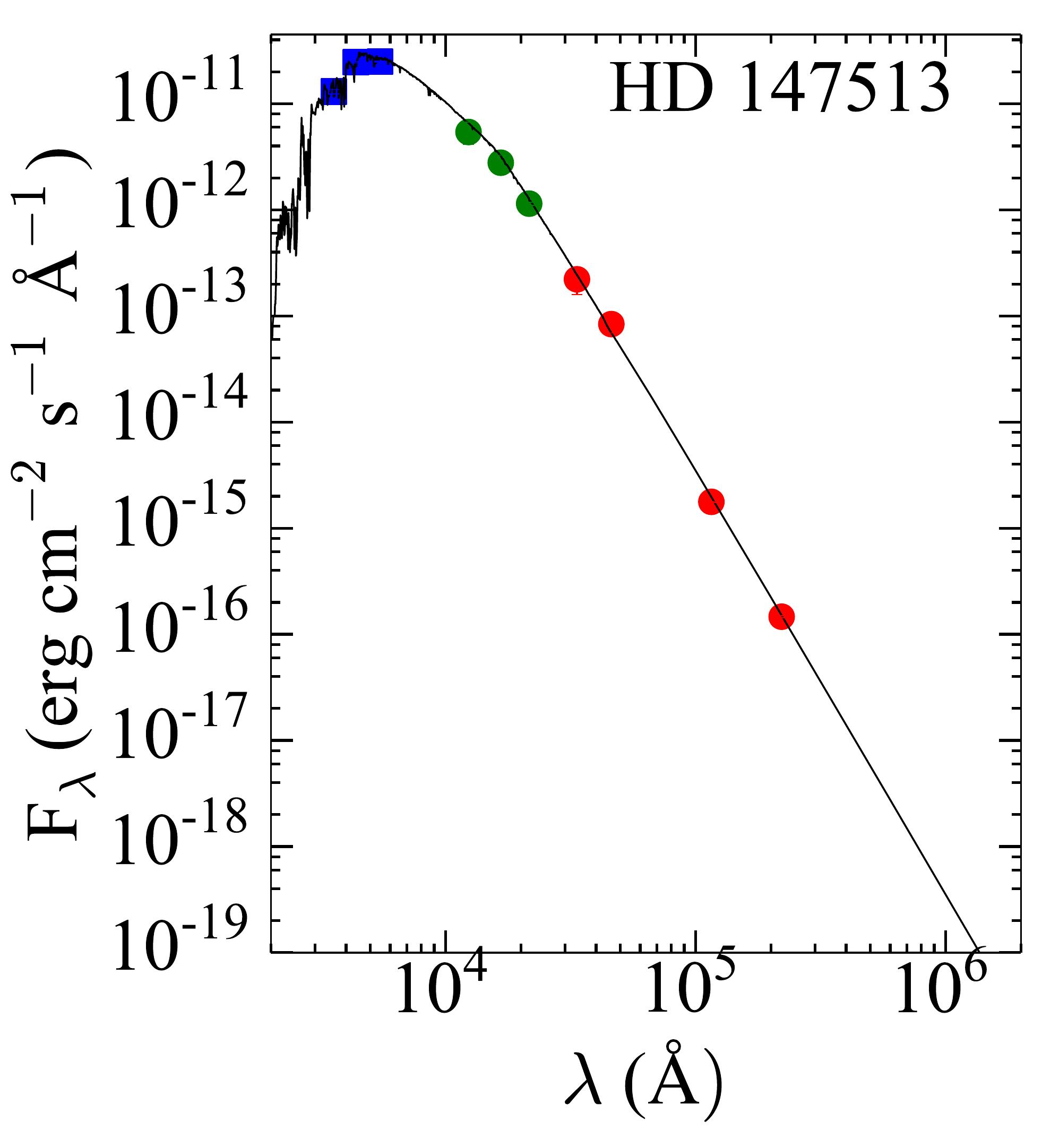}
\plotone{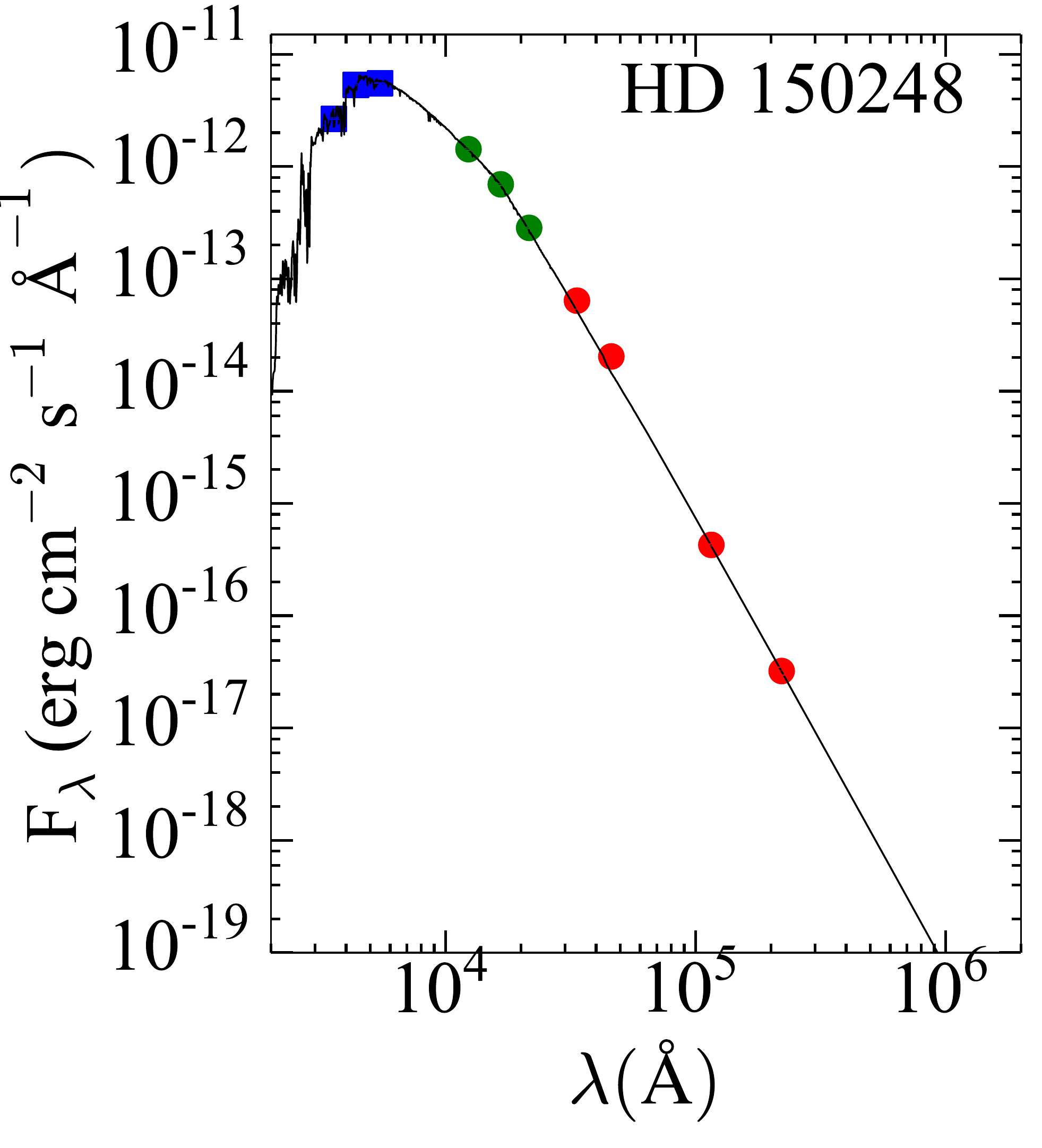}
\plotone{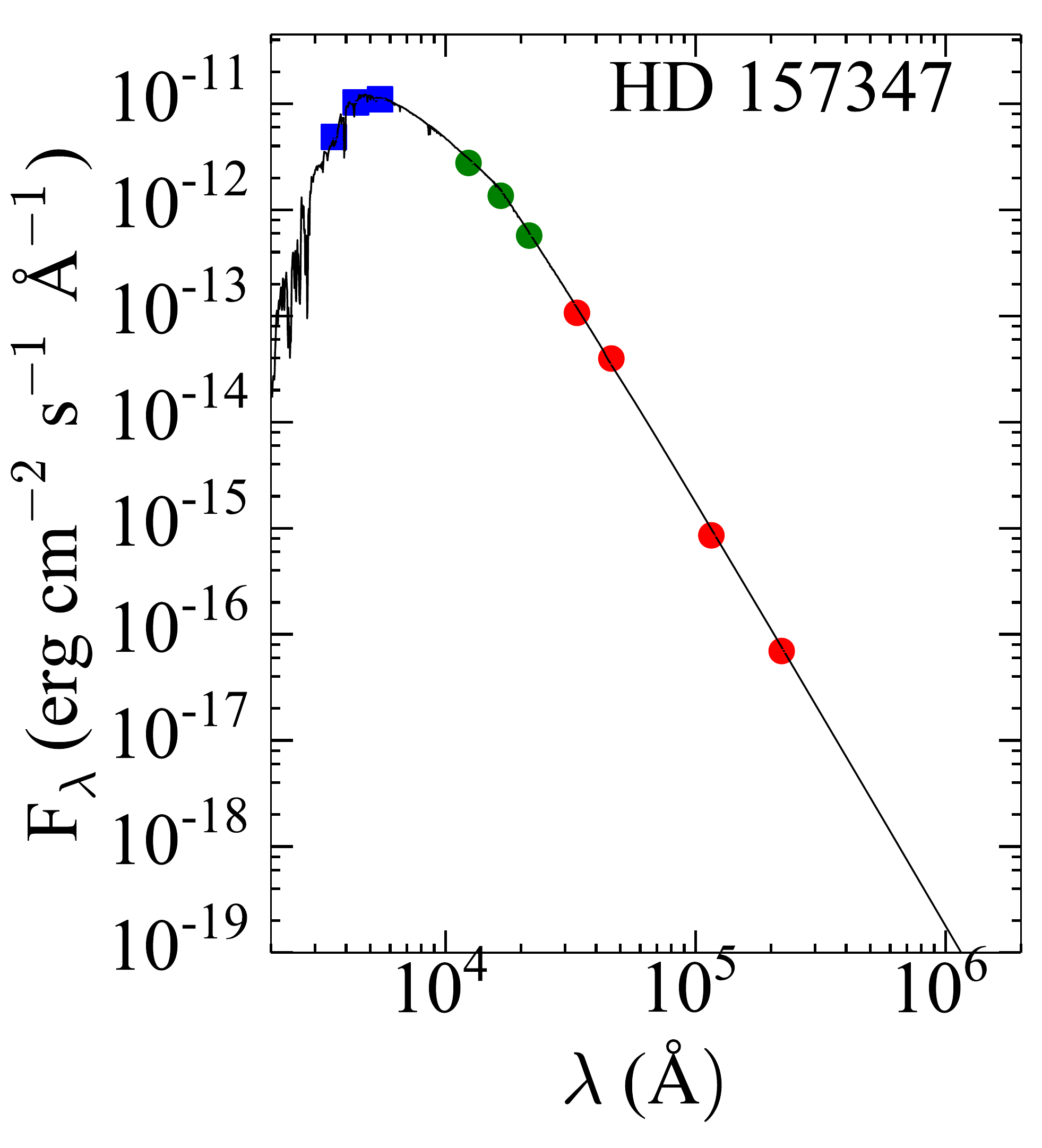}
\plotone{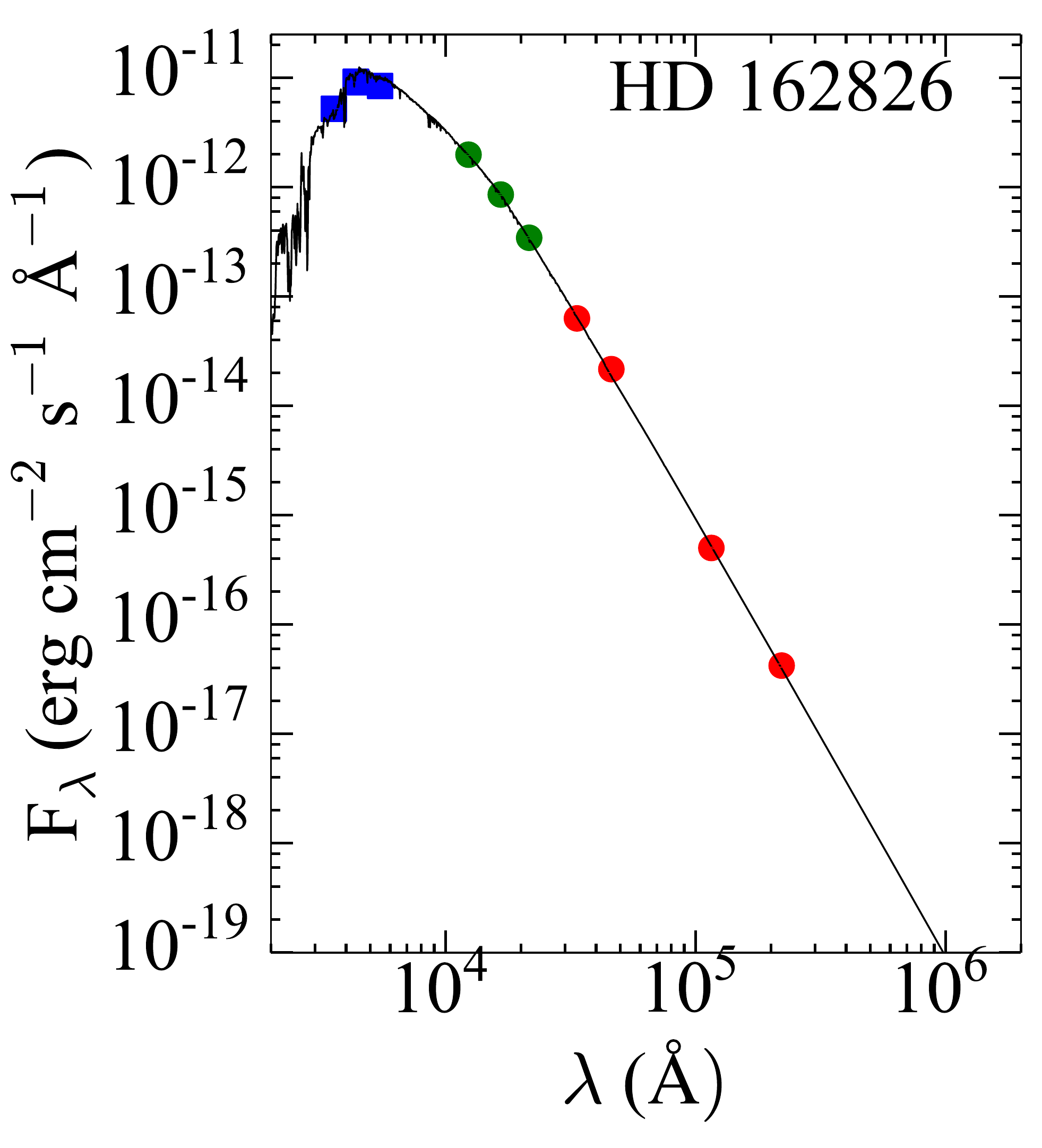}
\plotone{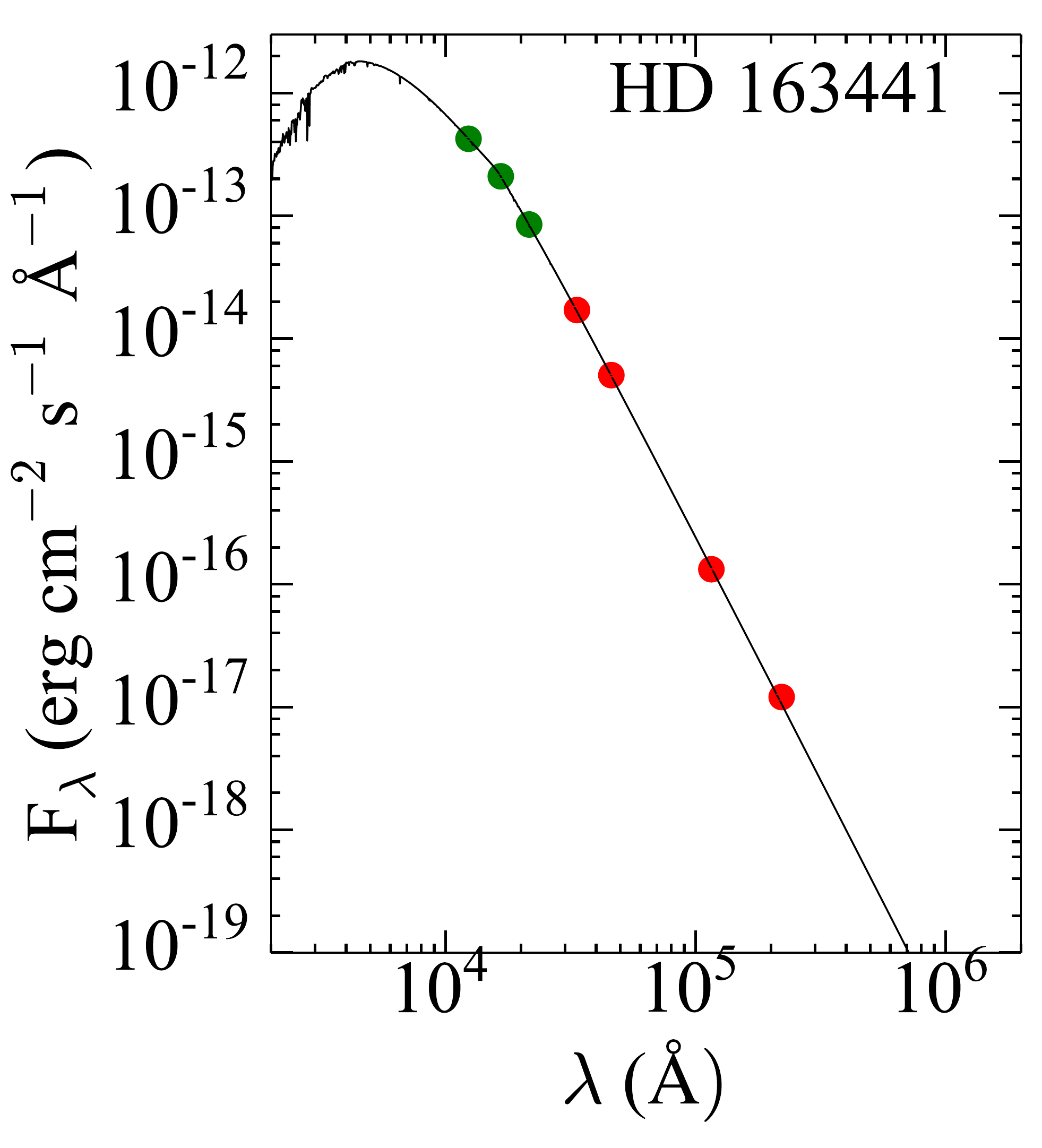}
\plotone{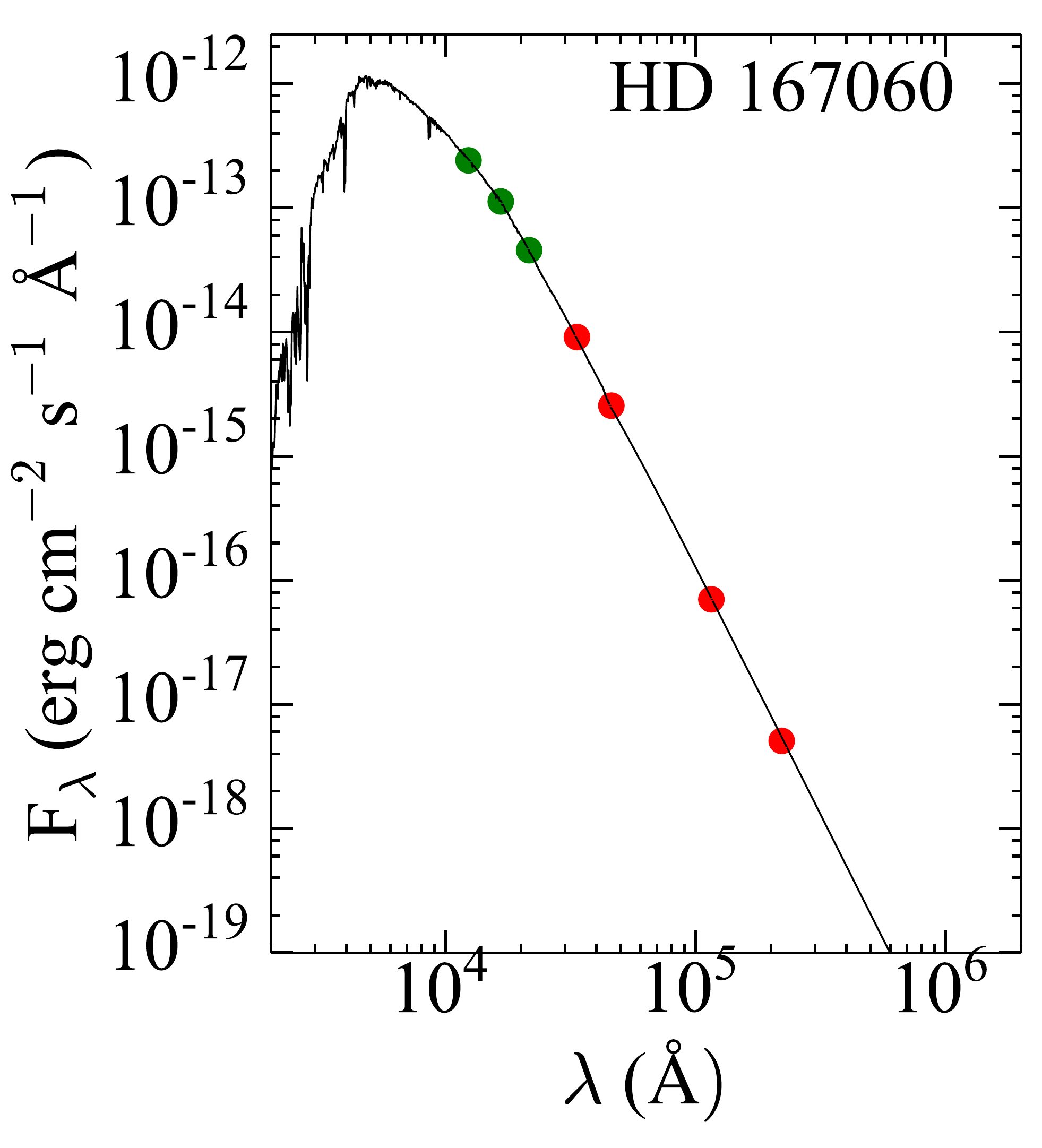}
\caption{Cont. Fig. 9}
\end{figure*}

\begin{figure*}[ht]
\centering
\epsscale{0.35}
\plotone{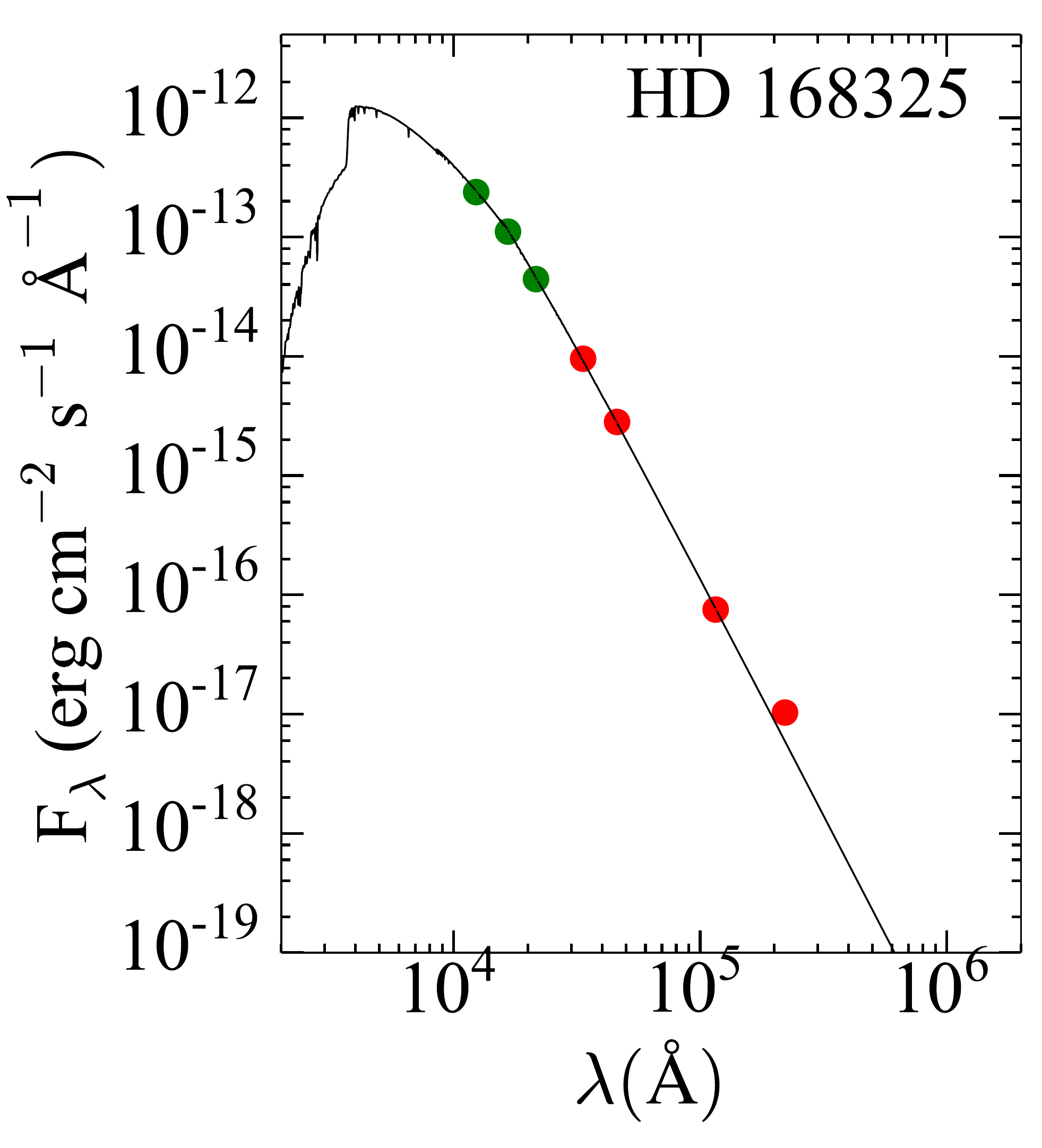}
\plotone{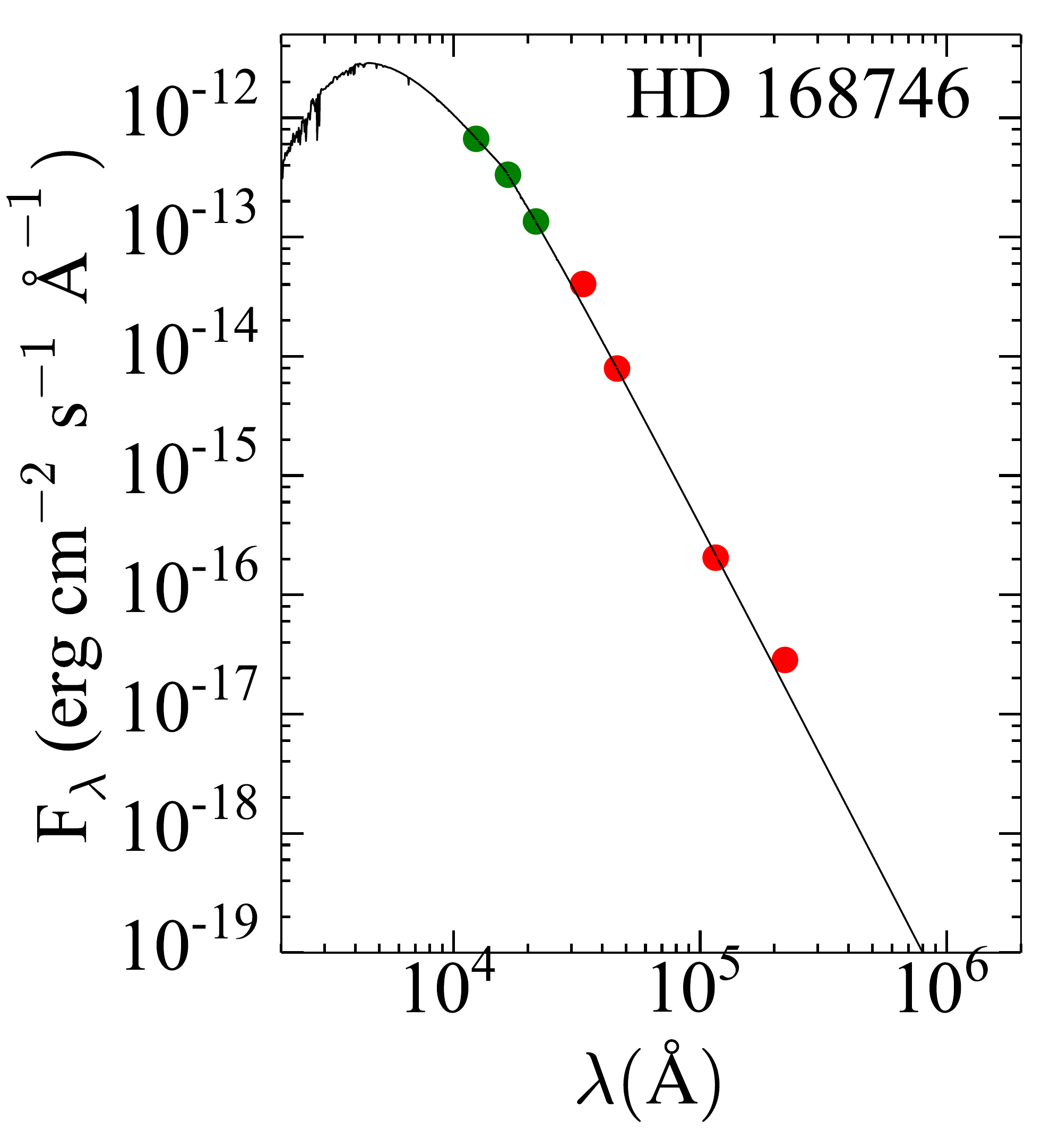}
\plotone{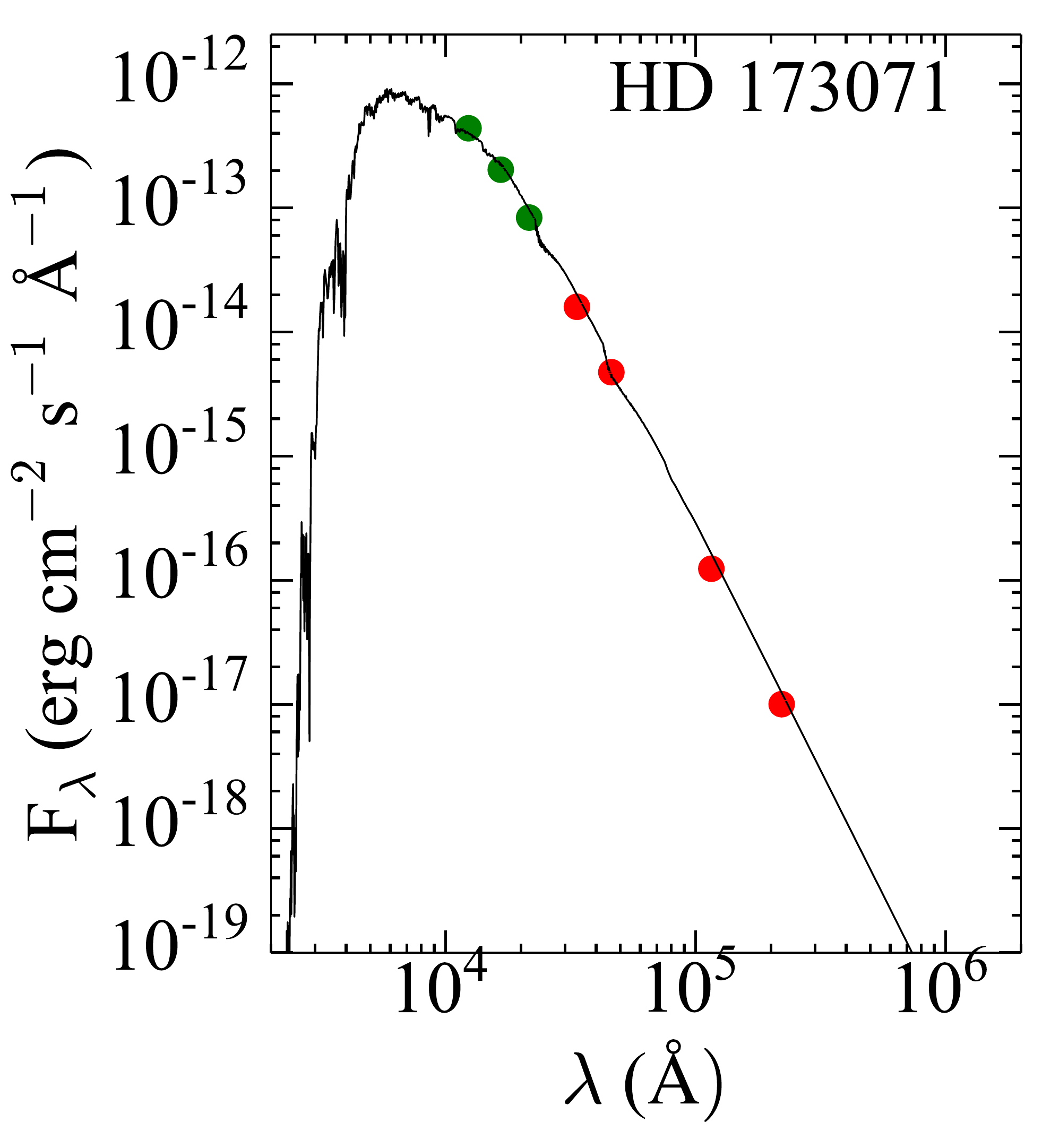}
\plotone{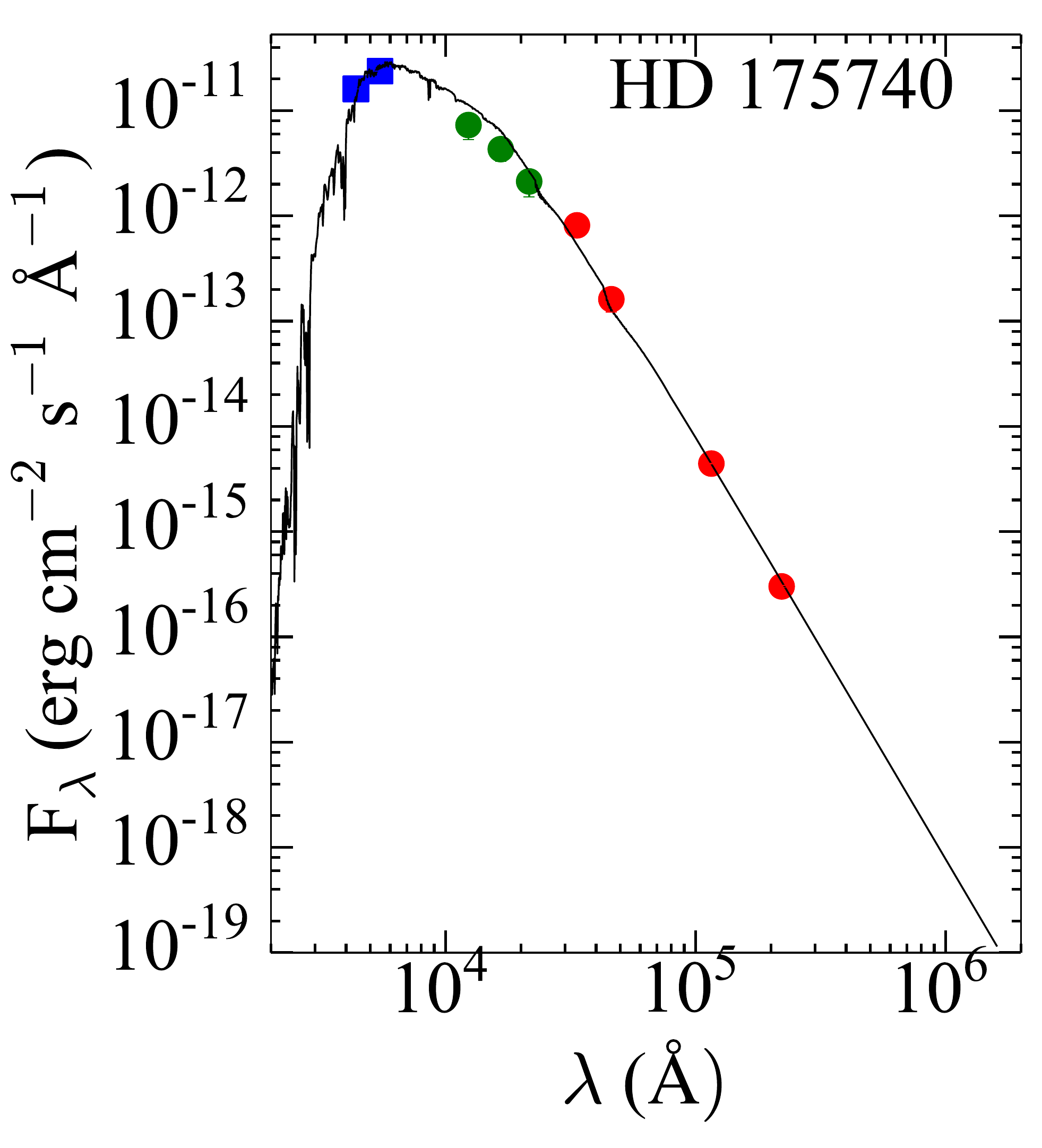}
\plotone{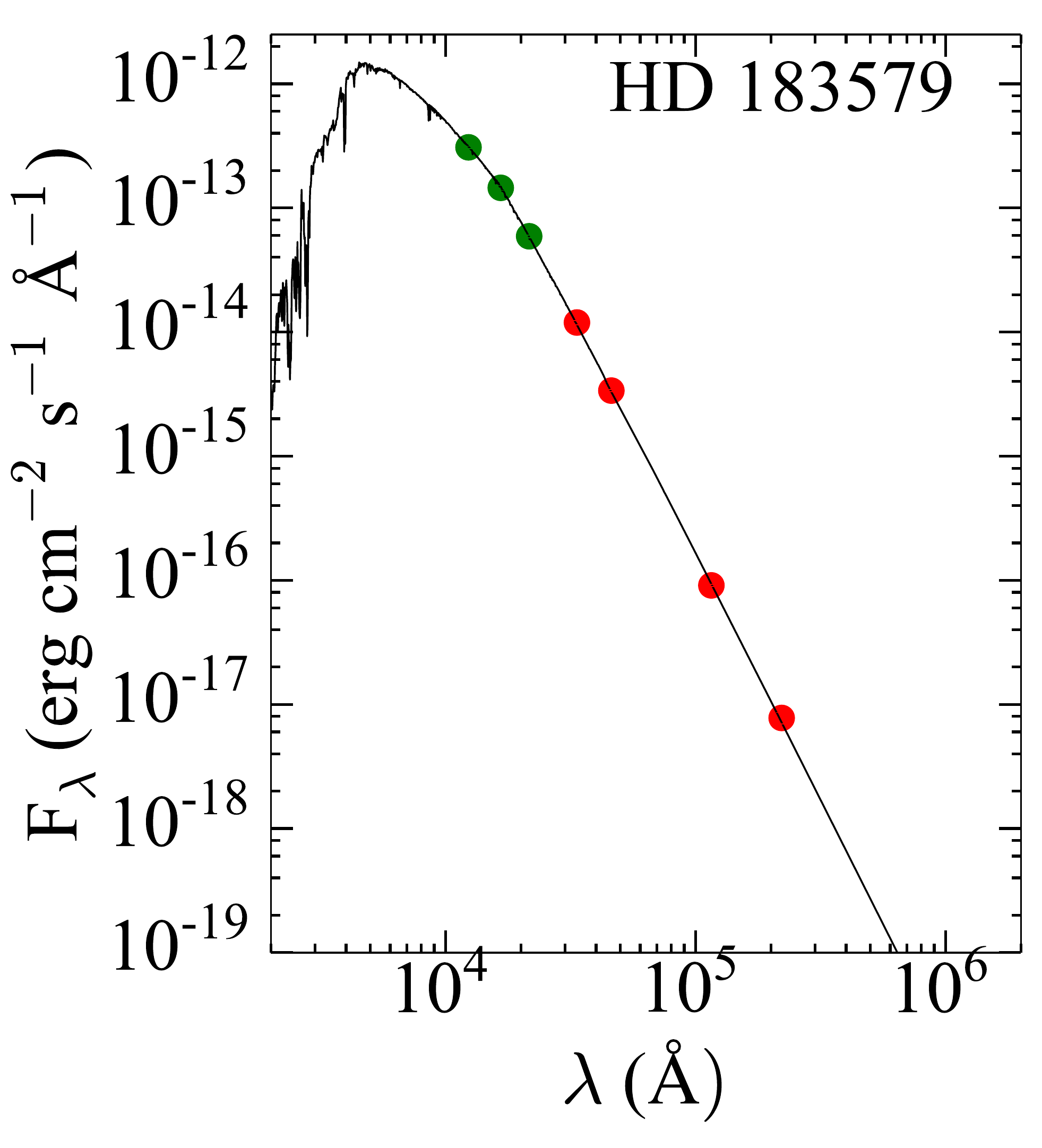}
\plotone{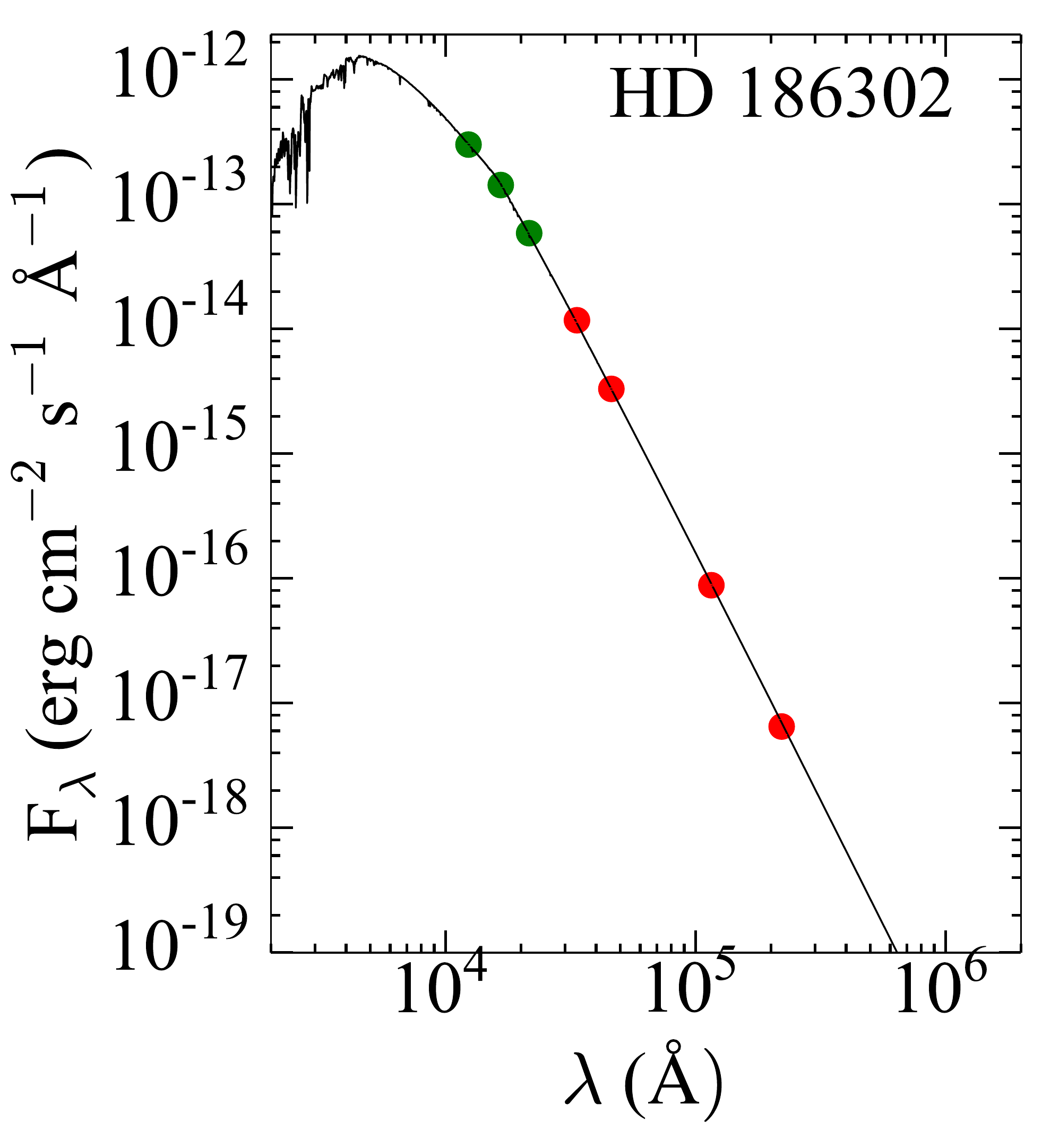}

\caption{Cont. Fig. 9}
\end{figure*}

\begin{figure*}[ht]
\centering
\epsscale{0.35}

\plotone{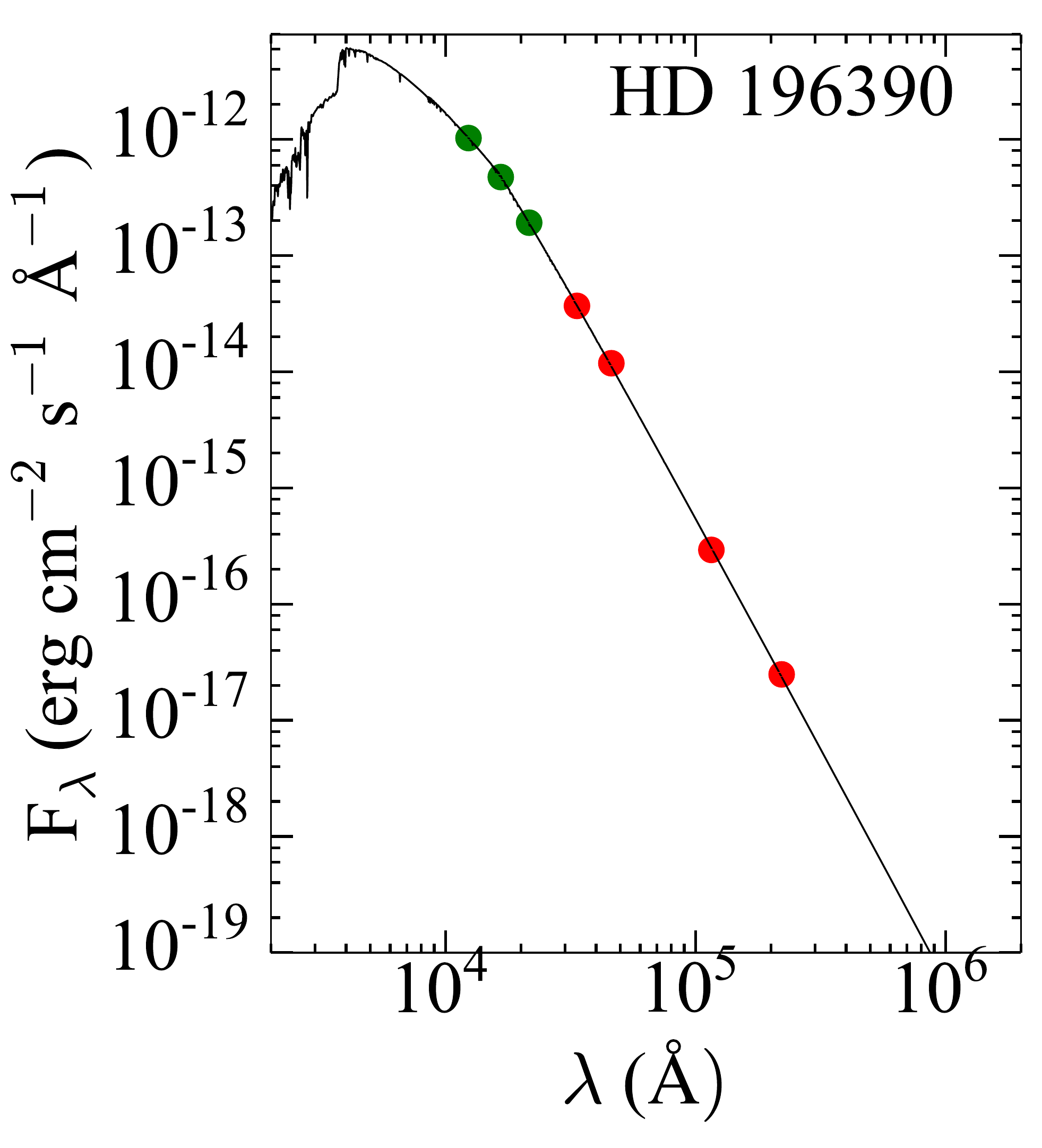}
\plotone{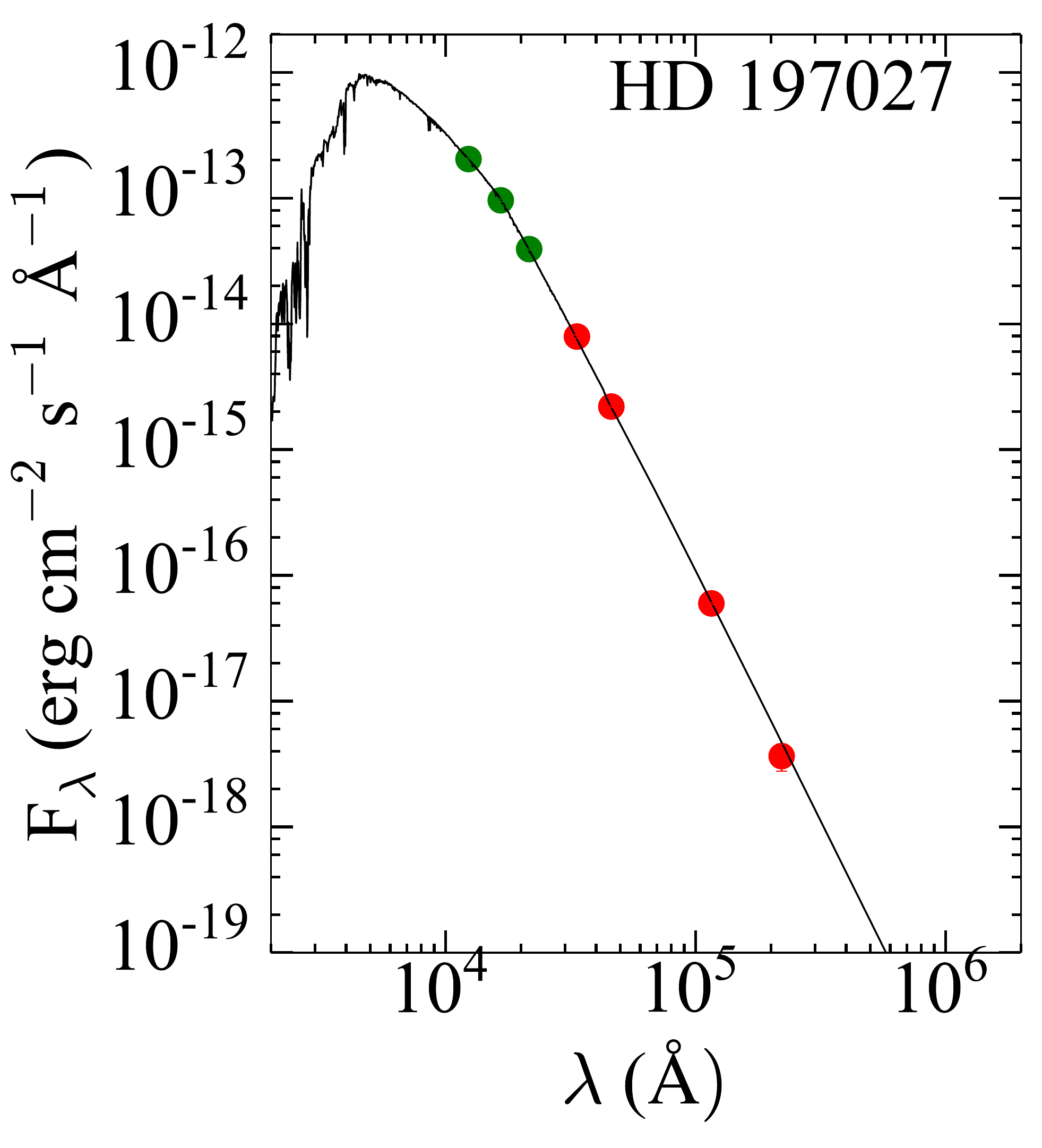}
\plotone{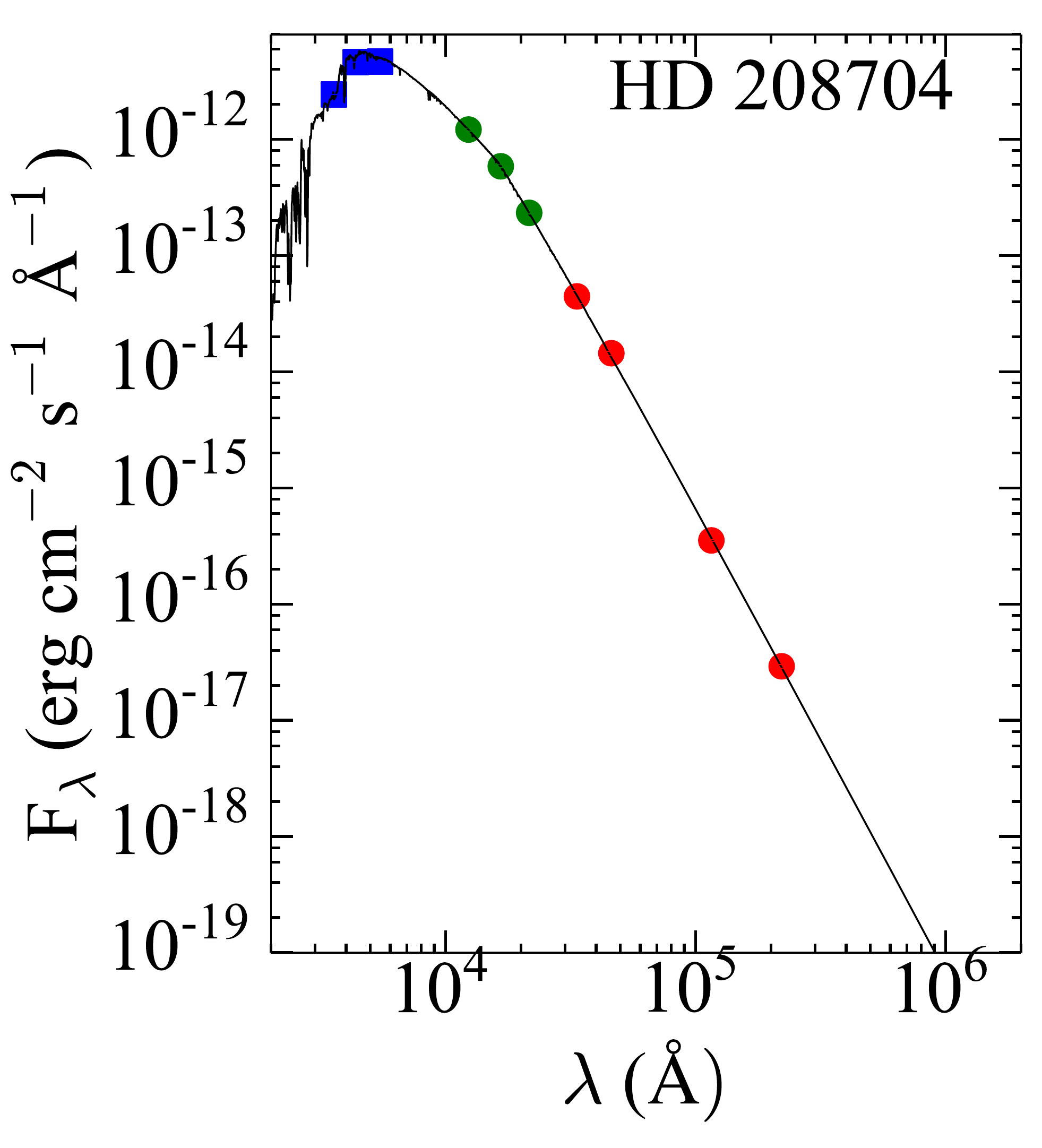}
\plotone{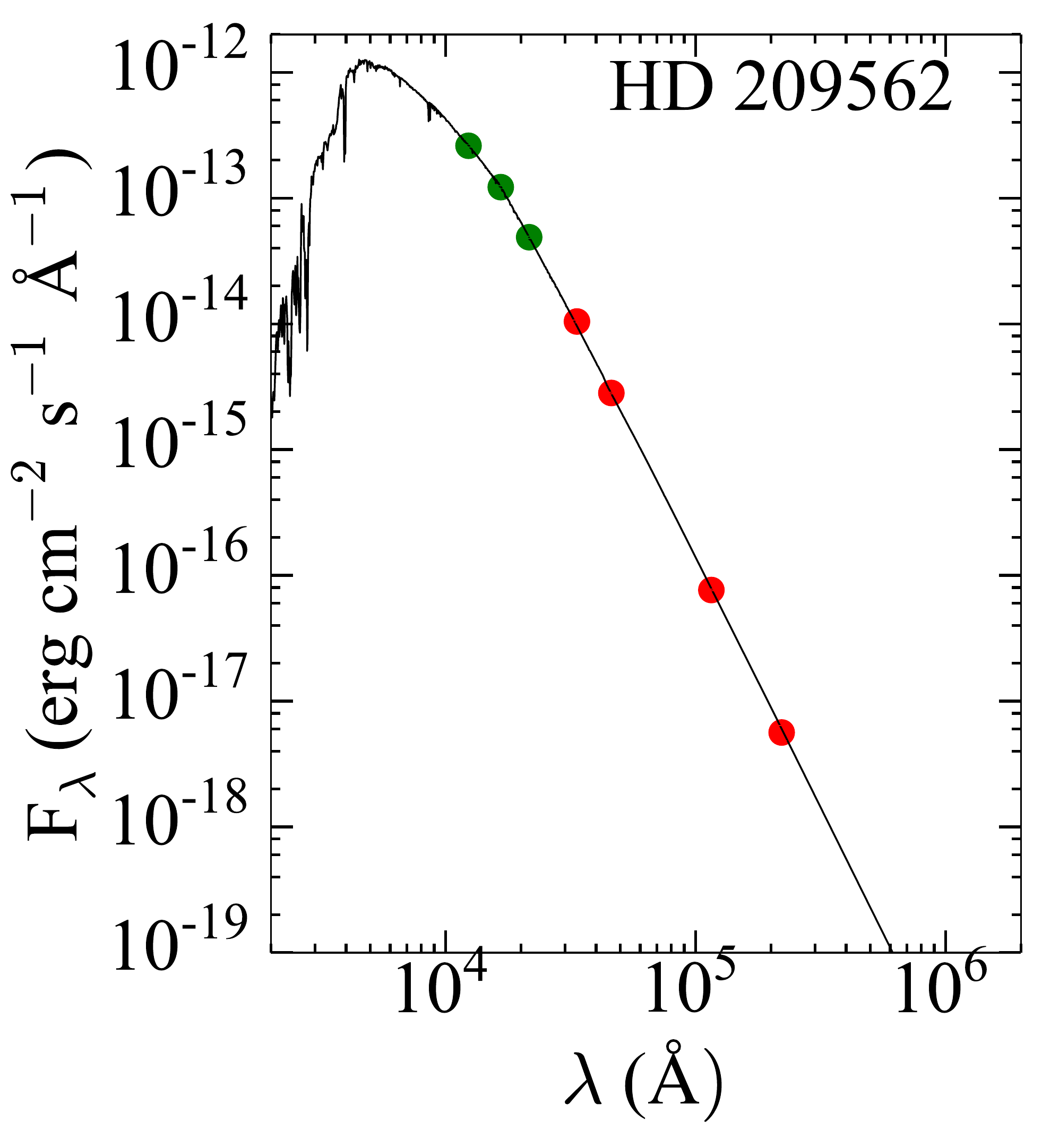}
\plotone{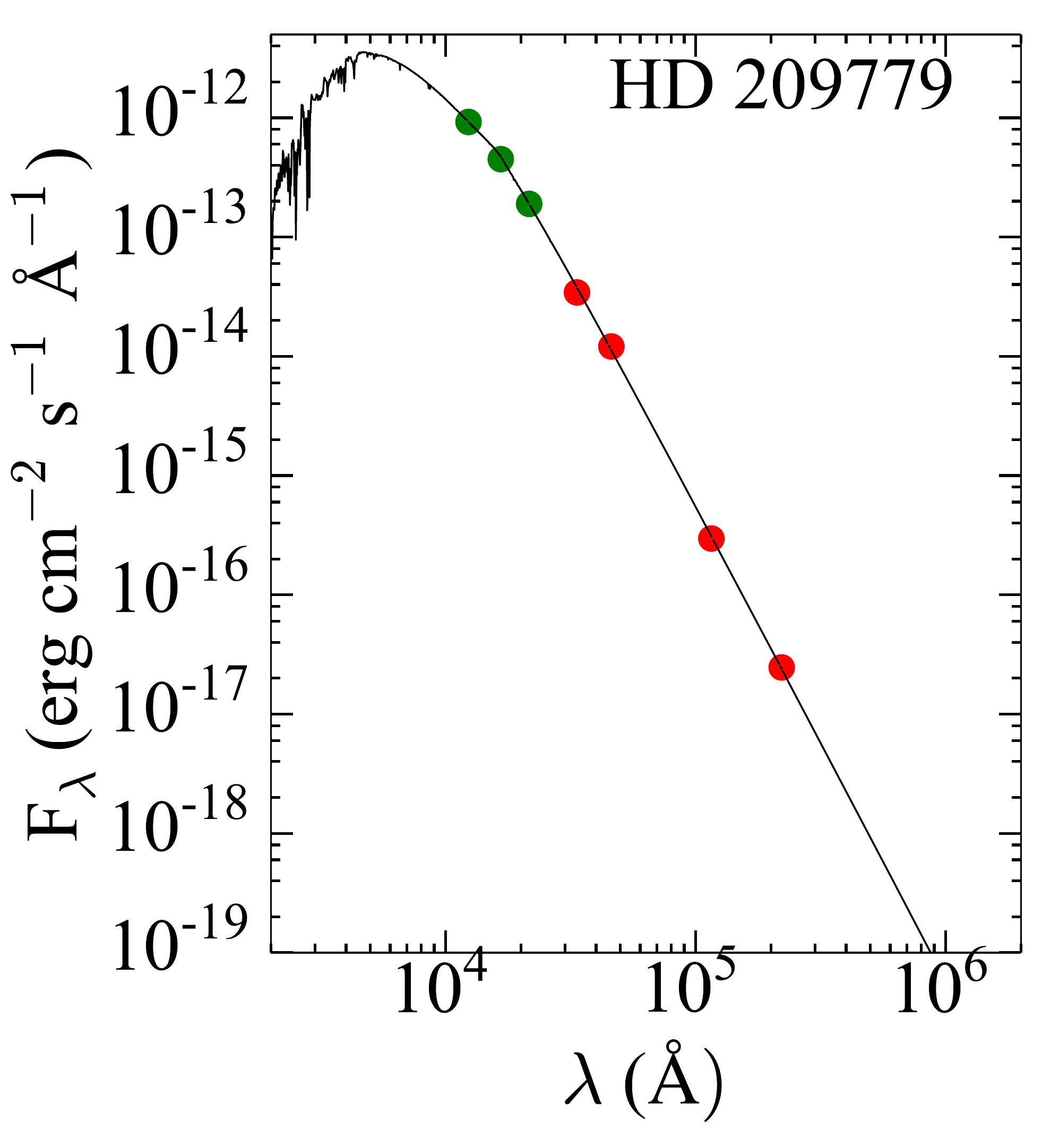}
\plotone{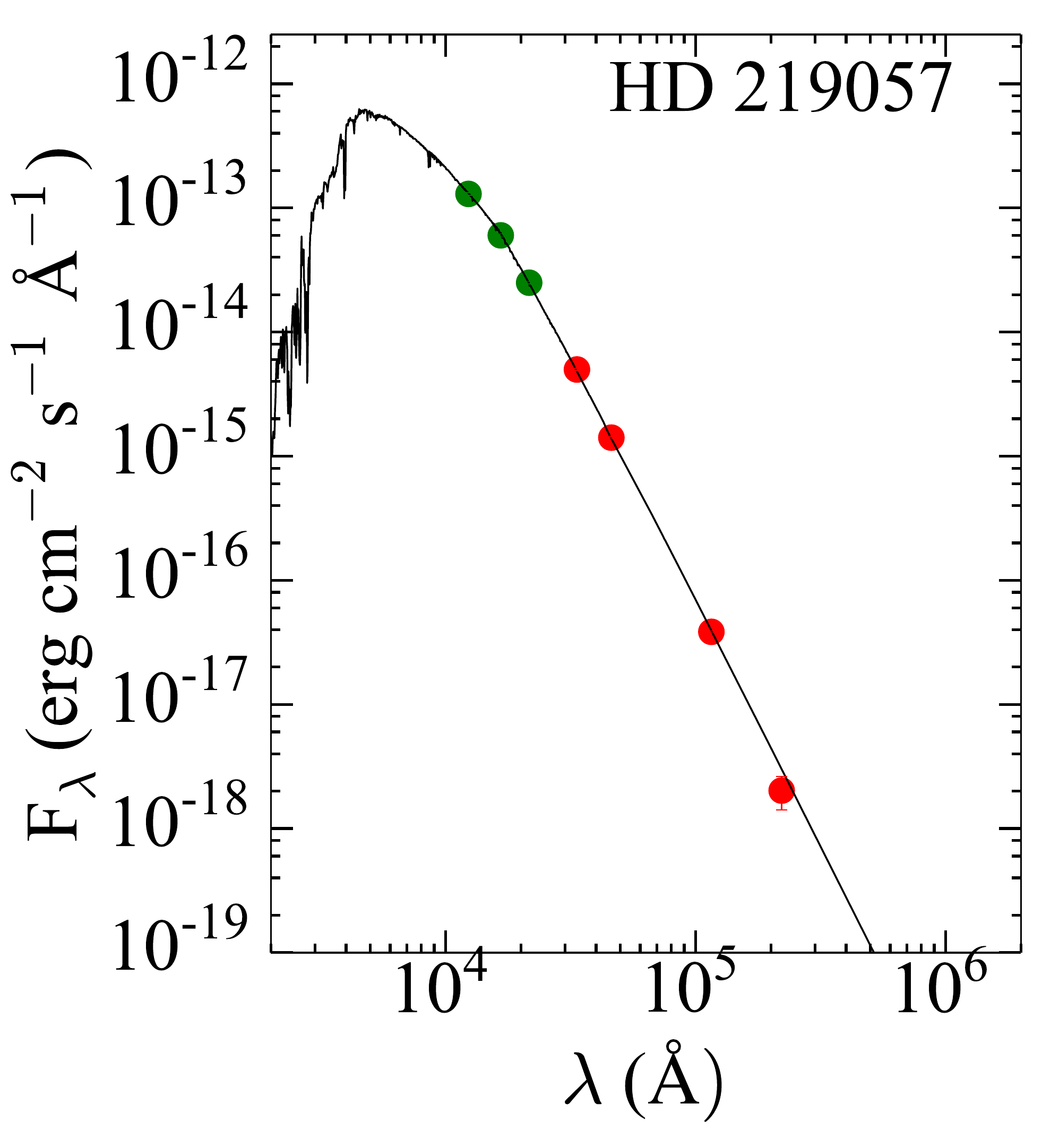}
\caption{Cont. Fig. 9}
\end{figure*}

\begin{figure*}[ht]
\centering
\epsscale{0.35}
\plotone{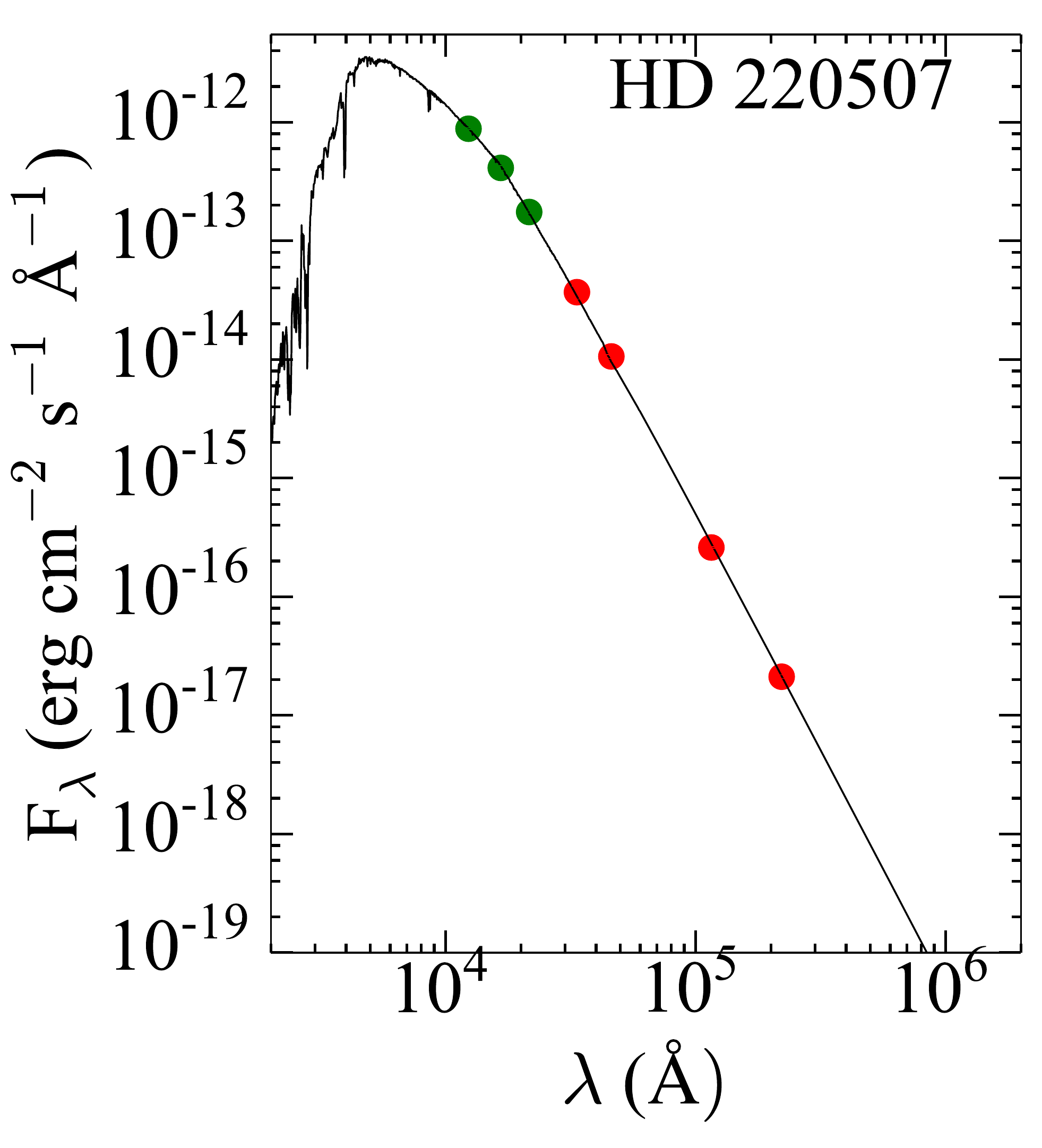}
\plotone{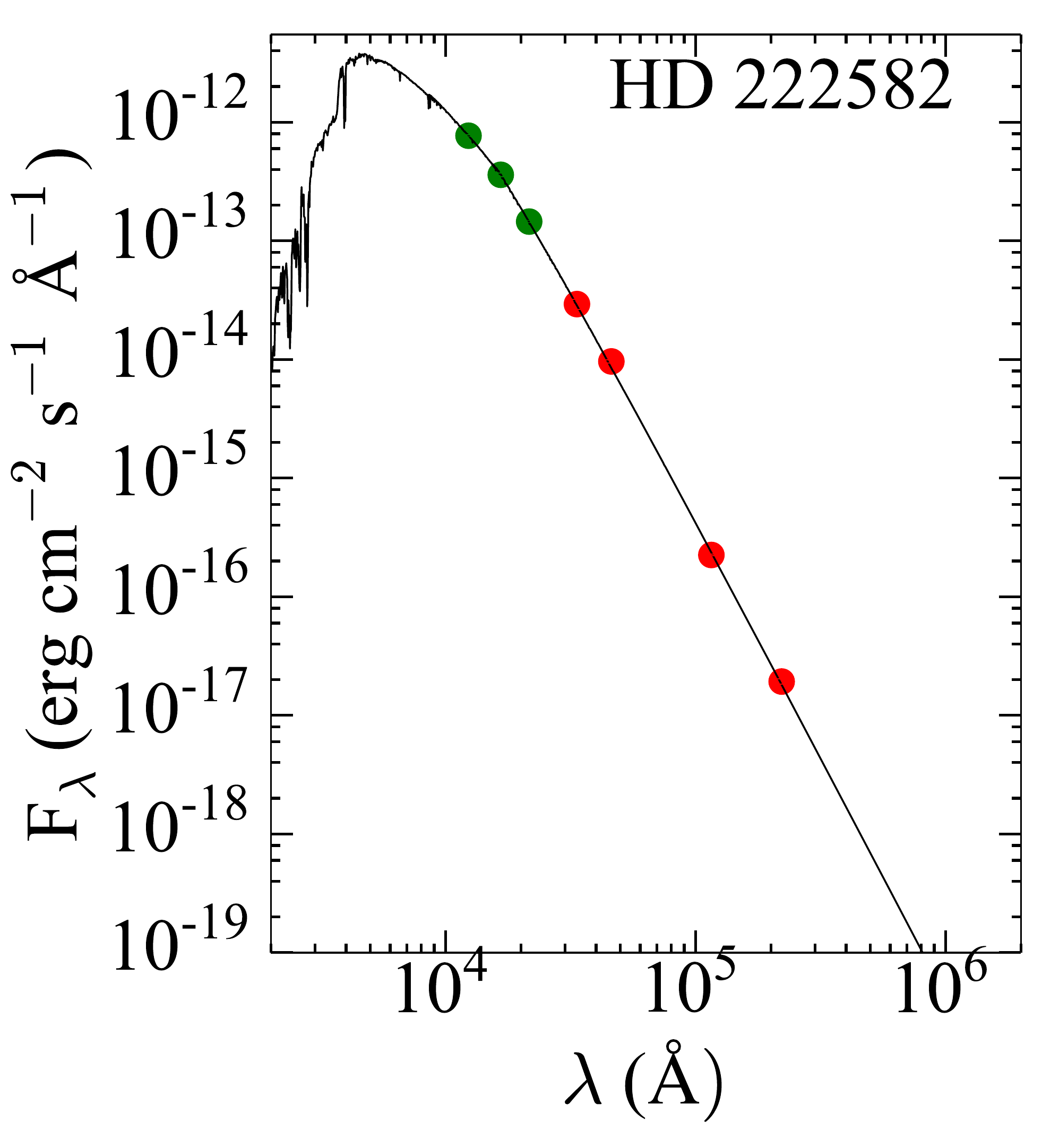}
\plotone{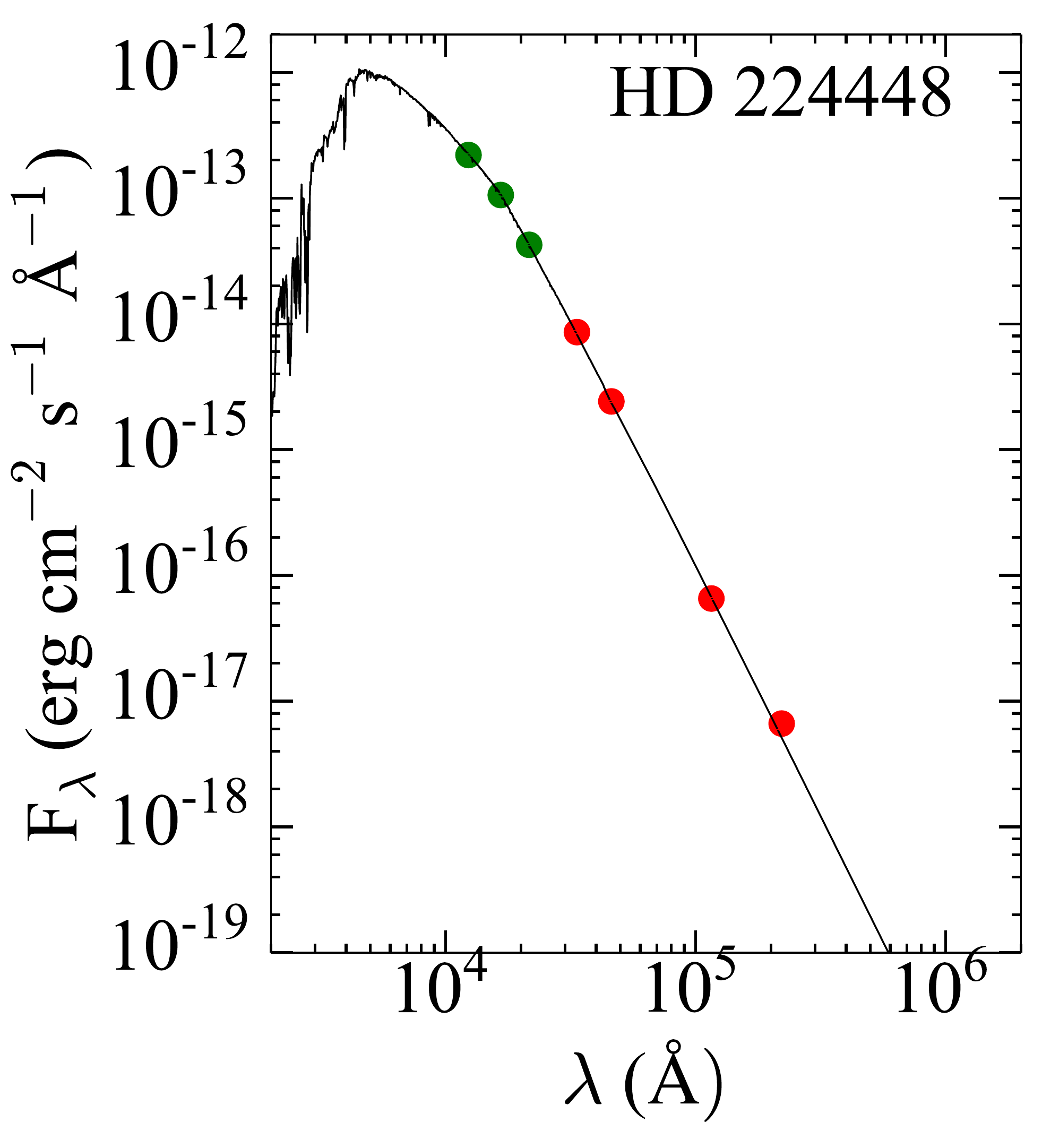}
\plotone{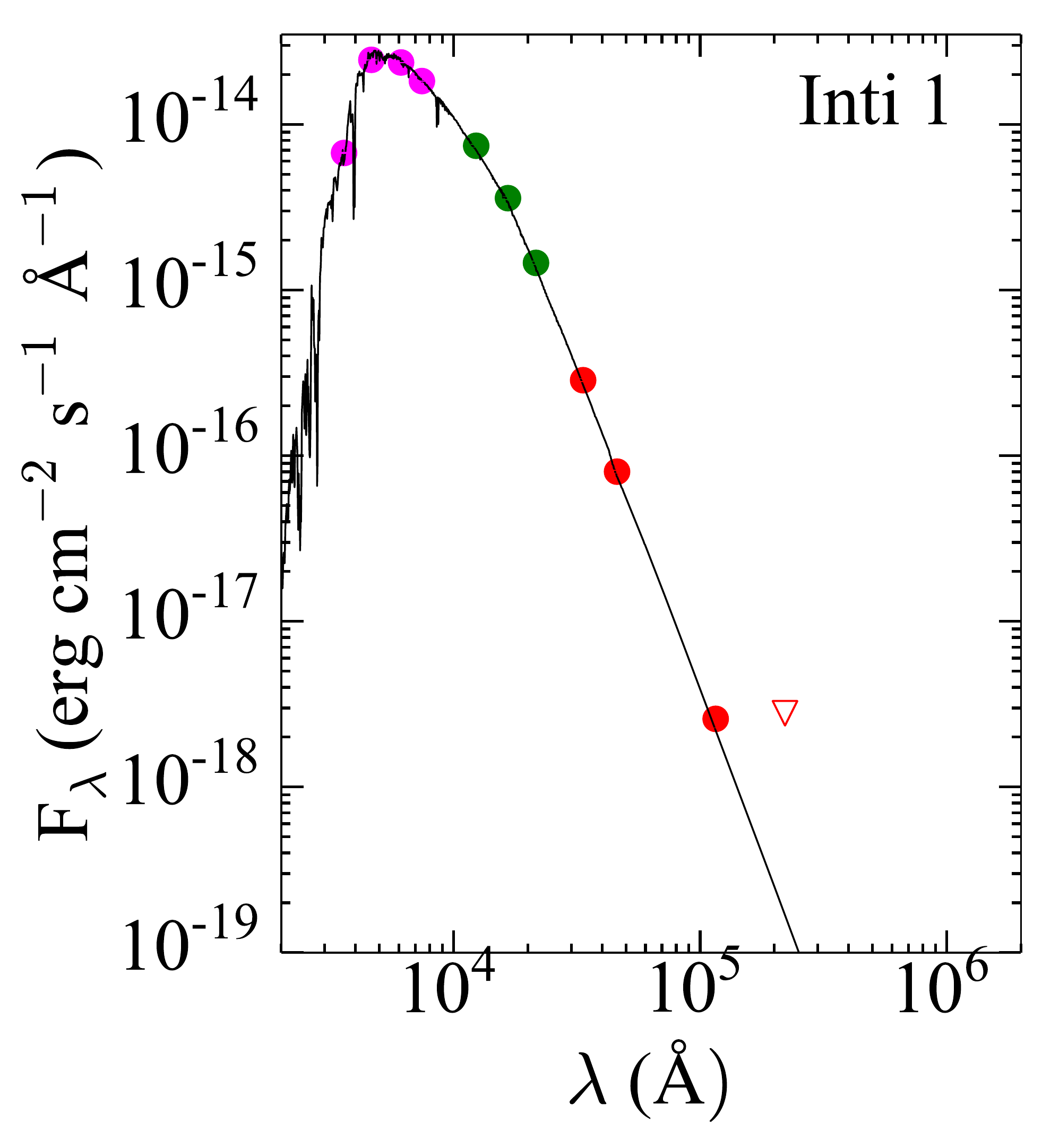}
\plotone{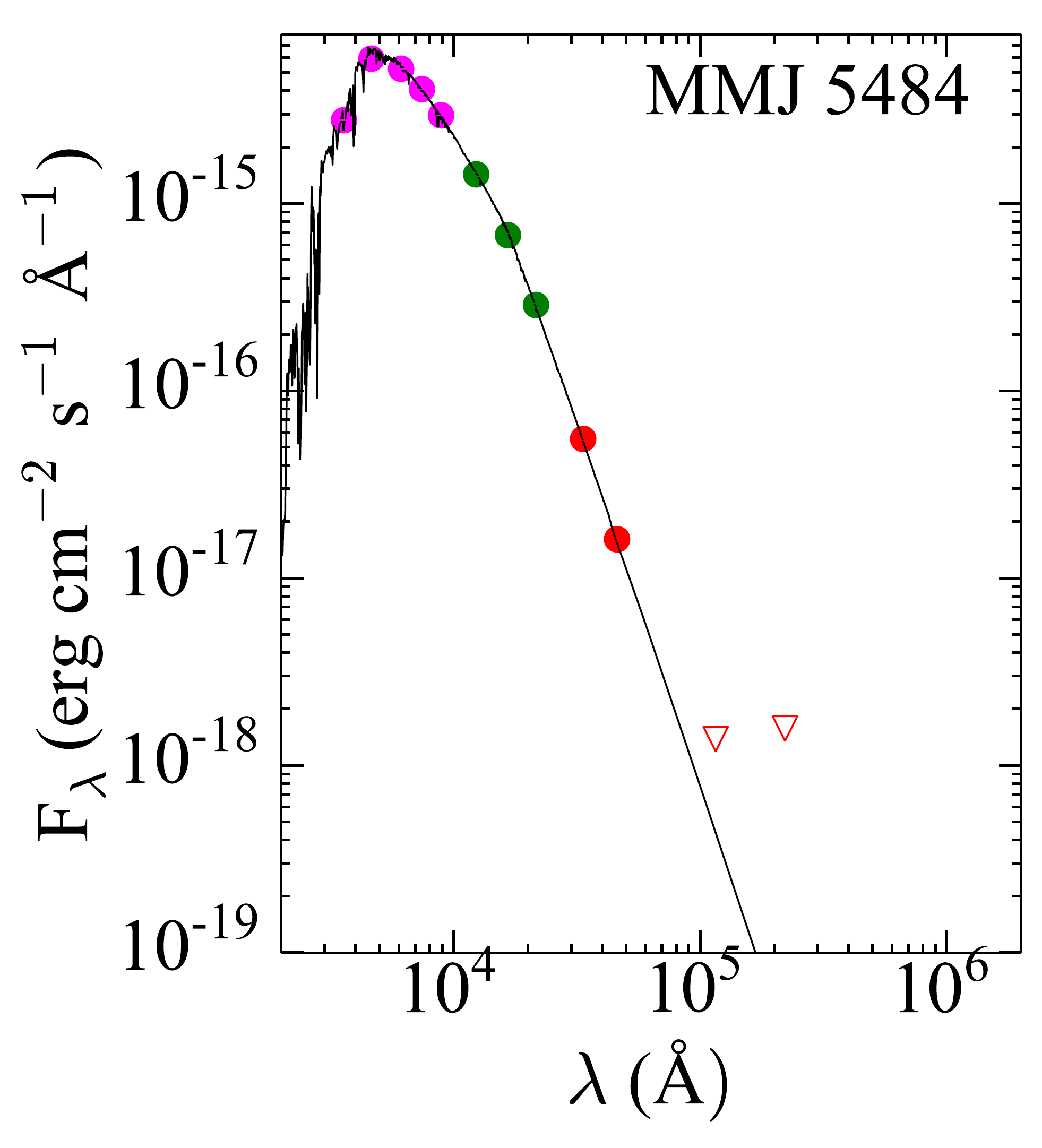}
\plotone{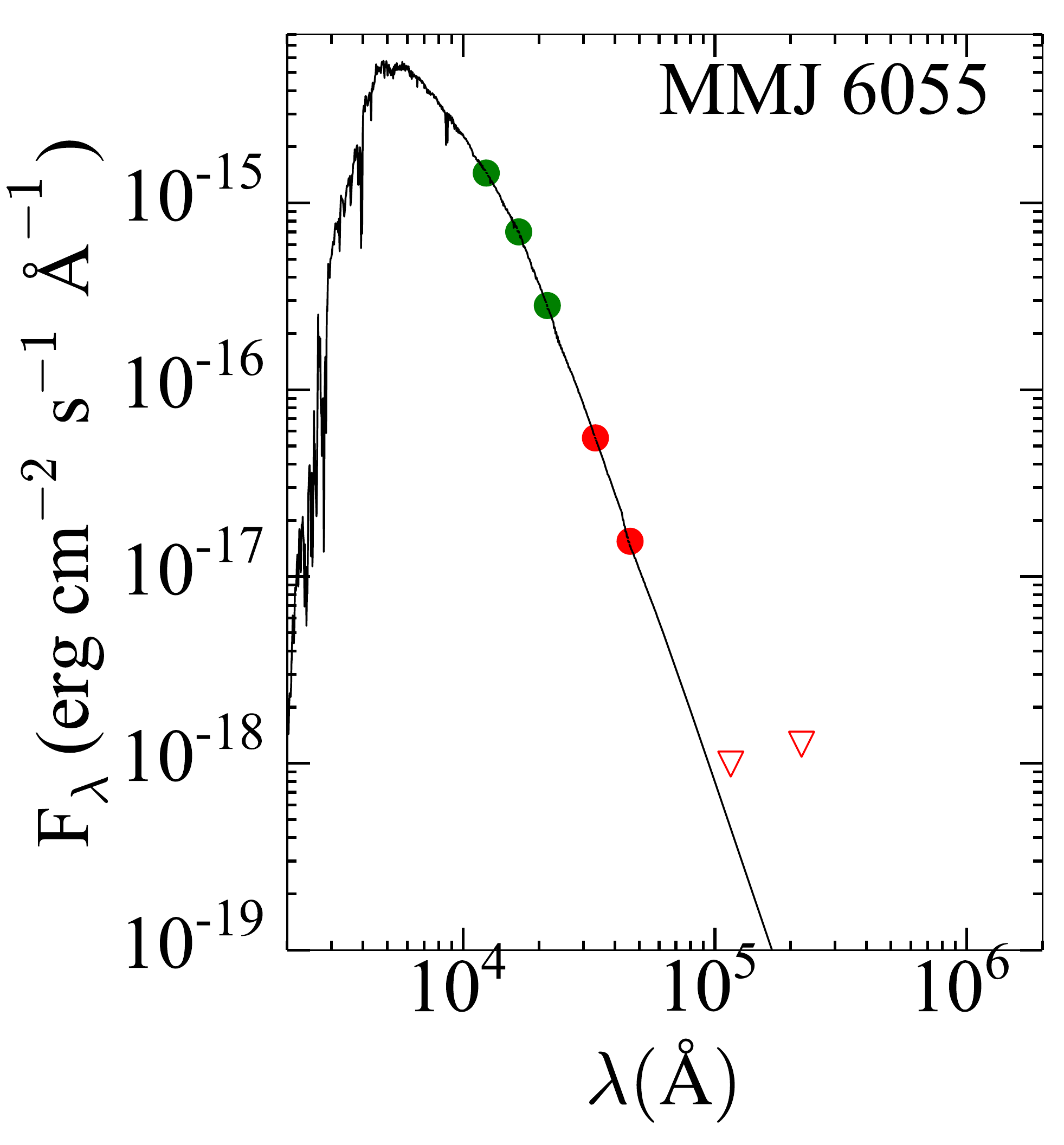}
\caption{Cont. Fig. 9}
\end{figure*}

\begin{figure*}[ht]
\centering
\epsscale{0.35}
\plotone{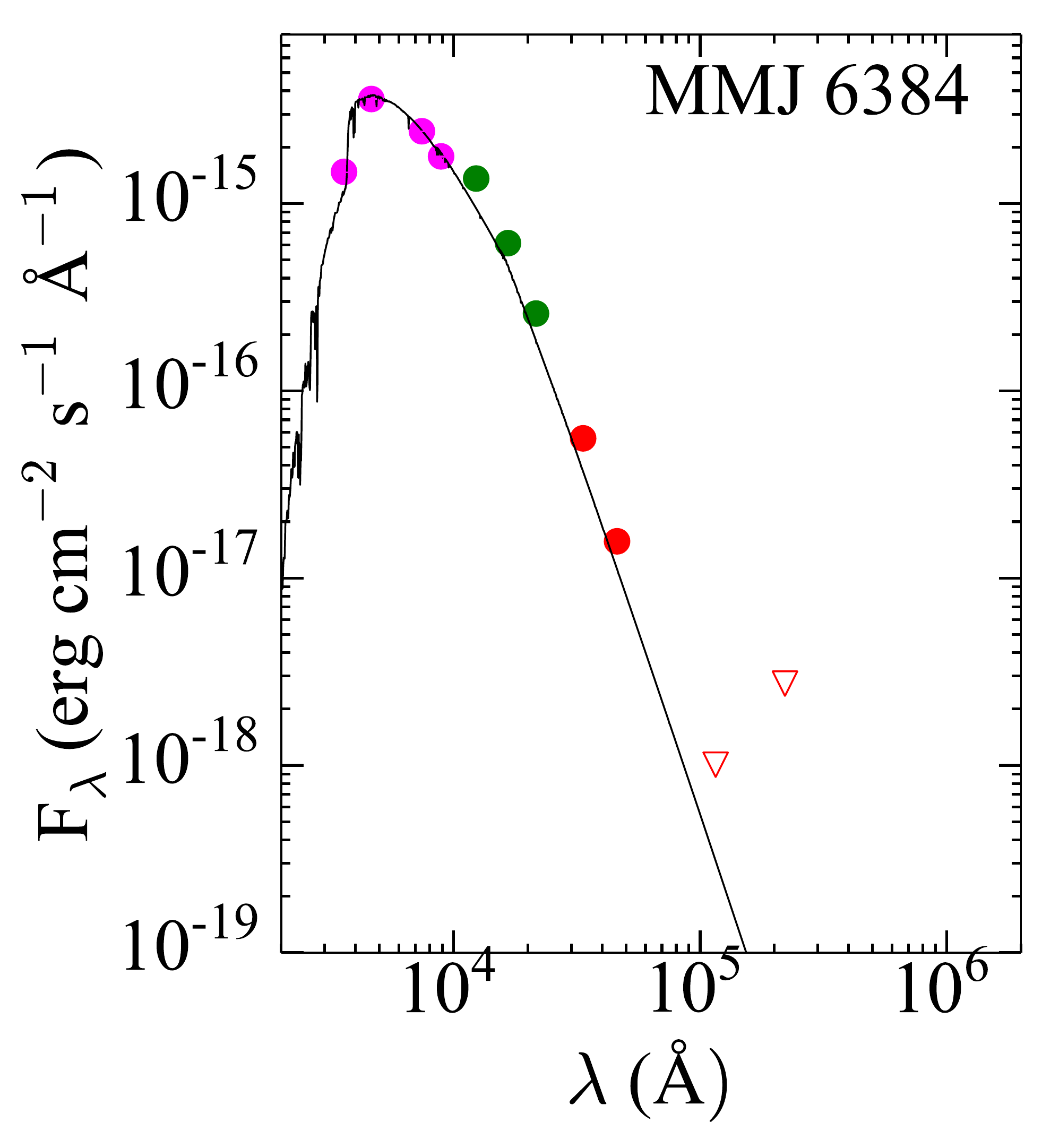}
\plotone{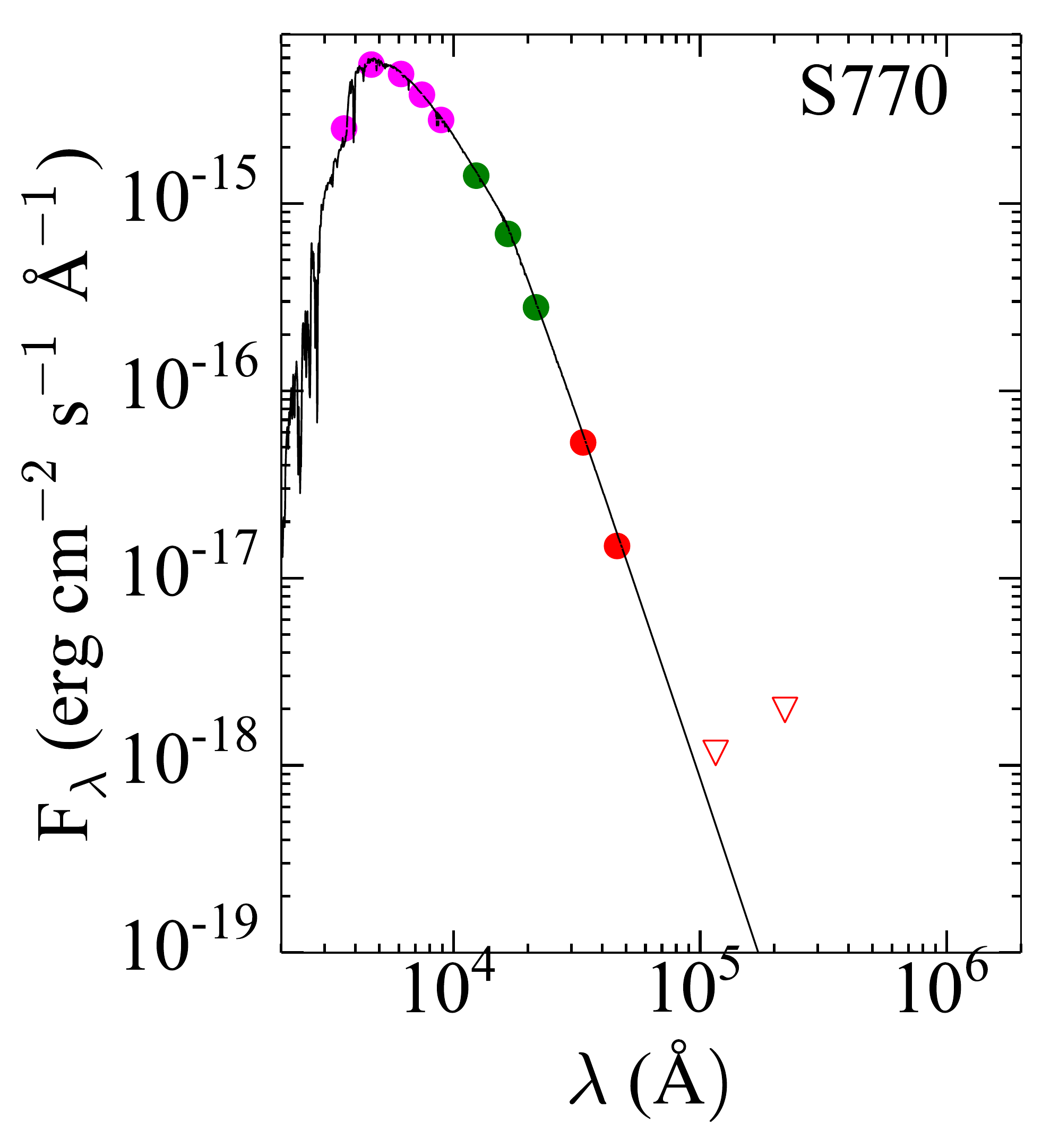}
\plotone{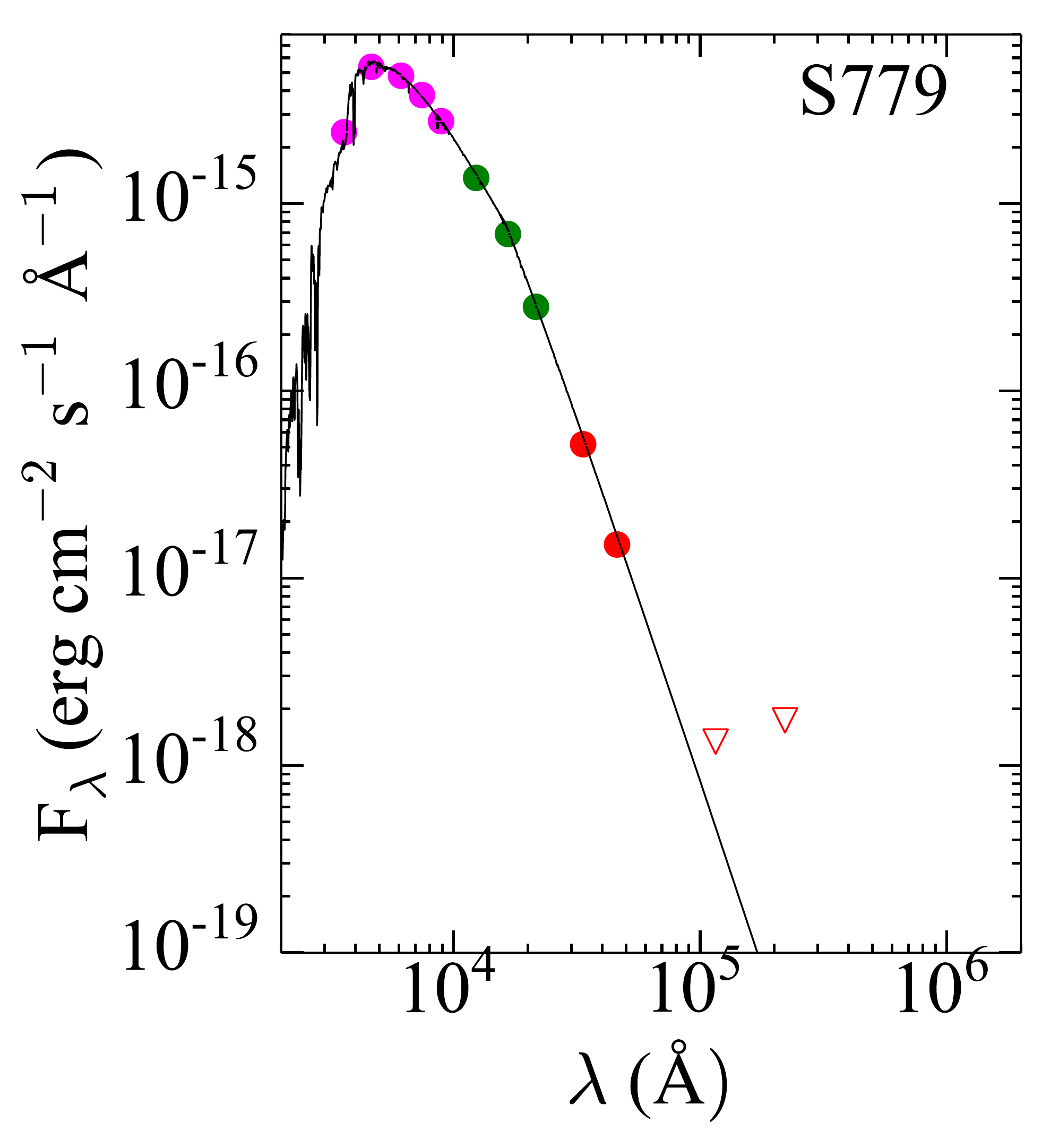}
\plotone{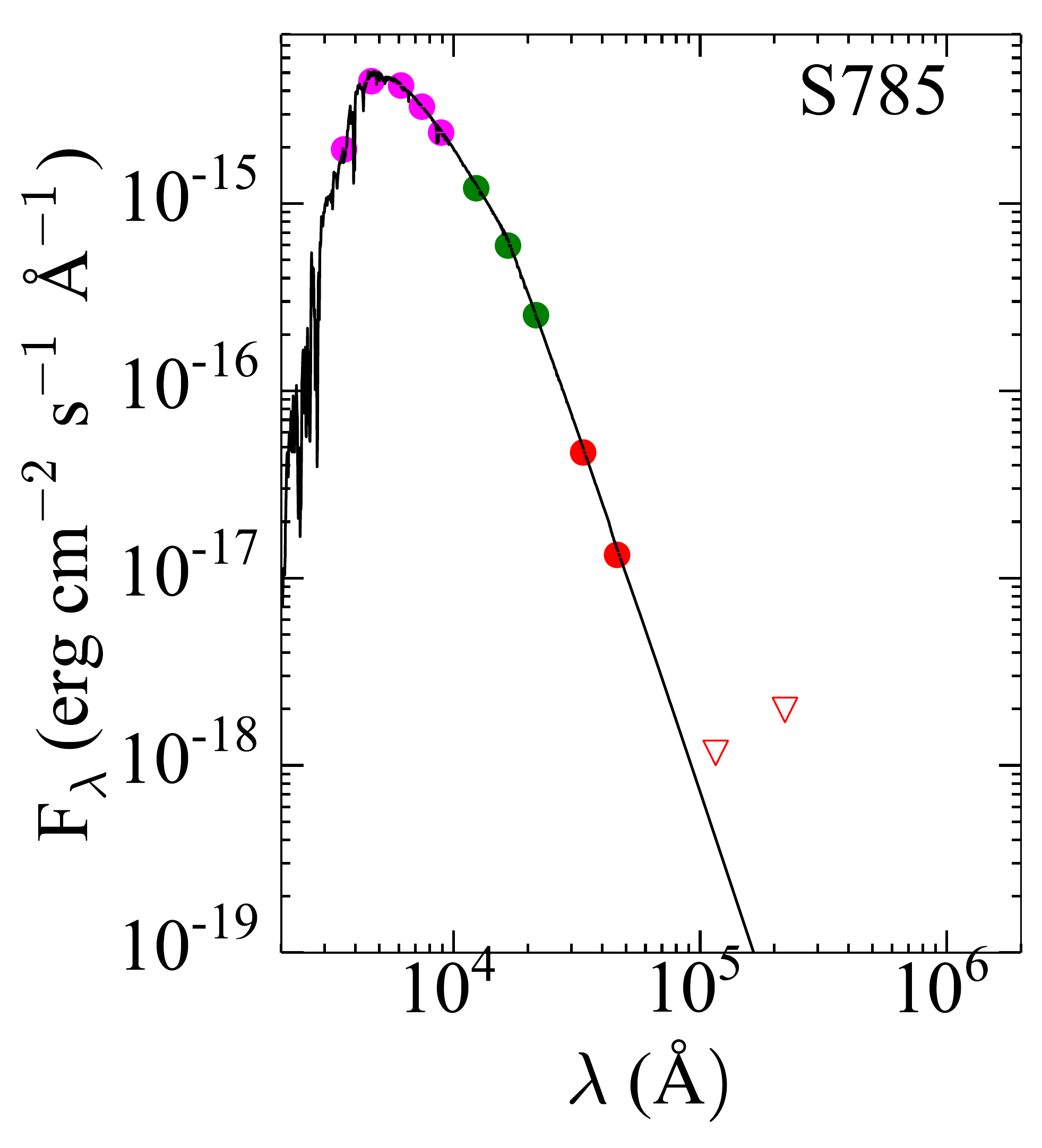}
\plotone{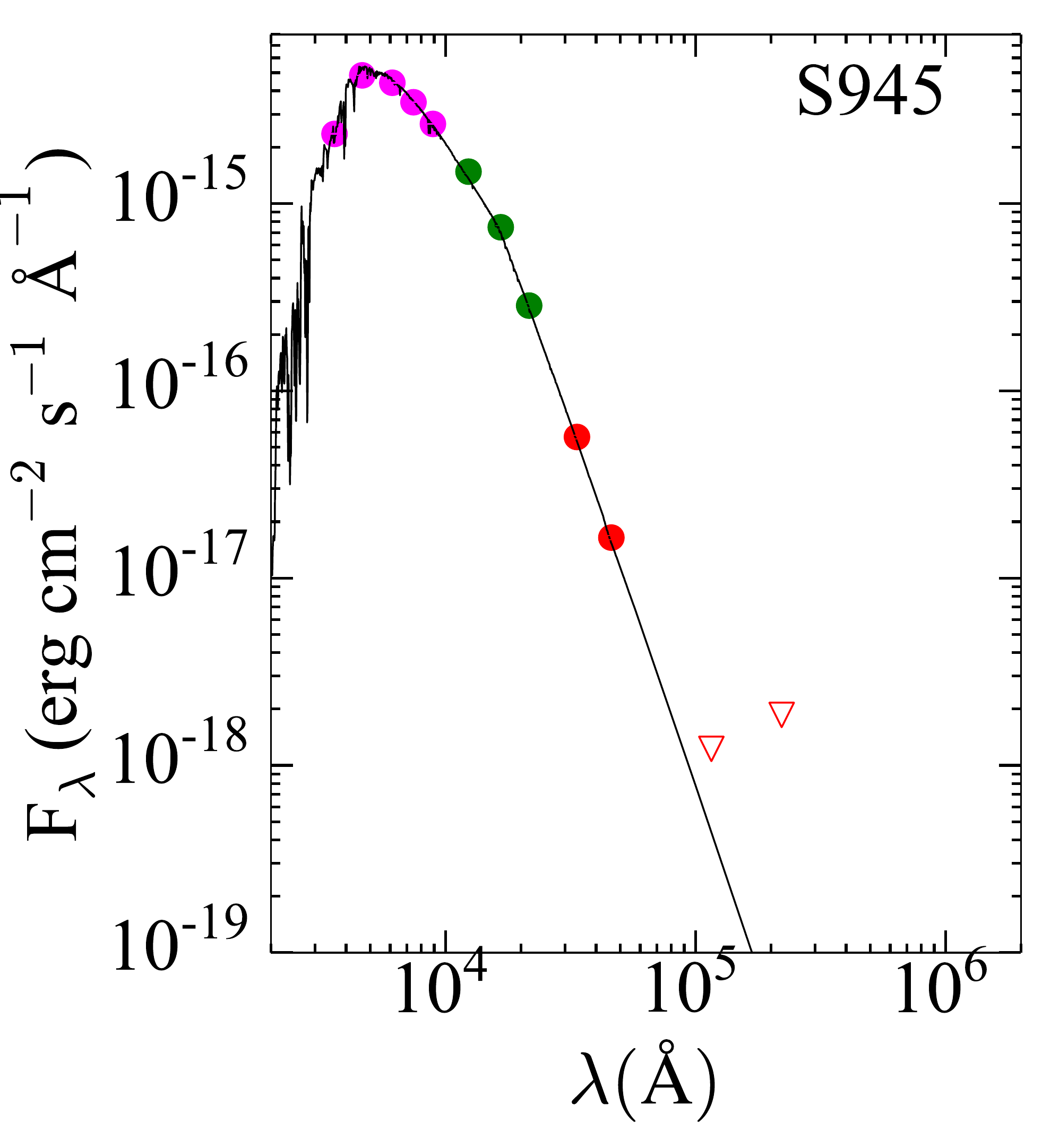}
\plotone{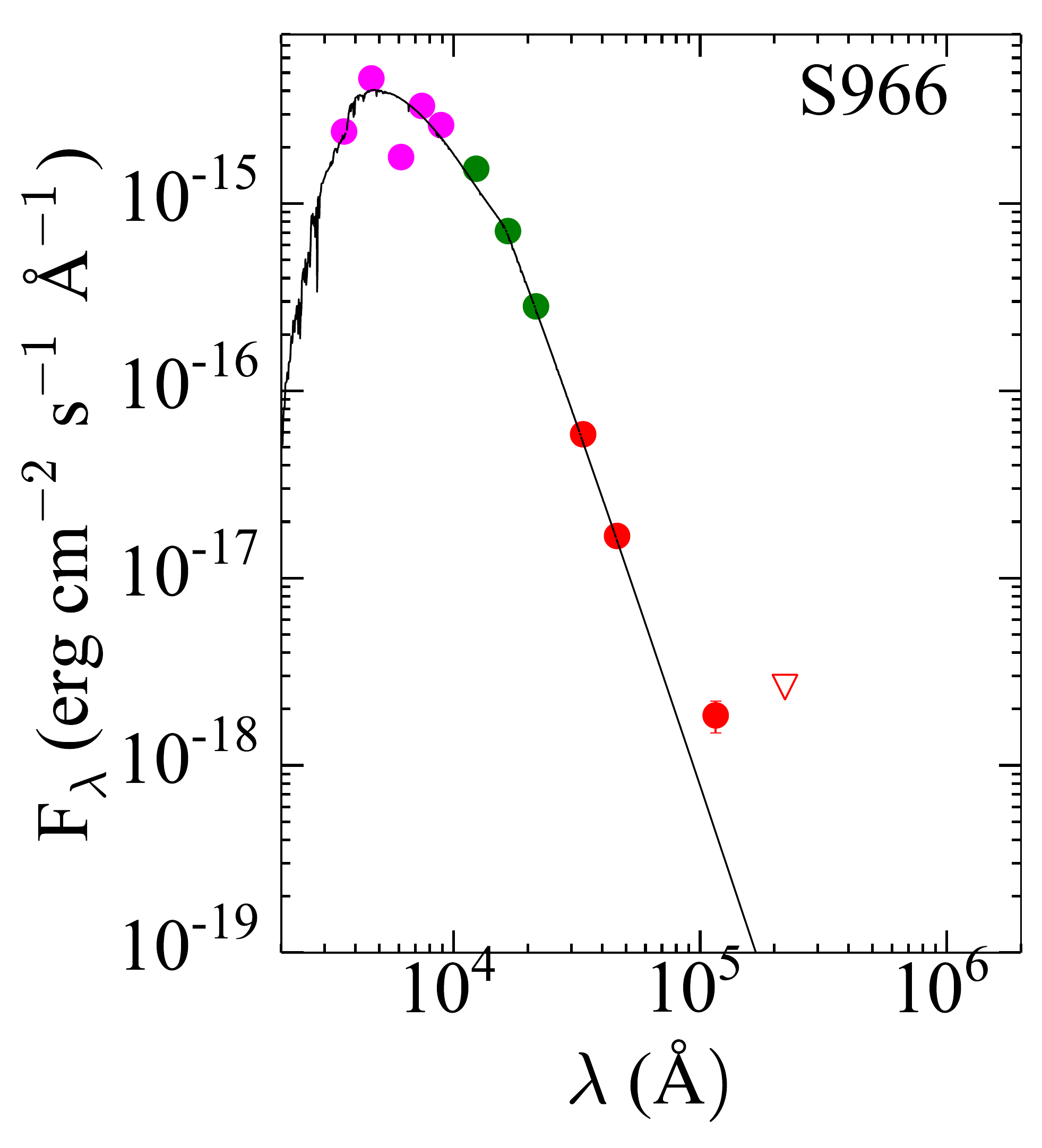}
\caption{Cont. Fig. 9}
\end{figure*}

\begin{figure*}[ht]
\centering
\epsscale{0.35}
\plotone{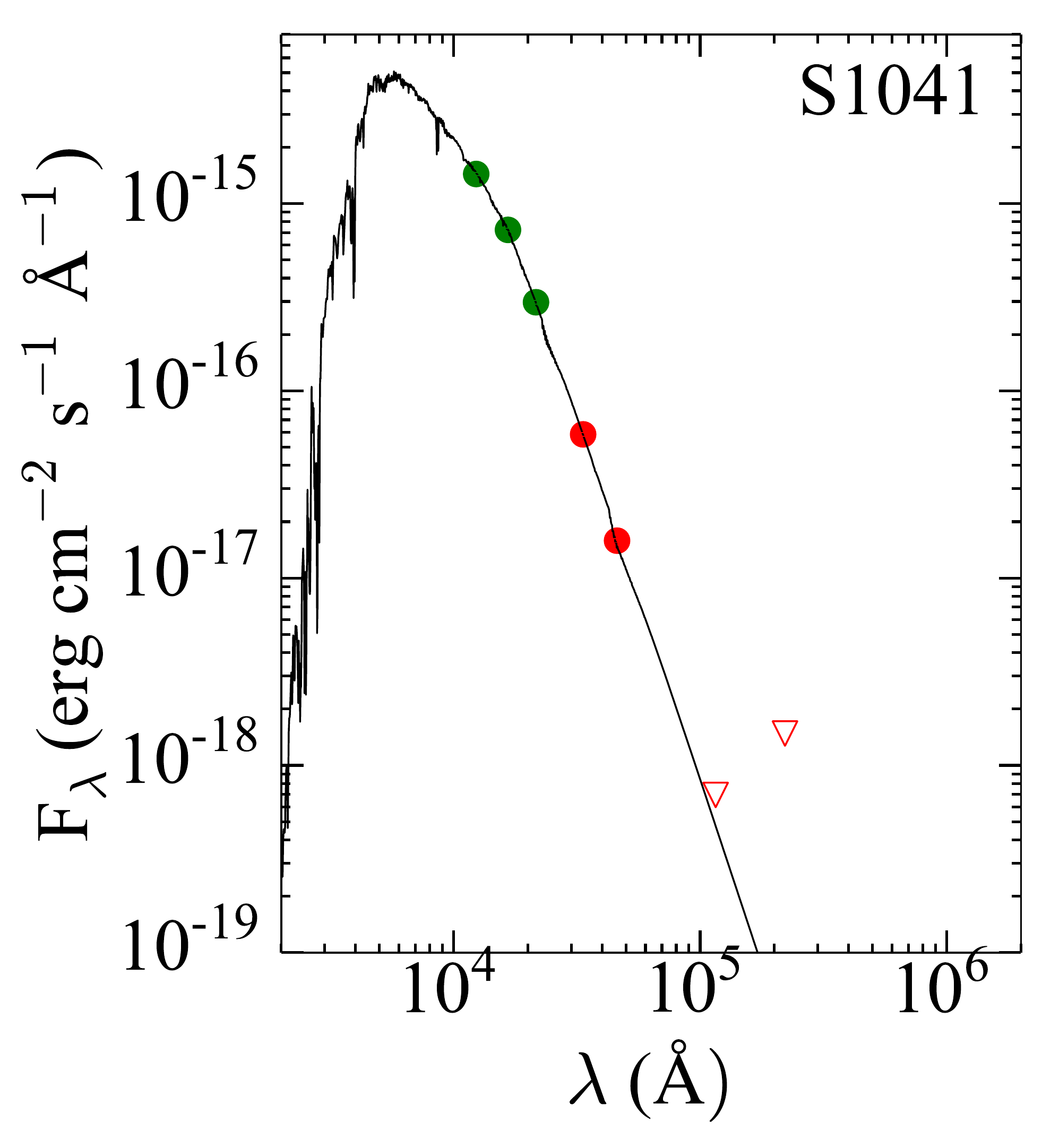}
\plotone{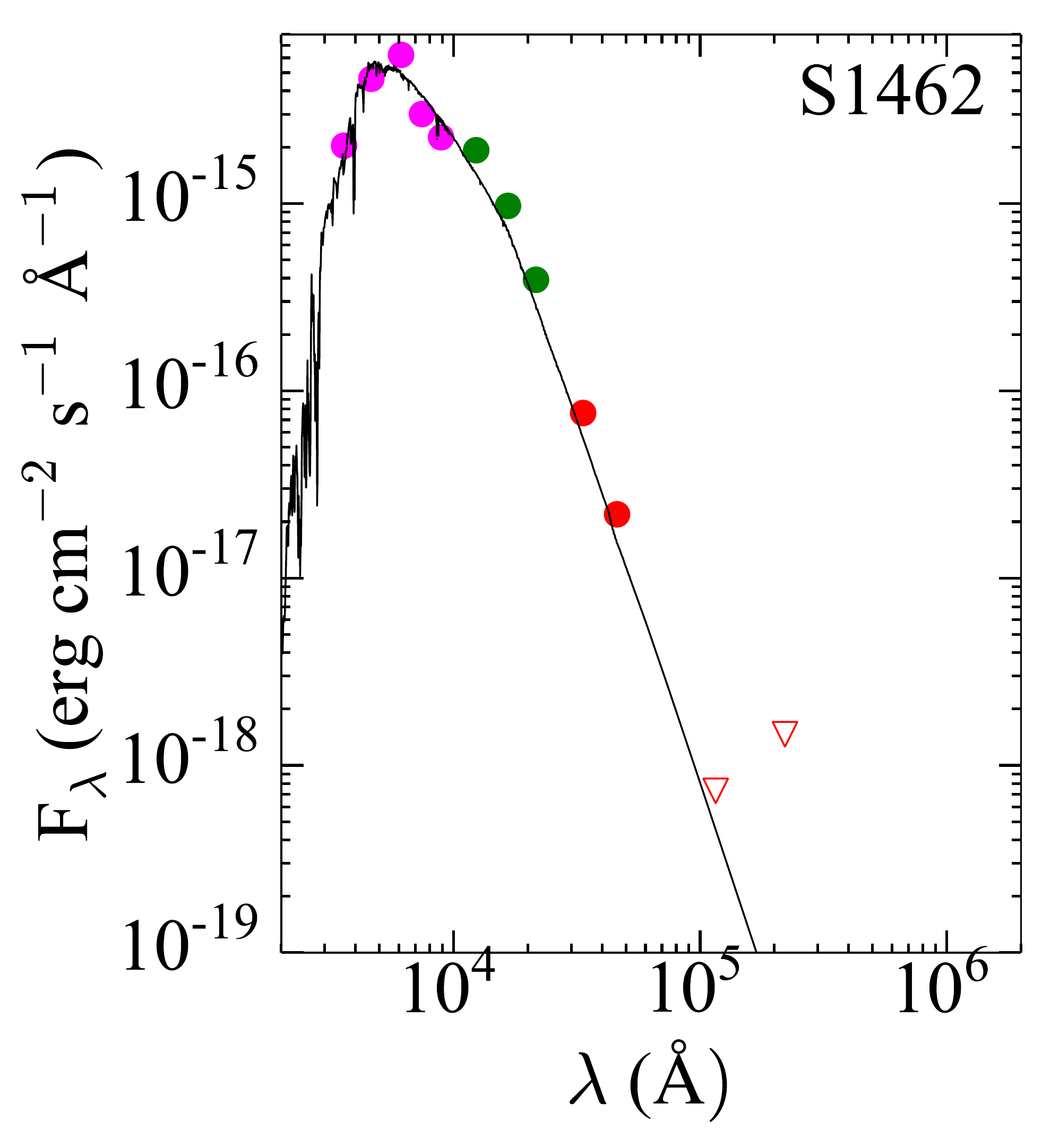}
\caption{Cont. Fig. 9}
\end{figure*}

\clearpage
\begin{sidewaystable}

\label{TableXRay}
\caption{Stellar parameters, including V, 2MASS and WISE magnitudes and excess significance $\chi_{\lambda}$ for the four WISE bands of the 216 Sun-like stars. }

\tiny
\begin{tabular}{lcclllccccccccrrrr}
\hline \hline
Star & V  &ST    & ~~T$_{eff}$   & ~~~log g       & ~~~[Fe/H] & Ref  & J     & H      & K     & [3.4] & [4.6] & [12]  & [22]  & $\chi_{3.4}$ & $\chi_{4.6}$	& $\chi_{12}$	&  $\chi_{22}$ \\
	 & (mag) & & ~~~(K)         & ~(cm s$^{-2})$ & ~~~~(dex)    &      & (mag) & (mag)  & (mag) & (mag) & (mag) & (mag) & (mag) &              &           	&           	&              \\

\hline
\multicolumn{18}{c}{Analogs} \\  
\hline

HD4392	&	7.88 &G5V&	5656			&	4.34			&		~0.050			&	1,1,2,2,2	&	6.66	&	6.37	&	6.25	&	6.25	&	6.21	&	6.26	&	6.26	&	-0.281	&	2.020	&	-0.553	&	0.341	\\
HD6470	&	8.97 &G5	&	5779	$\pm$	38	&	4.49	$\pm$	0.07	&		-0.190	$\pm$	0.023	&	5,1,5,5,5	&	7.78	&	7.54	&	7.43	&	7.31	&	7.38	&	7.41	&	7.82	&	3.145	&	0.145	&	-0.439	&	-1.050	\\
HD7678	&	8.25 &G3V	&	5844			&	4.52			&		~0.140			&	1,1,2,2,2	&	7.12	&	6.87	&	6.75	&	6.69	&	6.75	&	6.75	&	6.68	&	0.553	&	0.347	&	-0.333	&	0.571	\\
HD8076	&	7.65 & G2V	&	5893			&	4.43			&		~0.050			&	1,1,2,2,2	&	6.56	&	6.26	&	6.19	&	6.19	&	6.18	&	6.20	&	6.16	&	-1.812	&	2.156	&	-0.186	&	-0.113	\\
HD9986	&	6.77 &G5V	&	5828			&	4.49			&		~0.110			&	1,1,2,2,2	&	5.62	&	5.33	&	5.22	&	5.21	&	5.03	&	5.27	&	5.25	&	0.106	&	2.426	&	-0.709	&	0.459	\\
HD10180	&	7.33 & G1V	&	5891			&	4.36			&		~0.070			&	1,1,2,2,2	&	6.25	&	5.93	&	5.87	&	5.81	&	5.80	&	5.86	&	5.83	&	0.142	&	1.047	&	-0.379	&	0.747	\\
HD11131	&	6.72 &G0	&	5804			&	4.53			&		-0.090			&	1,1,2,2,2	&	5.54	&	5.29	&	5.15	&	4.99	&	4.79	&	5.11	&	5.07	&	0.744	&	4.339	&	1.399	&	2.197	\\
HD12264	&	7.99 &G5V	&	5819			&	4.47			&		~0.050			&	1,1,2,2,2	&	6.82	&	6.54	&	6.45	&	6.38	&	6.42	&	6.45	&	6.39	&	0.301	&	1.415	&	-0.638	&	1.102	\\
HD13724	&	7.89& G3/G5V	&	5809			&	4.54			&		~0.250			&	1,1,2,2,2	&	6.73	&	6.48	&	6.38	&	6.27	&	6.35	&	6.38	&	6.30	&	0.613	&	2.507	&	-0.567	&	1.399	\\
HD15632	&	8.03 &G0	&	5762			&	4.51			&		~0.030			&	1,1,2,2,2	&	6.81	&	6.55	&	6.47	&	6.47	&	6.47	&	6.49	&	6.39	&	0.174	&	1.023	&	-0.483	&	1.368	\\
HD18330	&	7.90 & F5	&	5989			&	4.52			&		-0.030			&	1,1,2,2,2	&	6.80	&	6.55	&	6.46	&	6.44	&	6.46	&	6.48	&	6.45	&	0.150	&	1.225	&	-0.492	&	0.645	\\
HD19617	&	8.69&G5	&	5727			&	4.38			&		~0.150			&	1,1,2,2,2	&	7.45	&	7.20	&	7.11	&	7.03	&	7.11	&	7.11	&	7.12	&	1.035	&	-0.030	&	-0.104	&	-0.285	\\
HD20201	&	7.27&G0V	&	5994			&	4.44			&		~0.100			&	1,1,2,2,2	&	6.21	&	5.98	&	5.88	&	5.83	&	5.79	&	5.90	&	5.86	&	0.197	&	1.432	&	-0.395	&	0.847	\\
HD20527	&	8.70	&G5&	5615			&	4.45			&		~0.110			&	1,1,2,2,2	&	7.44	&	7.12	&	7.08	&	6.92	&	7.00	&	7.02	&	6.92	&	1.629	&	0.378	&	-0.242	&	1.301	\\
HD24552	&	7.97& G0	&	5825			&	4.44			&		-0.060			&	1,1,2,2,2	&	6.86	&	6.63	&	6.49	&	6.49	&	6.49	&	6.52	&	6.47	&	-0.269	&	0.682	&	-0.512	&	0.797	\\
HD25710	&	8.11&G0	&	5994			&	4.42			&		-0.090			&	1,1,2,2,2	&	7.00	&	6.78	&	6.65	&	6.61	&	6.64	&	6.65	&	6.59	&	-5.823	&	1.102	&	-0.362	&	1.007	\\
HD25926	&	7.70&G2V	&	5915			&	4.38			&		-0.010			&	1,1,2,2,2	&	6.56	&	6.30	&	6.19	&	6.14	&	6.11	&	6.17	&	6.12	&	-0.021	&	1.981	&	-0.443	&	0.875	\\
HD26767	&	8.05&G0	&	5792			&	4.30			&		~0.080			&	1,1,2,2,2	&	6.86	&	6.61	&	6.53	&	6.50	&	6.55	&	6.53	&	6.37	&	-0.083	&	0.742	&	-0.713	&	1.100	\\
HD27685	&	7.86&G4V	&	5705			&	4.42			&		~0.070			&	1,1,2,2,2	&	6.58	&	6.33	&	6.20	&	6.18	&	6.18	&	6.22	&	6.14	&	-0.718	&	0.668	&	-1.388	&	0.731	\\
HD28068	&	8.04&G1V	&	5761			&	4.32			&		~0.070			&	1,1,2,2,2	&	6.84	&	6.56	&	6.44	&	6.36	&	6.39	&	6.42	&	6.41	&	-0.010	&	0.516	&	-0.586	&	0.635	\\
HD28099	&	8.10&G8V	&	5718			&	4.38			&		~0.060			&	1,1,2,2,2	&	6.89	&	6.64	&	6.55	&	6.54	&	6.55	&	6.58	&	6.66	&	-0.414	&	1.408	&	-0.446	&	0.021	\\
HD29150	&	7.58&G5	&	5755			&	4.48			&		-0.020			&	1,1,2,2,2	&	6.38	&	6.08	&	5.99	&	6.01	&	5.92	&	6.01	&	5.95	&	-0.677	&	1.293	&	-0.318	&	0.878	\\
HD29161	&	7.88&G0	&	5890			&	4.59			&		-0.030			&	1,1,2,2,2	&	6.77	&	6.49	&	6.41	&	6.34	&	6.38	&	6.41	&	6.39	&	0.238	&	0.740	&	-0.396	&	0.610	\\
HD29461	&	7.96&G5	&	5766			&	4.38			&		~0.160			&	1,1,2,2,2	&	6.81	&	6.52	&	6.44	&	6.37	&	6.42	&	6.44	&	6.44	&	0.464	&	1.595	&	-0.400	&	0.443	\\
HD30246	&	8.30&G5	&	5723			&	4.52			&		~0.080			&	1,1,2,2,2	&	7.10	&	6.84	&	6.74	&	6.75	&	6.75	&	6.76	&	6.69	&	0.190	&	1.764	&	0.326	&	1.213	\\
\hline	\hline
\end{tabular} 
 
\end{sidewaystable}



\begin{sidewaystable}
\label{TableXRay}

\tiny
\begin{tabular}{lcclllccccccccrrrr}
\hline \hline
Star & V &ST    & ~~T$_{eff}$   & ~~~log g       & ~~~[Fe/H] & Ref  & J     & H      & K     & [3.4] & [4.6] & [12]  & [22]  & $\chi_{3.4}$ & $\chi_{4.6}$	& $\chi_{12}$	&  $\chi_{22}$ \\
& (mag)  & &~~~(K)         & ~(cm s$^{-2})$ & ~~~~(dex)    &      & (mag) & (mag)  & (mag) & (mag) & (mag) & (mag) & (mag) &              &           	&           	&              \\
\hline
													
HD30306	&	7.76 &G6V	&	5557			&	4.53			&		~0.170			&	1,1,2,2,2	&	6.49	&	6.15	&	6.06	&	5.98	&	6.02	&	6.06	&	6.03	&	0.288	&	0.998	&	-0.387	&	0.624	\\
HD31222	&	8.66 &G5	&	5659			&	4.50			&		~0.120			&	3,14,2,2,2	&	7.41	&	7.17	&	7.08	&	7.04	&	7.12	&	7.10	&	7.10	&	1.064	&	0.151	&	-0.253	&	0.494	\\
HD31622	&	8.49 &G5	&	5823			&	4.52			&		-0.170			&	1,1,2,2,2	&	7.35	&	7.10	&	6.98	&	6.93	&	6.96	&	6.98	&	6.91	&	0.632	&	0.384	&	-0.213	&	1.270	\\
HD32963	&	7.60 &G5IV	&	5710			&	4.42			&		~0.070			&	1,1,2,2,2	&	6.42	&	6.10	&	6.02	&	5.96	&	6.00	&	6.06	&	6.05	&	0.283	&	1.852	&	-0.058	&	0.939	\\
HD33866	&	7.82 &G2V	&	5836			&	4.45			&		-0.030			&	1,1,2,2,2	&	6.57	&	6.24	&	6.13	&	6.05	&	6.08	&	6.12	&	6.11	&	0.235	&	2.016	&	-0.349	&	0.432	\\
HD33873	&	8.65 & G6V	&	5658			&	4.37			&		~0.190			&	1,1,2,2,2	&	7.42	&	7.14	&	7.02	&	6.98	&	7.04	&	7.02	&	7.05	&	0.648	&	-0.277	&	-0.063	&	0.804	\\
HD34239	&	7.10 &G0V 	&	5932			&	4.40			&	~0.040			&	4,14,2,2,2	&	6.01	&	5.75	&	5.63	&	5.62	&	5.51	&	5.65	&	5.58	&	-0.849	&	1.290	&	-0.389	&	0.994	\\
HD34386	&	8.54 &G5III	&	5689			&	4.25			&	~0.070			&	1,1,2,2,2	&	7.37	&	7.14	&	7.04	&	6.99	&	7.04	&	7.04	&	6.88	&	0.802	&	0.377	&	-0.255	&	0.967	\\
HD34599	&	8.28&G3V	&	5834			&	4.40			&	~0.080			&	1,1,2,2,2	&	7.13	&	6.83	&	6.78	&	6.71	&	6.76	&	6.76	&	6.73	&	0.334	&	-0.089	&	-0.122	&	0.719	\\
HD36553	&	5.46 &G3IV	&	6017			&	3.81			&	~0.320			&	1,1,2,2,2	&	4.39	&	4.29	&	4.11	&	4.02	&	3.52	&	4.06	&	4.04	&	0.088	&	2.049	&	0.787	&	1.679	\\
HD39060 &   3.86 &A3V	&   8052	&	4.15			&	~0.110			&	9,1,10,10,17	&	3.67	&	3.54	&	3.53	&	3.48	&	3.18	&	2.56	&	0.01	& -0.248		&	0.715		&	1.159	&	16.260	\\	
HD39833	&	7.65&G0III	&	5821			&	4.38			&	~0.160			&	1,1,2,2,2	&	6.49	&	6.23	&	6.15	&	6.13	&	6.12	&	6.16	&	6.08	&	-0.199	&	1.646	&	-0.497	&	1.018	\\
HD46090	&	7.14&G0	&	5778			&	4.46			&	~0.010			&	1,1,2,2,2	&	5.86	&	5.52	&	5.40	&	5.33	&	5.16	&	5.42	&	5.44	&	-0.253	&	1.925	&	-0.419	&	0.098	\\
HD55693	&	7.17&G1V	&	5855			&	4.40			&	~0.240			&	1,1,2,2,2	&	6.04	&	5.78	&	5.70	&	5.70	&	5.60	&	5.70	&	5.63	&	-0.747	&	1.360	&	-0.365	&	1.224	\\
HD58895	&	6.58&G5IV	&	5690			&	4.02			&	~0.240			&	1,1,2,2,2	&	5.40	&	5.13	&	5.02	&	5.00	&	4.72	&	5.01	&	4.90	&	0.031	&	1.651	&	-1.414	&	1.264	\\
HD59967	&	6.66&G3V	&	5860			&	4.58			&	-0.030			&	1,1,2,2,2	&	5.53	&	5.25	&	5.10	&	5.05	&	4.84	&	5.10	&	4.94	&	0.124	&	1.952	&	-0.418	&	2.033	\\
HD63433	&	6.90&G5IV	&	5693	$\pm$	58	&	4.52	$\pm$	0.07	&	~0.007	$\pm$	0.025	&	5,1,5,5,5	&	5.62	&	5.36	&	5.26	&	5.20	&	5.05	&	5.29	&	5.17	&	0.126	&	1.378	&	-0.662	&	1.612	\\
HD64474	&	7.83 &G6V	&	5656			&	4.25			&	-0.020			&	4,14,2,2,2	&	6.58	&	6.32	&	6.17	&	6.19	&	6.20	&	6.21	&	6.13	&	-0.289	&	1.245	&	-0.604	&	0.818	\\
HD64942	&	8.37&G5	&	5839			&	4.45			&	~0.040			&	1,1,2,2,2	&	7.25	&	6.97	&	6.87	&	6.82	&	6.84	&	6.87	&	6.76	&	0.433	&	0.436	&	-0.369	&	1.461	\\
HD67578	&	8.60 &G2V	&	5835	$\pm$	30	&	4.48	$\pm$	0.06	&	-0.200	$\pm$	0.030	&	5,1,5,5,5	&	7.42	&	7.15	&	7.08	&	7.03	&	7.06	&	7.06	&	7.05	&	1.924	&	0.111	&	-0.174	&	0.305	\\
HD72905	&	5.6 &G1.5Vb	&	5864	$\pm$	47	&	4.46	$\pm$	0.09	&	-0.052	$\pm$	0.026	&	5,1,5,5,5	&	4.35	&	4.28	&	4.17	&	4.14	&	3.70	&	4.17	&	4.05	&	-0.227	&	1.916	&	-1.008	&	1.098	\\
HD75288	&	8.51& G3/G5V	&	5775	$\pm$	30	&	4.37	$\pm$	0.06	&	~0.120	$\pm$	0.030	&	5,1,5,5,5	&	7.28	&	6.94	&	6.93	&	6.87	&	6.94	&	6.93	&	6.82	&	1.249	&	-0.114	&	-0.279	&	1.583	\\
HD76332	&	8.58&G2V	&	5827			&	4.34			&	-0.060			&	1,1,2,2,2	&	7.29	&	6.97	&	6.84	&	6.80	&	6.81	&	6.80	&	6.75	&	0.369	&	0.720	&	-0.025	&	-0.248	\\
HD76780	&	7.63&G5	&	5745			&	4.40			&	~0.110			&	5,1,5,5,5	&	6.41	&	6.17	&	6.07	&	5.97	&	5.96	&	6.05	&	5.93	&	0.382	&	2.868	&	-0.344	&	0.984	\\
HD77006	&	7.93 &G5	&	5934	$\pm$	49	&	4.51	$\pm$	0.06	&	-0.020	$\pm$	0.019	&	5,1,5,5,5	&	6.75	&	6.41	&	6.36	&	6.32	&	6.30	&	6.32	&	6.25	&	-0.346	&	1.454	&	-0.406	&	0.756	\\
\hline

\end{tabular} 

\end{sidewaystable}


\begin{sidewaystable}
\centering
\label{TableXRay}

\center

\tiny
\begin{tabular}{lcclllccccccccrrrr}
\hline \hline
Star & V &ST    & ~~T$_{eff}$   & ~~~log g       & ~~~[Fe/H] & Ref  & J     & H      & K     & [3.4] & [4.6] & [12]  & [22]  & $\chi_{3.4}$ & $\chi_{4.6}$	& $\chi_{12}$	&  $\chi_{22}$ \\
& (mag)  & & ~~~(K)         & ~(cm s$^{-2})$ & ~~~~(dex)    &      & (mag) & (mag)  & (mag) & (mag) & (mag) & (mag) & (mag) &              &           	&           	&              \\
\hline

HD77461	&	8.83 & G3V	&	5835			&	4.49			&	-0.020			&	3,19,2,2,2	&	7.49	&	7.19	&	7.12	&	7.07	&	7.13	&	7.11	&	7.00	&	1.798	&	0.084	&	-0.254	&	0.913	\\
HD78317	&	8.15 &F8	&	5848			&	4.43			&	~0.050			&	1,1,2,2,2	&	7.06	&	6.73	&	6.71	&	6.60	&	6.67	&	6.70	&	6.65	&	2.426	&	0.859	&	-0.713	&	1.045	\\
HD78538	&	8.15 &G5	&	5800			&	4.59			&	-0.020			&	1,1,2,2,2	&	6.99	&	6.74	&	6.63	&	6.58	&	6.61	&	6.61	&	6.70	&	0.203	&	1.135	&	-0.252	&	-0.312	\\
HD80533	&	8.93 &G0	&	5709	$\pm$	65	&	4.49	$\pm$	0.12	&	-0.070	$\pm$	0.039	&	5,1,5,5,5	&	7.70	&	7.39	&	7.34	&	7.28	&	7.31	&	7.31	&	7.36	&	1.588	&	-0.063	&	-0.182	&	0.016	\\
HD81659	&	7.91 &G6/G8V	&	5658			&	4.52			&	~0.230			&	1,1,2,2,2	&	6.69	&	6.41	&	6.31	&	6.24	&	6.32	&	6.32	&	6.29	&	0.521	&	2.231	&	-0.490	&	0.566	\\
HD81700	&	8.51 &G2V	&	5890	$\pm$	30	&	4.48	$\pm$	0.06	&	~0.140	$\pm$	0.030	&	5,1,5,5,5	&	7.35	&	7.10	&	7.01	&	6.80	&	6.88	&	6.88	&	6.89	&	1.337	&	0.929	&	0.051	&	0.694	\\
HD86087	&	5.71 &A0V	&	10500			&	4.00			&		0.080		&	1,1,11,11,17	&	5.70	&	5.77	&	5.74	&	5.71	&	5.61	&	5.66	&	4.56	&	-1.302	&	-1.924	&	-2.463	&	9.890	\\			
HD86226	&	7.93&G2V	&	5934			&	4.37			&	~0.010			&	1,1,2,2,2	&	6.84	&	6.58	&	6.46	&	6.44	&	6.45	&	6.45	&	6.40	&	0.002	&	1.307	&	-0.465	&	0.644	\\
HD89454	&	8.03&G5	&	5690			&	4.48			&	~0.100			&	1,1,2,2,2	&	6.81	&	6.54	&	6.46	&	6.42	&	6.44	&	6.45	&	6.46	&	0.231	&	1.921	&	-0.391	&	0.233	\\
HD90722	&	7.88&G5/G6IV	&	5720	$\pm$	25	&	4.23	$\pm$	0.05	&	~0.360	$\pm$	0.030	&	5,1,5,5,5	&	6.62	&	6.34	&	6.26	&	6.22	&	6.25	&	6.26	&	6.23	&	0.216	&	1.729	&	-0.446	&	0.227	\\
HD90936	&	8.37&G3V	&	5918			&	4.43			&	~0.070			&	1,1,2,2,2	&	7.24	&	6.99	&	6.91	&	6.87	&	6.91	&	6.93	&	6.96	&	1.161	&	0.411	&	-0.327	&	-0.030	\\
HD91489	&	8.44&G2/G3V	&	5884			&	4.48			&	-0.010			&	1,1,2,2,2	&	7.30	&	6.97	&	6.93	&	6.86	&	6.91	&	6.93	&	7.02	&	1.235	&	0.052	&	-0.207	&	0.492	\\
HD92074	&	8.64&G0	&	5842	$\pm$	69	&	4.56	$\pm$	0.08	&	~0.070	$\pm$	0.026	&	5,1,5,5,5	&	7.46	&	7.15	&	7.11	&	7.06	&	7.09	&	7.09	&	7.08	&	1.577	&	0.293	&	-0.182	&	1.150	\\
HD93215	&	8.05&G5V	&	5558			&	4.09			&	~0.040			&	1,1,2,2,2	&	6.89	&	6.60	&	6.55	&	6.48	&	6.52	&	6.56	&	6.54	&	0.725	&	1.623	&	-0.779	&	0.261	\\
HD93489	&	7.92&G3V	&	5904			&	4.41			&	~0.040			&	1,1,2,2,2	&	6.81	&	6.50	&	6.47	&	6.39	&	6.45	&	6.48	&	6.38	&	0.577	&	1.522	&	-0.530	&	1.101	\\

HD98618	&	7.65&G5V	&	5838	$\pm$	21	&	4.42	$\pm$	0.03	&	-0.010	$\pm$	0.020	&	5,1,5,5,5	&	6.45	&	6.14	&	6.06	&	5.97	&	5.99	&	6.05	&	5.99	&	0.355	&	1.818	&	-0.273	&	0.924	\\
HD101364	&	8.67&G5	&	5795	$\pm$	23	&	4.43	$\pm$	0.03	&	~0.023	$\pm$	0.014	&	5,1,5,5,5	&	7.48	&	7.20	&	7.16	&	7.09	&	7.14	&	7.13	&	7.12	&	1.215	&	0.360	&	-0.320	&	0.857	\\
HD101530	&	8.07&G2V	&	5839			&	4.41			&	-0.190			&	1,1,2,2,2	&	6.98	&	6.71	&	6.62	&	6.56	&	6.59	&	6.61	&	6.55	&	0.516	&	1.098	&	-0.415	&	0.320	\\
HD102117	&	7.47&G6V	&	5690	$\pm$	22.00	&	4.30	$\pm$	0.04	&	~0.304	$\pm$	0.030	&	5,1,5,5,5	&	6.22	&	5.95	&	5.83	&	5.80	&	5.73	&	5.83	&	5.69	&	0.102	&	1.536	&	-0.395	&	0.836	\\
HD107148	&	8.01&G5	&	5811	$\pm$	21.00	&	4.38	$\pm$	0.04	&	~0.315	$\pm$	0.030	&	5,1,5,5,5	&	6.86	&	6.60	&	6.47	&	6.46	&	6.47	&	6.50	&	6.41	&	0.076	&	0.971	&	-0.450	&	0.879	\\
HD107633	&	8.78&G6V	&	5874	$\pm$	72.00	&	4.52	$\pm$	0.10	&	~0.110	$\pm$	0.033	&	5,1,5,5,5	&	7.59	&	7.33	&	7.29	&	7.21	&	7.24	&	7.24	&	7.28	&	2.68	&	0.345	&	-0.239	&	0.232	\\

HD109573	&	5.78&A0V	&	10000			&	4.00			&	-0.030			&	1,1,12,12, 17	&	5.78	&	5.79	&	5.77	&	5.37	&	5.40	&	5.02	&	1.22	&	2.756	&	1.923	&	7.830	&	17.000\\																																					
HD110668	&	8.29&G3G5V	&	5819			&	4.42			&	~0.190			&	1,1,2,2,2	&	7.15	&	6.87	&	6.79	&	6.76	&	6.79	&	6.81	&	6.77	&	0.331	&	0.549	&	-0.302	&	0.547	\\
HD113766	&	7.48&F3/F5V	&	7250			&	3.50			&	~0.010			&	1,1,13,13,14	&	6.73	&	6.59	&	6.49	&	6.36	&	5.97	&	3.50	&	1.75	&	0.280	&	6.853	&	20.021	&	16.954	\\

HD114826	&	8.92&G5	&	5860	$\pm$	110	&	4.56	$\pm$	0.11	&	~0.120	$\pm$	0.037	&	5,1,5,5,5	&	7.74	&	7.53	&	7.42	&	7.37	&	7.41	&	7.41	&	7.37	&	1.880	&	0.051	&	-0.182	&	0.143	\\
HD115739	&	8.86&G3/G5V	&	5875	$\pm$	30	&	4.56	$\pm$	0.06	&	~0.090	$\pm$	0.030	&	5,1,5,5,5	&	7.72	&	7.40	&	7.34	&	7.31	&	7.35	&	7.36	&	7.16	&	1.523	&	0.024	&	-0.418	&	1.602	\\
\hline

\end{tabular}  
\end{sidewaystable}

\begin{sidewaystable}
\centering
\label{TableXRay}

\center

\tiny
\begin{tabular}{lcclllccccccccrrrr}
\hline \hline
Star & V  &ST   & ~~T$_{eff}$   & ~~~log g       & ~~~[Fe/H] & Ref  & J     & H      & K     & [3.4] & [4.6] & [12]  & [22]  & $\chi_{3.4}$ & $\chi_{4.6}$	& $\chi_{12}$	&  $\chi_{22}$ \\
& (mag)  & ~~~(K)         & ~(cm s$^{-2})$ & ~~~~(dex)    &      & (mag) & (mag)  & (mag) & (mag) & (mag) & (mag) & (mag) &              &           	&           	&              \\
\hline

HD119856	&	8.21 & G1V	&	5894			&	4.58			&	-0.140			&	1,1,2,2,2	&	7.08	&	6.83	&	6.74	&	6.67	&	6.72	&	6.73	&	6.62	&	0.418	&	0.614	&	-0.367	&	1.519	\\
HD122973	&	8.08 &G0	&	5982			&	4.51			&	~0.120			&	1,1,2,2,2	&	7.01	&	6.75	&	6.67	&	6.51	&	6.58	&	6.62	&	6.49	&	1.067	&	1.625	&	-0.489	&	0.299	\\
HD123152	&	8.88 &G2/G3V	&	5670	$\pm$	30	&	4.31	$\pm$	0.06	&	-0.450	$\pm$	0.030	&	5,1,5,5,5	&	7.67	&	7.34	&	7.25	&	7.22	&	7.28	&	7.31	&	7.31	&	1.015	&	0.375	&	-0.728	&	-0.724	\\
HD125612	&	8.31&G3V	&	5874			&	4.51			&	~0.220			&	1,1,2,2,2	&	7.18	&	6.95	&	6.84	&	6.80	&	6.83	&	6.84	&	6.88	&	0.365	&	0.162	&	-0.159	&	0.400	\\
HD134664	&	7.76 &G2V	&	5884			&	4.43			&	~0.130			&	1,1,2,2,2	&	6.60	&	6.32	&	6.25	&	6.18	&	6.24	&	6.27	&	6.15	&	0.595	&	1.032	&	-0.523	&	1.324	\\
HD134902	&	8.85&G0	&	5853	$\pm$	57	&	4.51	$\pm$	0.08	&	~0.090	$\pm$	0.026	&	5,1,5,5,5	&	7.64	&	7.39	&	7.32	&	7.27	&	7.33	&	7.32	&	7.30	&	1.877	&	0.311	&	-0.471	&	0.506	\\
HD138159	&	9.16 &G2V	&	5775	$\pm$	25	&	4.56	$\pm$	0.05	&	-0.020	$\pm$	0.020	&	5,1,5,5,5	&	7.98	&	7.70	&	7.63	&	7.57	&	7.61	&	7.60	&	7.55	&	0.526	&	0.023	&	-0.114	&	0.382	\\
HD141937	&	7.25 &G2/G3V	&	5900	$\pm$	19	&	4.45	$\pm$	0.04	&	~0.125	$\pm$	0.030	&	5,1,5,5,5	&	6.13	&	5.87	&	5.76	&	5.77	&	5.68	&	5.80	&	5.73	&	-4.369	&	1.676	&	-0.267	&	1.314	\\
HD142072	&	7.85 &G5V	&	5761			&	4.38			&	~0.140			&	1,1,2,2,2	&	6.71	&	6.43	&	6.32	&	6.31	&	6.30	&	6.33	&	6.20	&	-0.075	&	0.891	&	-0.302	&	1.074	\\
HD144270	&	8.21 &F8	&	5923	$\pm$	67	&	4.57	$\pm$	0.08	&	~0.000	$\pm$	0.027	&	5,1,5,5,5	&	7.05	&	6.75	&	6.74	&	6.68	&	6.71	&	6.70	&	6.62	&	0.461	&	-0.215	&	-0.049	&	1.427	\\
HD145478	&	8.67&G0	&	5945	$\pm$	30	&	4.53	$\pm$	0.06	&	~0.110	$\pm$	0.030	&	5,1,5,5,5	&	7.50	&	7.26	&	7.17	&	7.16	&	7.18	&	7.18	&	7.13	&	0.819	&	-0.010	&	-0.191	&	0.920	\\
HD145518	&	7.41&G1/G2V	&	5900			&	4.49			&	-0.060			&	1,1,2,2,2	&	6.30	&	6.02	&	5.96	&	5.88	&	5.85	&	5.97	&	5.85	&	0.195	&	1.491	&	-0.405	&	0.889	\\
HD146070	&	7.53&G1V	&	5821			&	4.43			&	-0.090			&	1,1,2,2,2	&	6.41	&	6.11	&	6.04	&	6.01	&	6.01	&	6.07	&	5.95	&	-0.065	&	1.008	&	-0.408	&	1.232	\\
HD150027	&	8.99&G5/G6V	&	5640	$\pm$	30	&	4.21	$\pm$	0.06	&	-0.180	$\pm$	0.030	&	5,1,5,5,5	&	7.75	&	7.47	&	7.43	&	7.33	&	7.39	&	7.38	&	7.26	&	1.118	&	-0.087	&	-0.113	&	1.501	\\
HD152322	&	8.01 &G3V	&	5949			&	4.57			&	~0.020			&	1,1,2,2,2	&	6.90	&	6.65	&	6.58	&	6.52	&	6.57	&	6.58	&	6.57	&	0.148	&	0.372	&	-0.230	&	0.330	\\
HD154221	&	8.60&G3V	&	5880	$\pm$	30	&	4.45	$\pm$	0.06	&	~0.150	$\pm$	0.030	&	5,1,5,5,5	&	7.45	&	7.17	&	7.10	&	7.00	&	7.02	&	7.09	&	7.06	&	1.735	&	0.145	&	-0.290	&	-0.425	\\
HD155114	&	7.53&G3V	&	5791			&	4.56			&	-0.070			&	1,1,2,2,2	&	6.35	&	6.06	&	6.00	&	5.99	&	5.94	&	6.02	&	5.96	&	-0.273	&	1.365	&	-0.462	&	0.765	\\
HD155968	&	8.41&G5	&	5720			&	4.47			&	~0.160			&	1,1,2,2,2	&	7.23	&	6.95	&	6.87	&	6.86	&	6.86	&	6.88	&	6.87	&	0.111	&	0.189	&	-0.216	&	1.037	\\
HD156922	&	9.10&G3V	&	5700	$\pm$	30	&	4.42	$\pm$	0.06	&	-0.340	$\pm$	0.030	&	5,1,5,5,5	&	7.90	&	7.58	&	7.53	&	7.47	&	7.52	&	7.49	&	7.39	&	0.566	&	0.286	&	-0.252	&	0.581	\\
HD157691	&	8.37&G3/G5V	&	5730	$\pm$	30	&	4.43	$\pm$	0.06	&	-0.390	$\pm$	0.030	&	5,1,5,5,5	&	7.18	&	6.89	&	6.78	&	6.75	&	6.79	&	6.82	&	6.79	&	0.303	&	0.614	&	-0.497	&	1.033	\\
HD165357	&	8.89&G3V	&	5835	$\pm$	30	&	4.32	$\pm$	0.06	&	~0.070	$\pm$	0.030	&	5,1,5,5,5	&	7.72	&	7.48	&	7.33	&	7.31	&	7.37	&	7.36	&	7.20	&	1.298	&	-0.140	&	-0.214	&	1.472	\\
HD168746	&	7.95&G5	&	5572			&	4.44			&	-0.070			&	1,1,2,2,2	&	6.68	&	6.33	&	6.25	&	6.02	&	6.13	&	6.14	&	5.43	&	5.059	&	-0.050	&	-0.883	&	5.211	\\
HD171918	&	7.98&G0	&	5775	$\pm$	25	&	4.20	$\pm$	0.05	&	~0.206	$\pm$	0.030	&	5,1,5,5,5	&	6.76	&	6.45	&	6.38	&	6.30	&	6.34	&	6.39	&	6.29	&	1.357	&	1.688	&	-2.526	&	-0.439	\\
HD181296    &	5.03& A0Vn	&	9500			&	4.00			&	0.170		&	1,1,11,11,17	&	5.09	&	5.14	&	5.00	&	5.05	&	4.82	&	4.72	&	3.26	&	-2.521		&	-1.224		&	0.562	&	12.635	\\
HD183505	&	8.16 &G3IV/V	&	5716			&	4.52			&	~0.140			&	1,1,2,2,2	&	6.98	&	6.67	&	6.59	&	6.51	&	6.59	&	6.61	&	6.51	&	0.294	&	0.557	&	-0.532	&	1.106	\\
\hline

\end{tabular}  
\end{sidewaystable}

\begin{sidewaystable}
\centering
\label{TableXRay}

\center

\tiny
\begin{tabular}{lcclllccccccccrrrr}
\hline \hline
Star & V  &ST   & ~~T$_{eff}$   & ~~~log g       & ~~~[Fe/H] & Ref  & J     & H      & K     & [3.4] & [4.6] & [12]  & [22]  & $\chi_{3.4}$ & $\chi_{4.6}$	& $\chi_{12}$	&  $\chi_{22}$ \\
& (mag)  & &~~~(K)         & ~(cm s$^{-2})$ & ~~~~(dex)    &      & (mag) & (mag)  & (mag) & (mag) & (mag) & (mag) & (mag) &              &           	&           	&              \\
\hline

HD192417	&	8.20 &G3V	&	5745			&	4.56			&	~0.050			&	1,1,2,2,2	&	6.98	&	6.75	&	6.60	&	6.57	&	6.57	&	6.63	&	6.55	&	0.272	&	0.791	&	-0.428	&	0.575	\\
HD201422	&	8.54 &G5V	&	5836	$\pm$	48	&	4.50	$\pm$	0.06	&	-0.160	$\pm$	0.022	&	5,1,5,5,5	&	7.37	&	7.11	&	7.02	&	6.96	&	7.00	&	7.01	&	7.02	&	1.592	&	0.566	&	-0.588	&	0.082	\\
HD207043	&	7.59  &G5V	&	5731			&	4.46			&	~0.030			&	1,1,2,2,2	&	6.42	&	6.10	&	6.05	&	6.02	&	5.98	&	6.04	&	5.98	&	-0.202	&	1.354	&	-0.315	&	1.029	\\
HD209096	&	8.94&G5V	&	5875	$\pm$	51	&	4.51	$\pm$	0.07	&	~0.150	$\pm$	0.024	&	5,1,5,5,5	&	7.77	&	7.49	&	7.44	&	7.36	&	7.40	&	7.41	&	7.25	&	1.993	&	0.207	&	-0.518	&	1.244	\\
HD209262	&	8.00&G5	&	5760			&	4.40			&	~0.130			&	1,1,2,2,2	&	6.82	&	6.60	&	6.45	&	6.42	&	6.44	&	6.46	&	6.36	&	0.670	&	0.980	&	-0.497	&	1.259	\\
HD213199	&	8.16 &G2V	&	5908			&	4.41			&	-0.040			&	1,1,2,2,2	&	7.05	&	6.80	&	6.75	&	6.68	&	6.72	&	6.74	&	6.67	&	0.604	&	0.583	&	-0.286	&	1.211	\\
HD214954	&	8.26 &G3/G5IV	&	5727			&	4.50			&	~0.180			&	1,1,2,2,2	&	7.12	&	6.81	&	6.70	&	6.67	&	6.68	&	6.69	&	6.67	&	0.138	&	0.543	&	-0.264	&	0.505	\\
HD215657	&	7.22&G3IV-V	&	5989			&	4.42			&	~0.070			&	1,1,2,2,2	&	6.13	&	5.87	&	5.78	&	5.73	&	5.66	&	5.79	&	5.75	&	-0.645	&	0.656	&	-1.217	&	0.017	\\
HD218205	&	7.67&G2V	&	5945			&	4.55			&	~0.120			&	1,1,2,2,2	&	6.54	&	6.29	&	6.20	&	6.17	&	6.19	&	6.24	&	6.24	&	0.225	&	2.819	&	-0.612	&	0.388	\\
HD218396	&	5.96 &A5V	&	7430	$\pm$	75	&	4.35	$\pm$	0.05	&	-0.470	$\pm$	~0.100	&	5,1,5,5,5	&	5.38	&	5.28	&	5.24	&	5.19	&	5.05	&	5.21	&	4.85	&	0.251	&	2.123	&	0.675	&	4.803	\\

HD221343	&	8.37 &G2V	&	5822			&	4.58			&	~0.120			&	1,1,2,2,2	&	7.21	&	6.92	&	6.83	&	6.81	&	6.85	&	6.85	&	6.72	&	0.224	&	0.318	&	-0.371	&	1.217	\\
HD222669	&	7.68&G2V	&	5877			&	4.46			&	~0.090			&	1,1,2,2,2	&	6.58	&	6.26	&	6.24	&	6.21	&	6.18	&	6.21	&	6.15	&	-0.313	&	1.666	&	-0.432	&	0.741	\\
HD224448	&	9.02&G0	&	5905	$\pm$	44	&	4.55	$\pm$	0.07	&	-0.010	$\pm$	0.022	&	5,1,5,5,5	&	7.88	&	7.58	&	7.51	&	7.49	&	7.50	&	7.51	&	7.23	&	0.947	&	0.159	&	-0.206	&	2.023	\\
HD225299	&	8.13&G5V	&	5754			&	4.56			&	~0.230			&	1,1,2,2,2	&	6.88	&	6.65	&	6.50	&	6.51	&	6.55	&	6.55	&	6.56	&	-0.051	&	1.319	&	-0.441	&	0.732	\\

HD238838	&	8.84&G5	&	5796	$\pm$	73	&	4.48	$\pm$	0.12	&	-0.020	$\pm$	0.038	&	5,1,5,5,5	&	7.59	&	7.30	&	7.23	&	7.19	&	7.24	&	7.24	&	7.27	&	0.657	&	0.516	&	-0.338	&	-0.142	\\
HIP2894~~	&	8.64&G0	&	5820	$\pm$	44	&	4.54	$\pm$	0.07	&	-0.030	$\pm$	0.025	&	5,1,5,5,5	&	7.49	&	7.18	&	7.11	&	7.04	&	7.11	&	7.10	&	7.14	&	1.576	&	-0.016	&	-0.286	&	0.244	\\
HIP8841~~ 	&	9.23&G5	&	5676	$\pm$	45	&	4.50	$\pm$	0.06	&	-0.120	$\pm$	0.021	&	5,1,5,5,5	&	8.03	&	7.68	&	7.64	&	7.55	&	7.62	&	7.62	&	7.50	&	1.762	&	-0.375	&	-0.231	&	0.974	\\
HIP49572 	&	9.27&G0	&	5831	$\pm$	52	&	4.33	$\pm$	0.06	&	~0.010	$\pm$	0.021	&	5,1,5,5,5	&	8.06	&	7.80	&	7.73	&	7.65	&	7.68	&	7.69	&	7.60	&	1.688	&	-0.016	&	-0.436	&	0.569	\\

HIP78028	&	9.23&G5	&	5790	$\pm$	58	&	4.46	$\pm$	0.07	&	-0.020	$\pm$	0.025	&	5,1,5,5,5	&	8.03	&	7.73	&	7.67	&	7.60	&	7.64	&	7.63	&	7.97	&	0.595	&	-0.158	&	0.066	&	-0.699	\\
HIP81512	&	8.63&G5	&	5879	$\pm$	98	&	4.57	$\pm$	0.12	&	-0.030	$\pm$	0.041	&	5,1,5,5,5	&	7.44	&	7.16	&	7.09	&	7.07	&	7.12	&	7.11	&	7.12	&	0.326	&	0.265	&	-0.372	&	0.898	\\

\hline 
\multicolumn{18}{c}{Twins}\\  
\hline

HD3821	&	7.02 &G0 	&	5850	$\pm$	10	&	4.52	$\pm$	0.02	&	-0.087	$\pm$	0.008	&	6,1,6,6,6	&	5.91	&	5.60	&	5.52	&	5.53	&	5.29	&	5.55	&	5.470	&	0.082	&	1.889	&	-0.311	&	1.463	\\
HD6204	&	8.52 &G0	&	5854	$\pm$	10	&	4.50	$\pm$	0.02	&	~0.028	$\pm$	0.008	&	6,1,6,6,6	&	7.35	&	7.08	&	6.96	&	6.88	&	6.97	&	7.00	&	7.220	&	2.101	&	0.260	&	-0.282	&	-1.684	\\
HD8291	&	8.61&G5	&	5764	$\pm$	8	&	4.52	$\pm$	0.01	&	-0.068	$\pm$	0.007	&	6,1,6,6,6	&	7.42	&	7.14	&	7.07	&	6.98	&	7.02	&	7.04	&	6.920	&	0.954	&	-0.046	&	-0.231	&	1.251	\\
HD11195	&	8.89&G5V	&	5725	$\pm$	6	&	4.49	$\pm$	0.02	&	-0.096	$\pm$	0.006	&	6,1,6,6,6	&	7.67	&	7.37	&	7.27	&	7.19	&	7.22	&	7.23	&	7.110	&	2.552	&	-0.194	&	-0.239	&	1.154	\\
HD13357	&	7.63&G5+	&	5738	$\pm$	7	&	4.51	$\pm$	0.01	&	-0.007	$\pm$	0.005	&	6,1,6,6,6	&	6.91	&	6.66	&	6.58	&	6.56	&	6.58	&	6.63	&	6.510	&	0.253	&	0.700	&	-0.420	&	0.689	\\
\hline

\end{tabular} 

\end{sidewaystable} 


\begin{sidewaystable}
\label{TableXRay}

\center

\tiny

\begin{tabular}{lcclllccccccccrrrr}
\hline \hline
Star & V &ST    & ~~T$_{eff}$   & ~~~log g       & ~~~[Fe/H] & Ref  & J     & H      & K     & [3.4] & [4.6] & [12]  & [22]  & $\chi_{3.4}$ & $\chi_{4.6}$	& $\chi_{12}$	&  $\chi_{22}$ \\
& (mag)  & &~~~(K)         & ~(cm s$^{-2})$ & ~~~~(dex)    &      & (mag) & (mag)  & (mag) & (mag) & (mag) & (mag) & (mag) &              &           	&           	&              \\
\hline

HD26990	&	7.50&G0	&	5764	$\pm$	12	&	4.47	$\pm$	0.04	&	-0.070	$\pm$	0.011	&	6,1,6,6,6	&	6.25	&	5.95	&	5.85	&	5.84	&	5.76	&	5.87	&	5.760	&	-0.027	&	1.442	&	-0.382	&	0.620	\\

HD28904	&	8.26&G5V	&	5846	$\pm$	11	&	4.50	$\pm$	0.03	&	-0.070	$\pm$	0.008	&	6,1,6,6,6	&	7.00	&	6.69	&	6.58	&	6.48	&	6.51	&	6.50	&	6.410	&	0.067	&	0.458	&	-0.196	&	1.382	\\
HD30774	&	7.88&G3/G5V	&	5789	$\pm$	8	&	4.43	$\pm$	0.02	&	~0.084	$\pm$	0.008	&	6,1,6,6,6	&	6.58	&	6.23	&	6.13	&	6.04	&	6.02	&	6.09	&	6.000	&	0.861	&	0.056	&	-0.236	&	1.456	\\
HD35769	&	8.67&G5	&	5631			&	4.34			&	-0.130			&	1,1,2,2,2	&	7.41	&	7.08	&	7.04	&	7.03	&	7.05	&	7.05	&	7.150	&	0.225	&	2.054	&	-0.613	&	1.118	\\
HD36152	&	8.27&G5V	&	5771	$\pm$	5	&	4.44	$\pm$	0.02	&	~0.057	$\pm$	0.005	&	6,1,6,6,6	&	7.08	&	6.87	&	6.73	&	6.80	&	6.82	&	6.81	&	6.830	&	0.659	&	-0.022	&	-0.307	&	1.397	\\
HD39649	& 8.52&G5V	&		5740			&	4.40			&	~0.010			&	3,19,2,2,2	&	7.37	&	7.05	&	6.99	&	6.93	&	6.98	&	6.98	&	6.830	&	0.497	&	-0.111	&	-0.197	&	1.497	\\
HD41708	&	8.02&G0V	&	5928			&	4.45			&	~0.080			&	1,1,2,2,2	&	6.88	&	6.63	&	6.58	&	6.54	&	6.53	&	6.58	&	6.440	&	-0.148	&	0.682	&	-0.938	&	1.430	\\
HD44821	&	7.37 &K0/1V(+G)	&	5750	$\pm$	9	&	4.50	$\pm$	0.02	&	~0.063	$\pm$	0.007	&	6,1,6,6,6	&	6.18	&	5.91	&	5.81	&	5.72	&	5.68	&	5.78	&	5.700	&	0.161	&	1.730	&	-0.230	&	1.364	\\
HD45021	&	9.16&G5V	&	5668	$\pm$	5	&	4.42	$\pm$	0.01	&	-0.011	$\pm$	0.004	&	6,1,6,6,6	&	7.95	&	7.68	&	7.59	&	7.48	&	7.54	&	7.53	&	7.570	&	-0.845	&	-0.102	&	0.035	&	0.276	\\
HD45346	&	8.66&G5V	&	5721	$\pm$	6	&	4.41	$\pm$	0.02	&	-0.076	$\pm$	0.006	&	6,1,6,6,6	&	7.47	&	7.14	&	7.07	&	7.04	&	7.07	&	7.08	&	7.110	&	0.341	&	0.127	&	-0.223	&	0.446	\\
HD59711	&	7.73&G5V	&	5737	$\pm$	4	&	4.41	$\pm$	0.01	&	-0.117	$\pm$	0.004	&	6,1,6,6,6	&	6.52	&	6.21	&	6.15	&	6.15	&	6.13	&	6.18	&	6.160	&	-0.136	&	1.798	&	-0.371	&	0.876	\\
HD59967	&	6.66&G3V	&	5847	$\pm$	12	&	4.54	$\pm$	0.02	&	-0.021	$\pm$	0.009	&	6,1,6,6,6	&	5.53	&	5.25	&	5.10	&	5.05	&	4.84	&	5.10	&	4.940	&	0.124	&	1.952	&	-0.418	&	2.033	\\
HD63487	&	9.20&G2V	&	5849	$\pm$	8	&	4.49	$\pm$	0.02	&	~0.058	$\pm$	0.007	&	6,1,6,6,6	&	8.03	&	7.75	&	7.67	&	7.63	&	7.68	&	7.63	&	7.230	&	0.796	&	0.260	&	-0.426	&	1.092	\\
HD67010	&	8.51&G5	&	5613			&	4.48			&	-0.130			&	1,1,2,2,2	&	7.30	&	6.96	&	6.86	&	6.79	&	6.85	&	6.87	&	6.870	&	0.771	&	0.238	&	-0.364	&	0.447	\\
HD75302	&	7.45&G0	&	5702	$\pm$	5	&	4.46	$\pm$	0.01	&	~0.083	$\pm$	0.006	&	6,1,6,6,6	&	6.24	&	5.95	&	5.84	&	5.89	&	5.74	&	5.81	&	5.810	&	-2.542	&	-0.173	&	1.322	&	-0.256	\\

HD78660	&	8.34&G5	&	5776			&	4.47	$\pm$	0.02	&	-0.023	$\pm$	0.005	&	6,1,6,6,6	&	7.11	&	6.89	&	6.76	&	6.74	&	6.79	&	6.78	&	6.650	&	0.266	&	0.158	&	-0.234	&	1.388	\\

HD96116	&	8.65 &G3V	&	5820	$\pm$	9	&	4.51	$\pm$	0.02	&	-0.014	$\pm$	0.007	&	6,1,6,6,6	&	7.50	&	7.20	&	7.08	&	7.07	&	7.11	&	7.11	&	7.320	&	1.913	&	-0.191	&	-0.086	&	0.196	\\
HD96423	&	7.23 &G5V	&	5727	$\pm$	4	&	4.36	$\pm$	0.01	&	~0.118	$\pm$	0.004	&	6,1,6,6,6	&	6.02	&	5.74	&	5.63	&	5.70	&	5.49	&	5.65	&	5.610	&	-2.888	&	3.034	&	-0.548	&	0.471	\\
HD97356	&	8.15&G5	&	5805			&	4.35			&	~0.020			&	1,1,2,2,2	&	6.97	&	6.70	&	6.62	&	6.55	&	6.61	&	6.64	&	6.570	&	0.561	&	1.178	&	-0.535	&	1.005	\\

HD98618	&	8.15&G5V	&	5805			&	4.35			&	~0.020			&	1,1,2,2,2	&	6.97	&	6.70	&	6.62	&	6.55	&	6.61	&	6.64	&	6.570	&	0.561	&	1.178	&	-0.535	&	1.005	\\

HD115169	&	9.26&G3V	&	5767	$\pm$	8	&	4.46	$\pm$	0.02	&	-0.067	$\pm$	0.007	&	6,1,6,6,6	&	8.09	&	7.77	&	7.71	&	7.65	&	7.70	&	7.69	&	7.640	&	1.143	&	0.347	&	-0.510	&	0.342	\\

HD122194	&	9.39&G3V	&	5845	$\pm$	6	&	4.37	$\pm$	0.02	&	~0.054	$\pm$	0.005	&	6,1,6,6,6	&	8.16	&	7.93	&	7.77	&	7.74	&	7.80	&	7.78	&	7.780	&	1.016	&	0.132	&	-0.234	&	1.088	\\
HD129814	&	7.52&G5V	&	5842	$\pm$	8	&	4.35	$\pm$	0.02	&	-0.034	$\pm$	0.007	&	6,1,6,6,6	&	6.33	&	6.07	&	6.00	&	5.93	&	5.90	&	5.98	&	5.910	&	0.234	&	1.221	&	-0.398	&	1.095	\\
HD134664	&	7.76&G2V	&	5844	$\pm$	5	&	4.49	$\pm$	0.01	&	~0.077	$\pm$	0.004	&	6,1,6,6,6	&	6.60	&	6.32	&	6.25	&	6.18	&	6.24	&	6.27	&	6.150	&	0.595	&	1.032	&	-0.523	&	1.324	\\
HD138573	&	7.22&G5IV-V	&	5777			&	4.46			&	-0.037	$\pm$	0.006	&	6,1,6,6,6	&	6.03	&	5.74	&	5.66	&	5.61	&	5.53	&	5.68	&	5.620	&	0.119	&	2.563	&	-0.680	&	0.960	\\
HD140538	&	5.86&G5V	&	5683	$\pm$	5	&	4.48	$\pm$	0.02	&	~0.036	$\pm$	0.006	&	6,1,6,6,6	&	4.59	&	4.05	&	4.30	&	4.13	&	3.55	&	4.17	&	4.140	&	-0.129	&	1.311	&	-0.450	&	0.597	\\
\hline
\end{tabular}
\end{sidewaystable}

\begin{sidewaystable}
\label{TableXRay}

\center

\tiny

\begin{tabular}{lcclllccccccccrrrr}
\hline \hline
Star & V  &ST   & ~~T$_{eff}$   & ~~~log g       & ~~~[Fe/H] & Ref  & J     & H      & K     & [3.4] & [4.6] & [12]  & [22]  & $\chi_{3.4}$ & $\chi_{4.6}$	& $\chi_{12}$	&  $\chi_{22}$ \\
& (mag)  & & ~~~(K)         & ~(cm s$^{-2})$ & ~~~~(dex)    &      & (mag) & (mag)  & (mag) & (mag) & (mag) & (mag) & (mag) &              &           	&           	&              \\
\hline

HD142331	&	8.72&G0	&	5690	$\pm$	6	&	4.40	$\pm$	0.02	&	-0.006	$\pm$	0.006	&	6,1,6,6,6	&	7.48	&	7.18	&	7.13	&	7.09	&	7.12	&	7.13	&	7.130	&	0.976	&	-0.026	&	-0.096	&	0.451	\\
HD142415	&	7.33&G1V	&	5904			&	4.40			&	~0.080			&	1,1,2,2,2	&	6.24	&	5.99	&	5.89	&	5.83	&	5.80	&	5.90	&	5.840	&	-0.519	&	0.419	&	-1.382	&	-0.128	\\
HD145927	&	8.35&G2III	&	5803	$\pm$	6	&	4.38	$\pm$	0.02	&	-0.041	$\pm$	0.005	&	6,1,6,6,6	&	7.23	&	6.87	&	6.68	&	6.48	&	6.59	&	6.56	&	6.030	&	1.927	&	-0.164	&	0.228	&	2.148	\\
HD147513	&	5.37&G3/G5V	&	5885			&	4.53			&	~0.040			&	1,1,2,2,2	&	4.41	&	4.03	&	3.93	&	3.92	&	3.38	&	3.90	&	3.830	&	-0.076	&	1.138	&	-0.568	&	0.970	\\
HD150248	&	7.03&G3V	&	5715	$\pm$	5	&	4.40	$\pm$	0.02	&	-0.086	$\pm$	0.004	&	6,1,6,6,6	&	5.85	&	5.53	&	5.45	&	5.42	&	5.21	&	5.38	&	5.040	&	1.152	&	4.074	&	0.729	&	0.373	\\
HD157347	&	6.28&G5IV	&	5694	$\pm$	5	&	4.41	$\pm$	0.02	&	~0.015	$\pm$	0.004	&	6,1,6,6,6	&	5.13	&	4.80	&	4.69	&	4.68	&	4.24	&	4.71	&	4.660	&	-0.558	&	0.964	&	-2.846	&	-0.851	\\
HD163441	&	8.43&G5	&	5795			&	4.43			&	~0.041	$\pm$	0.006	&	6,1,6,6,6	&	7.17	&	6.83	&	6.76	&	6.66	&	6.74	&	6.73	&	6.590	&	0.312	&	0.331	&	-0.351	&	1.782	\\
HD167060	&	8.93&G3V	&	5841	$\pm$	5	&	4.44	$\pm$	0.02	&	-0.037	$\pm$	0.005	&	6,1,6,6,6	&	7.78	&	7.51	&	7.43	&	7.38	&	7.44	&	7.43	&	7.530	&	0.742	&	-0.023	&	-0.124	&	-0.345	\\
HD173071	&	8.19&G0	&	6044			&	4.49			&		0.250		&	1,1,2,2,2	&	7.13	&	6.87	&	6.77	&	6.74	&	6.78	&	6.82	&	6.920	&	0.050	&	0.390	&	-0.445	&	0.492	\\
HD183579	&	8.67&G5V	&	5781	$\pm$	8	&	4.50	$\pm$	0.02	&	-0.053	$\pm$	0.007	&	6,1,6,6,6	&	7.52	&	7.23	&	7.15	&	7.09	&	7.13	&	7.11	&	7.060	&	0.800	&	0.219	&	-0.244	&	0.905	\\
HD196390	&	7.33&G3V	&	5890	$\pm$	6	&	4.47	$\pm$	0.02	&	~0.057	$\pm$	0.006	&	6,1,6,6,6	&	6.21	&	5.95	&	5.87	&	5.81	&	5.78	&	5.87	&	5.790	&	-0.034	&	1.371	&	-0.270	&	1.168	\\
HD197027	&	9.18&G3V	&	5718	$\pm$	5	&	4.40	$\pm$	0.02	&	-0.020	$\pm$	0.005	&	6,1,6,6,6	&	7.96	&	7.68	&	7.59	&	7.55	&	7.59	&	7.61	&	7.920	&	1.185	&	-0.040	&	-0.247	&	-1.005	\\
HD208704	&	7.16&G5/G6V	&	5829	$\pm$	7	&	4.33	$\pm$	0.02	&	-0.111	$\pm$	0.006	&	6,1,6,6,6	&	6.03	&	5.72	&	5.66	&	5.59	&	5.58	&	5.67	&	5.620	&	-0.101	&	1.552	&	-0.546	&	0.850	\\
HD209562	&	8.88&G3V	&	5847	$\pm$	17	&	4.53	$\pm$	0.03	&	~0.064	$\pm$	0.013	&	6,1,6,6,6	&	7.70	&	7.42	&	7.36	&	7.30	&	7.34	&	7.33	&	7.520	&	2.127	&	-0.064	&	-0.316	&	-0.256	\\
HD209779	&	7.57&G0	&	5787	$\pm$	17	&	4.50	$\pm$	0.04	&	~0.035	$\pm$	0.014	&	6,1,6,6,6	&	6.33	&	6.00	&	5.88	&	5.87	&	5.74	&	5.86	&	5.810	&	-0.807	&	1.620	&	-0.419	&	0.854	\\
HD219057	&	9.60&G3V	&	5816	$\pm$	9	&	4.52	$\pm$	0.02	&	-0.077	$\pm$	0.008	&	6,1,6,6,6	&	8.46	&	8.19	&	8.09	&	8.03	&	8.07	&	8.08	&	8.700	&	0.964	&	-0.017	&	-0.176	&	-1.446	\\
HD220507	&	7.59&G5V	&	5699	$\pm$	9	&	4.25	$\pm$	0.03	&	~0.036	$\pm$	0.008	&	6,1,6,6,6	&	6.38	&	6.10	&	5.97	&	5.88	&	5.90	&	6.00	&	5.970	&	0.803	&	1.427	&	-0.502	&	0.514	\\
HD222582	&	7.68&G5	&	5792	$\pm$	6	&	4.37	$\pm$	0.02	&	~0.010	$\pm$	0.005	&	6,1,6,6,6	&	6.52	&	6.24	&	6.17	&	6.08	&	6.09	&	6.16	&	6.000	&	0.238	&	2.667	&	-0.416	&	1.158	\\

Inti 1	&	12.86 &$\cdots$	&	5837 $\pm$ 11	&	 4.42 $\pm$ 003	&	0.07 $\pm$ 0.01 	&	18,7,7,7	&	11.56	&	11.25	&	11.17	&	11.14	&	11.20	&	11.00	&	$\cdots$	&	1.850	&	0.955	&	0.131	&	$\cdots$	\\

MMJ5484	&	14.54&$\cdots$	&	5768 $\pm$ 70	&	$\cdots$ 	&	$\cdots$ 	&	7,7	&	13.35	&	13.06	&	12.93	&	12.92	&	12.94	&	11.68	&	8.78	&	-3.267	&	-4.639&		$<$ 10.992	&	$<$ 15.309	\\
MMJ6055	&	14.56&$\cdots$	&	5693 $\pm$ 74	&	$\cdots$ 	&	$\cdots$ 	&	7,7	&	13.34	&	13.02	&	12.95	&	12.93	&	12.98	&	12.05	&	9.01	&	-3.435	&	-3.455&		$<$ 12.112	&	$<$ 15.277	\\
MMJ6384	&	14.65&$\cdots$	&	5756 $\pm$ 60	&	$\cdots$ 	&	$\cdots$ 	&	7,7	&	13.41	&	13.16	&	13.05	&	12.92	&	12.97	&	12.03	&	8.18	&	0.165	&	-0.084	&	$<$ 6.165	&	$<$ 15.443	\\

S770	&	14.61&$\cdots$	&	5766 $\pm$ 64	&	$\cdots$	&$\cdots$&	7,7	&	13.37	&	13.04	&	12.97	&	12.97	&	13.03	&	11.87	&	8.53	&	0.697	&	0.913	&	$<$ 12.954	&	$<$ 15.383	\\
S779	&	14.64&$\cdots$	&	5716 $\pm$ 64	&	$\cdots$ 	&	$\cdots$ 	&	7,7	&	13.40	&	13.04	&	12.96	&	13.00	&	13.01	&	11.72	&	8.67	&	1.929	&	0.967	&	$<$ 11.096	&	$<$ 15.430	\\
\hline

\end{tabular}
\end{sidewaystable}

\begin{sidewaystable}
\label{TableXRay}

\center

\tiny

\begin{tabular}{lcclllccccccccrrrr}
\hline \hline
Star & V &ST    & ~~T$_{eff}$   & ~~~log g       & ~~~[Fe/H] & Ref  & J     & H      & K     & [3.4] & [4.6] & [12]  & [22]  & $\chi_{3.4}$ & $\chi_{4.6}$	& $\chi_{12}$	&  $\chi_{22}$ \\
& (mag)  & &~~~(K)         & ~(cm s$^{-2})$ & ~~~~(dex)    &      & (mag) & (mag)  & (mag) & (mag) & (mag) & (mag) & (mag) &              &           	&           	&              \\
\hline

S785	&	14.81&$\cdots$&	5716 $\pm$ 63	&	$\cdots$ 	&	$\cdots$ 	&	7,7	&	13.53	&	13.20	&	13.07	&	13.11	&	13.15	&	12.25	&	8.93	&	7.849	&	6.021	&	$<$ 11.252	&	$<$ 13.939	\\
S945	&	14.46&$\cdots$	&	5836  $\pm$ 67	&	$\cdots$ 	&	$\cdots$ 	&	7,7	&	13.31	&	12.95	&	12.94	&	12.90	&	12.92	&	11.82	&	8.59	&	1.519	&	1.218	&	$<$ 12.003	&	$<$ 15.318	\\
S966	&	14.49&$\cdots$	&	5806 $\pm$ 65	&	$\cdots$ 	&	$\cdots$ 	&	7,7	&	13.27	&	13.00	&	12.95	&	12.86	&	12.90	&	11.37	&	8.22	&	2.791	&	1.296	&	3.831	&	$<$ 4.462	\\
S1041	&	14.73&$\cdots$	&	5704  $\pm$ 64	&	$\cdots$ 	&	$\cdots$ 	&	7,7	&	13.34	&	12.99	&	12.90	&	12.86	&	12.96	&	12.43	&	8.84	&	8.279	&	7.411	&	$<$ 7.494	&	$<$ 15.561	\\
S1462	&	14.29&	$\cdots$&	5874 $\pm$ 58	&	$\cdots$ 	&	$\cdots$ 	&	7,7	&	13.03	&	12.67	&	12.60	&	12.58	&	12.61	&	12.38	&	8.85	&	-1.969	&	-2.447	&	$<$ 9.416	&	$<$ 15.135	\\

\hline 
\multicolumn{18}{c}{Sibling candidates}\\  
\hline
  
HD10157	&	8.83&G0	&	6046	&	4.59	&	~0.110	&	8,1,8,8,8	&	7.74	&	7.47	&	7.43	&	7.35	&	7.42	&	7.41	&	7.55	&	1.194	&	0.042	&	-0.116	&	-0.814	\\
HD11124	&	8.76&F6V	&	6111	&	4.43	&	-0.100	&	8,1,8,8,8	&	7.82	&	7.57	&	7.50	&	7.38	&	7.48	&	7.47	&	7.32	&	1.764	&	0.240	&	-0.380	&	1.325	\\

HD14706	&	8.71&G0V	&	5995	&	4.45	&	-0.070	&	8,1,8,8,8	&	7.60	&	7.33	&	7.20	&	7.07	&	7.16	&	7.16	&	7.03	&	1.887	&	0.197	&	-0.216	&	1.577	\\
HD19573	&	7.62& G0	&	6461	&	4.13	&	-0.090	&	8,1,8,8,8	&	6.72	&	6.52	&	6.48	&	6.47	&	6.39	&	6.48	&	6.35	&	-0.067	&	1.565	&	-0.479	&	1.419	\\
HD21216	&	8.44&F8	&	6184	&	4.39	&	-0.190	&	8,1,8,8,8	&	7.53	&	7.33	&	7.30	&	7.15	&	7.23	&	7.23	&	6.63	&	1.708	&	-0.272	&	-0.353	&	5.000	\\
HD30059	&	9.31&G8/KV&	5481	&	4.54	&	~0.160	&	8,1,8,8,8	&	8.00	&	7.67	&	7.57	&	7.53	&	7.60	&	7.58	&	7.61	&	0.831	&	0.408	&	-0.324	&	-0.033	\\
HD33503	&	7.64&F8III	&	6531	&	4.20	&	~0.030	&	8,1,8,8,8	&	6.78	&	6.56	&	6.52	&	6.53	&	6.47	&	6.55	&	6.47	&	-0.161	&	1.186	&	-0.460	&	0.943	\\
HD35874	&	8.19&F6V	&	6526	&	4.23	&	~0.240	&	8,1,8,8,8	&	7.25	&	7.12	&	7.02	&	6.97	&	6.99	&	7.00	&	6.96	&	0.673	&	0.016	&	-0.157	&	0.865	\\
HD37574	&	6.74&F8	&	6344	&	3.90	&	-0.050	&	8,1,8,8,8	&	5.79	&	5.59	&	5.54	&	5.55	&	5.47	&	5.66	&	5.55	&	0.141	&	1.660	&	-1.074	&	1.268	\\
HD44821	&	7.37&K0/1V(+G)	&	5744	&	4.59	&	~0.070	&	8,1,8,8,8	&	6.18	&	5.91	&	5.81	&	5.76	&	5.67	&	5.78	&	5.69	&	0.161	&	1.730	&	-0.230	&	1.364	\\
HD50867	&	7.60&F8	&	6218	&	4.32	&	-0.090	&	8,1,8,8,8	&	6.65	&	6.45	&	6.33	&	6.35	&	6.28	&	6.36	&	6.33	&	-0.213	&	1.251	&	-0.415	&	0.363	\\
HD52242	&	7.39&F2/F3V	&	6752	&	4.23	&	-0.060	&	8,1,8,8,8	&	6.61	&	6.46	&	6.39	&	6.38	&	6.32	&	6.39	&	6.29	&	-0.119	&	1.129	&	-0.398	&	1.142	\\
HD84843	&	8.58&F8	&	6264	&	4.34	&	-0.160	&	8,1,8,8,8	&	7.62	&	7.48	&	7.39	&	7.25	&	7.36	&	7.35	&	7.32	&	2.316	&	-0.351	&	-0.046	&	0.519	\\
HD91320	&	8.43&G1V	&	5981	&	4.13	&	-0.050	&	8,1,8,8,8	&	7.34	&	7.12	&	7.01	&	6.89	&	7.00	&	7.02	&	7.00	&	1.773	&	0.280	&	-0.342	&	0.337	\\
HD95915	&	7.25&F6V	&	6305	&	4.04	&	-0.140	&	8,1,8,8,8	&	6.32	&	6.06	&	6.00	&	6.06	&	5.93	&	6.02	&	5.94	&	-1.194	&	1.565	&	-0.515	&	0.970	\\
HD101197	&	8.73&G5	&	5672	&	4.19	&	~0.000	&	8,1,8,8,8	&	7.57	&	7.30	&	7.21	&	7.02	&	7.20	&	7.18	&	7.19	&	2.941	&	0.541	&	-0.567	&	0.083	\\
HD105000	&	7.91&F2	&	6509	&	4.32	&	~0.010	&	8,1,8,8,8	&	7.09	&	6.91	&	6.82	&	6.79	&	6.79	&	6.84	&	6.89	&	0.343	&	0.757	&	-0.388	&	-0.297	\\
HD105678	&	6.34&F6IV	&	6166	&	3.66	&	-0.100	&	8,1,8,8,8	&	5.42	&	5.17	&	5.10	&	5.09	&	4.89	&	5.12	&	5.07	&	0.117	&	2.872	&	-0.331	&	0.782	\\
HD108271	&	8.31&F8	&	6101	&	4.03	&	-0.240	&	8,1,8,8,8	&	7.30	&	7.12	&	7.05	&	6.97	&	7.05	&	7.04	&	7.17	&	1.164	&	-0.297	&	-0.015	&	-0.829	\\
HD133815	&	9.06&G5	&	5985	&	4.43	&	~0.100	&	8,1,8,8,8	&	7.94	&	7.70	&	7.64	&	7.53	&	7.63	&	7.60	&	7.67	&	1.629	&	-0.150	&	-0.272	&	-0.349	\\
\hline

\end{tabular}
\end{sidewaystable}

\begin{sidewaystable}
\label{TableXRay}

\center

\tiny

\begin{tabular}{lcclllccccccccrrrr}
\hline \hline
Star & V &ST    & ~~T$_{eff}$   & ~~~log g       & ~~~[Fe/H] & Ref  & J     & H      & K     & [3.4] & [4.6] & [12]  & [22]  & $\chi_{3.4}$ & $\chi_{4.6}$	& $\chi_{12}$	&  $\chi_{22}$ \\
& (mag)  & &~~~(K)         & ~(cm s$^{-2})$ & ~~~~(dex)    &      & (mag) & (mag)  & (mag) & (mag) & (mag) & (mag) & (mag) &              &           	&           	&              \\
\hline

HD168325	&	9.66&F8	&	3925	&	5.37	&	-0.030	&	8,1,8,8,8	&	7.08	&	6.46	&	6.28	&	6.20	&	6.16	&	6.20	&	6.13	&	-0.042	&	1.826	&	-1.128	&	0.506	\\
HD168442	&	7.49&K7V	&	6809	&	4.12	&	~0.140	&	8,1,8,8,8	&	6.68	&	6.53	&	6.44	&	6.42	&	6.42	&	6.48	&	6.61	&	0.526	&	0.980	&	-0.273	&	-1.243	\\
HD176118	&	8.24&F8	&	6321	&	4.19	&	-0.350	&	8,1,8,8,8	&	7.42	&	7.28	&	7.22	&	7.19	&	7.20	&	7.21	&	7.20	&	0.222	&	-0.179	&	-0.071	&	0.393	\\
HD193549	&	6.46&F2IV/V	&	4828	&	3.23	&	-0.020	&	8,1,8,8,8	&	4.98	&	4.27	&	4.05	&	4.05	&	3.79	&	4.05	&	4.01	&	0.499	&	1.187	&	-0.050	&	0.060	\\
HD196676	&	7.54&K0	&	6669	&	4.18	&	~0.010	&	8,1,8,8,8	&	6.75	&	6.63	&	6.56	&	6.51	&	6.47	&	6.55	&	6.49	&	0.312	&	1.411	&	-0.419	&	0.879	\\
HD207164	&	8.04&F2	&	5808	&	4.12	&	~0.080	&	8,1,8,8,8	&	6.87	&	6.61	&	6.53	&	6.46	&	6.48	&	6.54	&	6.45	&	0.775	&	1.063	&	-0.460	&	1.140	\\
HD219828	&	9.12&G0IV	&	5934	&	4.65	&	-0.040	&	8,1,8,8,8	&	7.90	&	7.68	&	7.58	&	7.48	&	7.52	&	7.52	&	7.48	&	1.167	&	0.151	&	-0.135	&	0.659	\\
HIP76300	&	9.00&G5	&	6000	&	4.34	&	-0.110	&	8,1,8,8,8	&	7.80	&	7.52	&	7.46	&	7.33	&	7.33	&	7.34	&	6.74	&	1.139	&	0.547	&	-0.035	&	4.493	\\
HIP112584	&	9.93&G0	&	5685	&	4.65	&	0.180	&	8,1,8,8,8	&	8.67	&	8.38	&	8.31	&	8.24	&	8.30	&	8.27	&	8.20	&	0.597	&	0.185	&	-0.381	&	0.532	\\

\hline 
\multicolumn{18}{c}{Siblings}\\  
\hline

HD68314	&	8.76& A2/5II/III	&	6111	&	4.43	&	-0.100	&	1,19,8,8,8	&	7.82	&	7.57	&	7.50	&	7.38	&	7.48	&	7.47	&	7.32	&	1.764	&	0.240	&	-0.380	&	1.325	\\  
HD28676	&	8.71 &F5	&	5995	&	4.45	&	-0.070	&	1,1,16,16,16	&	7.60	&	7.33	&	7.20	&	7.07	&	7.16	&	7.16	&	7.03	&	1.887	&	0.197	&	-0.216	&	1.577	\\
HD83423	&	7.62&F8V	&	6461	&	4.13	&	-0.090	&	1,1,16,16,16	&	6.72	&	6.52	&	6.48	&	6.47	&	6.39	&	6.48	&	6.35	&	-0.067	&	1.565	&	-0.479	&	1.419	\\
HD162826	&	8.83 & F8V	&	6046	&	4.59	&	0.110	&	1,1,6,6,6	&	7.74	&	7.47	&	7.43	&	7.35	&	7.42	&	7.41	&	7.55	&	1.194	&	0.042	&	-0.116	&	-0.814	\\
HD175840	&	9.31&K2	&	5481	&	4.54	&	0.160	&	1,1,16,16,16	&	8.00	&	7.67	&	7.57	&	7.53	&	7.60	&	7.58	&	7.61	&	0.831	&	0.408	&	-0.324	&	-0.033	\\
HD186302	&	7.64&G3/G5V	&	6531	&	4.20	&	0.030	&	1,16,16,16	&	6.78	&	6.56	&	6.52	&	6.53	&	6.47	&	6.55	&	6.47	&	-0.161	&	1.186	&	-0.460	&	0.943	\\

\hline


\end{tabular}

\begin{tablenotes}
 Notes. The references to the V magnitude, spectral type, effective temperature, surface gravity and metallicity are indicated by the numbers in column 7: (1) Hipparcos (ESA 1997); (2)
 Datson et al. (2015); (3) Hog et al. (2000); (4) Houk \& Swift (1999) (5) Ramírez et al. (2012); (6) Ramirez et al. (2014); (7) Pasquini et al. (2008); (8) Liu et al. (2015); (9) Ducati (2002);  (10) Gray et al. (2006); (11) Chen et al. (2014); (12) Stauffer et al (1995); (13) Chen et al. (2011); (14)  Nordstr{\"o}m et al. (2004); (15) Chen et al. (2006); (16) Batista et al. (2014); (17) Saffe et al. (2008); (18) Henden et al (2016); (19) Houk \& Smith-Moore (1998).

 \end{tablenotes}
 \end{sidewaystable}

\end{document}